\documentclass{article}

\usepackage{PRIMEarxiv}
\usepackage[utf8]{inputenc} 
\usepackage[T1]{fontenc}    
\usepackage{hyperref}       
\usepackage{url}            
\usepackage{booktabs}       
\usepackage{amsfonts}       
\usepackage{nicefrac}       
\usepackage{microtype}      
\usepackage{lipsum}
\usepackage{fancyhdr}       
\usepackage{graphicx}       
\usepackage{bm}
\usepackage{makecell}
\usepackage{amsmath}
\usepackage{multirow}
\usepackage{leftindex}
\usepackage{xcolor}
\usepackage{mathrsfs}
\usepackage[lined,boxed]{algorithm2e}
\usepackage{amsthm}
\usepackage{amssymb}
\graphicspath{{media/}}     
\newtheorem{remark}{Remark}
\newtheorem{problem}{Problem}
\newtheorem{difference}{Diff.}
\usepackage[]{hyperref}

\title{Finite-PINN: A Physics-Informed Neural Network with Finite Geometric Encoding for Solid Mechanics
}

\author{
  Haolin Li \\
  Department of Aeronautics \\
  Imperial College London \\
  London\\
  \texttt{haolin.li20@imperial.ac.uk} \\
  \texttt{feliz19981004@gmail.com} \\
   \And
  Yuyang Miao \\
  Department of Electrical and Electronic Engineering \\
  Imperial College London \\
  London\\
  \texttt{yuyang.miao20@imperial.ac.uk} \\
  \AND
  Zahra Sharif Khodaei \\
  Department of Aeronautics \\
  Imperial College London \\
  London\\
  \texttt{z.sharif-khodaei@imperial.ac.uk} \\
  \And
  M.H. Aliabadi \\
  Department of Aeronautics \\
  Imperial College London \\
  London\\
  \texttt{m.h.aliabadi@imperial.ac.uk} \\
}

\begin{document}
\maketitle

\begin{abstract}
PINN models have demonstrated capabilities in addressing fluid PDE problems, and their potential in solid mechanics is beginning to emerge. This study identifies two key challenges when using PINN to solve general solid mechanics problems. These challenges become evident when comparing the limitations of PINN with the well-established numerical methods commonly used in solid mechanics, such as the finite element method (FEM). Specifically: a) PINN models generate solutions over an infinite domain, which conflicts with the finite boundaries typical of most solid structures; and b) the solution space utilised by PINN is Euclidean, which is inadequate for addressing the complex geometries often present in solid structures.

This work presents a PINN architecture for general solid mechanics problems, referred to as the Finite-PINN model. The model is designed to effectively tackle two key challenges, while retaining as much of the original PINN framework as possible. To this end, the Finite-PINN incorporates finite geometric encoding into the neural network inputs, thereby transforming the solution space from a conventional Euclidean space into a hybrid Euclidean–topological space. {\color{black}The model is comprehensively trained using both strong-form and weak-form loss formulations, enabling its application to a wide range of forward and inverse problems in solid mechanics.} For forward problems, the Finite-PINN model efficiently approximates solutions to solid mechanics problems when the geometric information of a given structure has been preprocessed. For inverse problems, it effectively reconstructs full-field solutions from very sparse observations by embedding both physical laws and geometric information within its architecture.

\end{abstract}


\section{Introduction}

Physics-informed neural networks (PINNs) have shown progress in solving various problems involving partial differential equations (PDEs) \cite{cai2021physics,huang2022applications,lawal2022physics,cuomo2022scientific}. Initially, the core idea of PINNs was to incorporate physical information into neural network training, enabling them to learn effectively from sparse data and observations, which is often needed when dealing with fields governed by known physical laws, typically represented by PDEs \cite{raissi2019physics,karniadakis2021physics}. Over time, this physics-informed concept has been extended to a broader range of applications. For forward PDE problems, where the goal is to solve specific partial differential equations, PINNs provide a fundamentally different approach compared to most other numerical methods. Due to their unique solution representation, PINNs express the solution as a continuous function (a trained neural network) rather than as discrete values at specific locations, which is the norm for traditional numerical methods \cite{raissi2019physics,karniadakis2021physics,yu2022gradient,cai2021physics,mao2020physics}. On the other hand, this continuous approximation ability, combined with solving PDEs through learning-based methods (i.e., formulating the problem as an optimisation objective involving loss functions), allows PINNs to address inverse PDE problems that conventional numerical methods often struggle with \cite{raissi2019physics,karniadakis2021physics,yu2022gradient,lu2021physics,jagtap2022physics}.

Theoretically, a sufficiently wide multi-layer perceptron (MLP) is capable of representing any solution field, due to the universal approximation theory \cite{cybenko1989approximation,hornik1989multilayer}. This makes PINNs a highly generalised method compared to most other numerical approaches for solving PDEs. Their implementation is also simpler, as traditional methods often require discretisation of the domain and derivation of specific steps such as weak formulations to convert complex PDEs into a solvable form \cite{larsson2003partial,karniadakis2005spectral,zienkiewicz1971finite,eymard2000finite,li2002meshfree}. Moreover, traditional numerical methods frequently face challenges with nonlinear or high-order PDEs due to complications introduced by domain discretisation \cite{karniadakis2005spectral}. In contrast, PINNs usually don't present with these issues.  They provide the solution as a continuous function, and their optimisation-based approach makes them particularly effective for inverse problems \cite{raissi2019physics,karniadakis2021physics}.

Significant progress has been made by researchers in applying PINNs to solve PDE problems \cite{huang2022applications,cuomo2022scientific,xu2023transfer,krishnapriyan2021characterizing,henkes2022physics,almajid2022prediction,lu2021physics,mao2020physics,cai2021physics2}. The core idea behind PINNs, which involves incorporating physical laws directly into the neural network training process, makes them a powerful tool for approximation and solution of PDEs. The early models of PINNs focused on utilising this concept effectively. PINNs have demonstrated excellent performance in fluid mechanics \cite{cai2021physics,mao2020physics,jagtap2022physics}. Their suitability for these fluid problems is due to the continuity properties inherent in both the PDEs and the PINN methodology. Additionally, the smoothness of solutions in fluid mechanics PDEs supports effective training of deep learning models by reducing differentiation residual losses, thereby facilitating smoother optimisation \cite{noll1955continuity,dong2007local}. While some equations have not been mathematically proven to possess smooth solutions \cite{chorin1968numerical,temam2024navier}, the actual behaviour of fluids in practical scenarios typically exhibits continuous and smooth characteristics. This advantage is also evident when applying PINNs to transient PDE problems, where the approximation along the time dimension tends to be successful, given that physical fields are generally continuous and smooth over time. A notable example is the use of PI-DeepONet to solve the Allen-Cahn equation: the model performs well along the time dimension but encounters difficulties with high irregularities along spatial coordinates \cite{li2024architectural}. 

Compared to the rapid progress of PINN in solving fluid mechanics problems, the use of PINN in solid mechanics remains in its beginning. The significant discontinuities and irregularities of the solution fields inherent in general solid mechanics problems make employing PINN for these tasks sometimes more challenging. An early attempt to use PINN for general solid mechanics is presented in \cite{haghighat2021physics}, where simple linear elasticity problems within regular (rectangular and square) domains were solved using the traditional PINN model. Since then, researchers have been exploring the application of PINN to various solid mechanics problems. However, due to the complex solid structures and geometries, the research often either focuses on simple problems or modifies the initial PINN model to make it more suitable for general solid mechanics. For instance, PINN has been employed in inverse problems such as identifying material properties \cite{bai2023physics,wu2023effective,song2024identifying,liao2024physics,zhang2020physics}, detecting specific material states \cite{shukla2020physics,dourado2020physics}, characterising internal structures \cite{zhang2022analyses}, and solving design/optimisation problems which can also be regarded as inverse problems \cite{jeong2023physics,jeong2023complete,hu2024physics}. On the other hand, researchers are striving to improve the initial PINN models or apply more advanced methods to solve different specific problems. {\color{black}For example, \cite{goswami2020transfer} utilises transfer learning models for phase-field modelling of fracture, while \cite{zhu2024transfer} applies transfer learning for parameter identification in soft materials. }Additionally, \cite{rezaei2022mixed, ning2023physics, gu2023enriched, manav2024phase} employ energy-based loss functions to incorporate physics into neural networks, enhancing the model's capability to address a broader range of solid mechanics problems by using variational losses rather than strong form losses. Some studies have also aimed to develop geometry-aware deep learning models to extend the application of PINN to problems involving complex geometries. For example, \cite{diao2023solving} decomposes complex domains into regular subdomains to perform neural network approximations within each subdomain. Moreover, specific PINN methods have been developed to model particular problems, such as plasticity \cite{niu2023modeling,arora2022physics}, plates \cite{gu2022physics}, shells \cite{bastek2023physics}, and 3D elasticity problems \cite{abueidda2021meshless}.

With a particular focus on the energy-based loss function used in \cite{rezaei2022mixed, ning2023physics, gu2023enriched, manav2024phase}, and in work such as the Deep Ritz Method \cite{yu2018deep} to solve variational problems, using the energy-based loss function to inform physics to a neural network is similar to the formulation step in traditional numerical methods \cite{finlayson1966method,larsson2003partial,karniadakis2005spectral,zienkiewicz1971finite,eymard2000finite,li2002meshfree}. These numerical methods perform a discrete conduction combined with the weak formulation to solve solid mechanics PDEs. The finite element method (FEM) \cite{zienkiewicz1971finite} is one of the most well-known methods, which has been extensively developed, packaged into commercial software, and widely utilised in engineering across many fields, demonstrating the capability of FEM in solving general solid mechanics problems \cite{cutress2010analysis}. In this context, it appears that the PINN architecture falls significantly short compared to FEM in solving general solid mechanics problems. The reason lies in the aforementioned challenges concerning the nature of solid mechanics problems and the characteristics of PINN: solid mechanics is inherently concerned with complex solid structures that introduce discontinuities and irregularities, which contrasts with the PINN's approach of producing continuous and smooth functional solutions. Similar discussions are presented in \cite{tu2022physics} and \cite{wang2024m}. At a simple glance at our world, the objects that solid mechanics problems focus on — actual solid structures — often exhibit complex inherent geometries, which is a key distinction between solid media and fluid media. The finite element method is particularly adept at approximating these solutions since its solution is represented in discrete form, with the finite element mesh embedding all the geometric information of the structures, which is prepared prior to calculation \cite{kaveh2002hybrid, sadayappan1987nearest, paulino1994node}.

To develop an effective PINN implementation for solving general solid mechanics problems, comparable to what can be achieved with FEM, specialised neural network models may be required to address the challenges posed by discontinuities and irregularities arising from complex geometries. The physics-informed graph neural network (PIGNN) proposed in \cite{gao2022physics} is one such solution, adopting a discrete deep learning architecture to approximate the solution field. The work shows potential in solving general solid mechanics problems; however, the purely discrete conduction mechanism embedded in PIGNN makes it less distinct from the finite element method, as it abandons the continuous functional approximation of traditional PINNs. Consequently, it also loses some of the advantages of traditional PINNs, such as lower computational costs (in terms of both memory and time), superior capability in solving inverse problems, and, most importantly, the ability to fuse arbitrary data and physics information to create a physics-informed, data-driven model. Another approach is to impose exact boundary conditions via the output of neural networks, thereby achieving a geometry-aware deep learning method \cite{sukumar2022exact}. Some studies have applied this method to solve certain solid mechanics problems by enforcing exact Dirichlet boundary conditions \cite{wang2023exact}. The results indicate that this method is relatively effective for assigning Dirichlet boundary conditions; however, enforcing exact Neumann boundary conditions on a solid structure is extremely challenging. Both free boundaries and boundaries subjected to loads must be assigned corresponding Neumann boundary conditions due to the infinite domain represented by a PINN solution. Moreover, the geometry-aware mechanism of the exact boundary condition imposition method provides a weak/inadequate incorporation of geometry into the PINN architecture: the method only accounts for geometrical (outer surface) information within the neural network, while the topological (interior structure) information remains absent. From these observations, it becomes evident that, as of now, there are no traditional PINN models or methods capable of solving more general types of solid mechanics problems with complex geometries.

In this context, this work proposes a physics-informed neural network architecture, called the Finite-PINN model, which incorporates the most significant properties of finite element methods into the PINN computations. The method retains the simplest and most effective implementation scheme of PINN for solving solid mechanics problems, based on a novel neural network structure designed to integrate the advantages of FEM into the computations. The method is presented as a highly general solution for applying PINN to solid mechanics problems. The paper provides a comprehensive introduction to the method. Section \ref{sec2} introduces general solid mechanics PDE problems and details the challenges encountered when using PINN to solve these problems. Section \ref{sec3} introduces the proposed Finite-PINN and outlines some of its basic mathematical properties. Section \ref{sec4} describes the implementation of Finite-PINN for solving general solid mechanics problems. Section \ref{sec5} presents several case studies and their results. Section \ref{sec6} includes discussions, followed by concluding remarks in Section \ref{sec7}.

\section{Solid mechanical problem and PINNs}
\label{sec2}
\subsection{Solid mechanics problems}
\begin{figure}[htbp]
    \centering
    \includegraphics[width=0.28\linewidth]{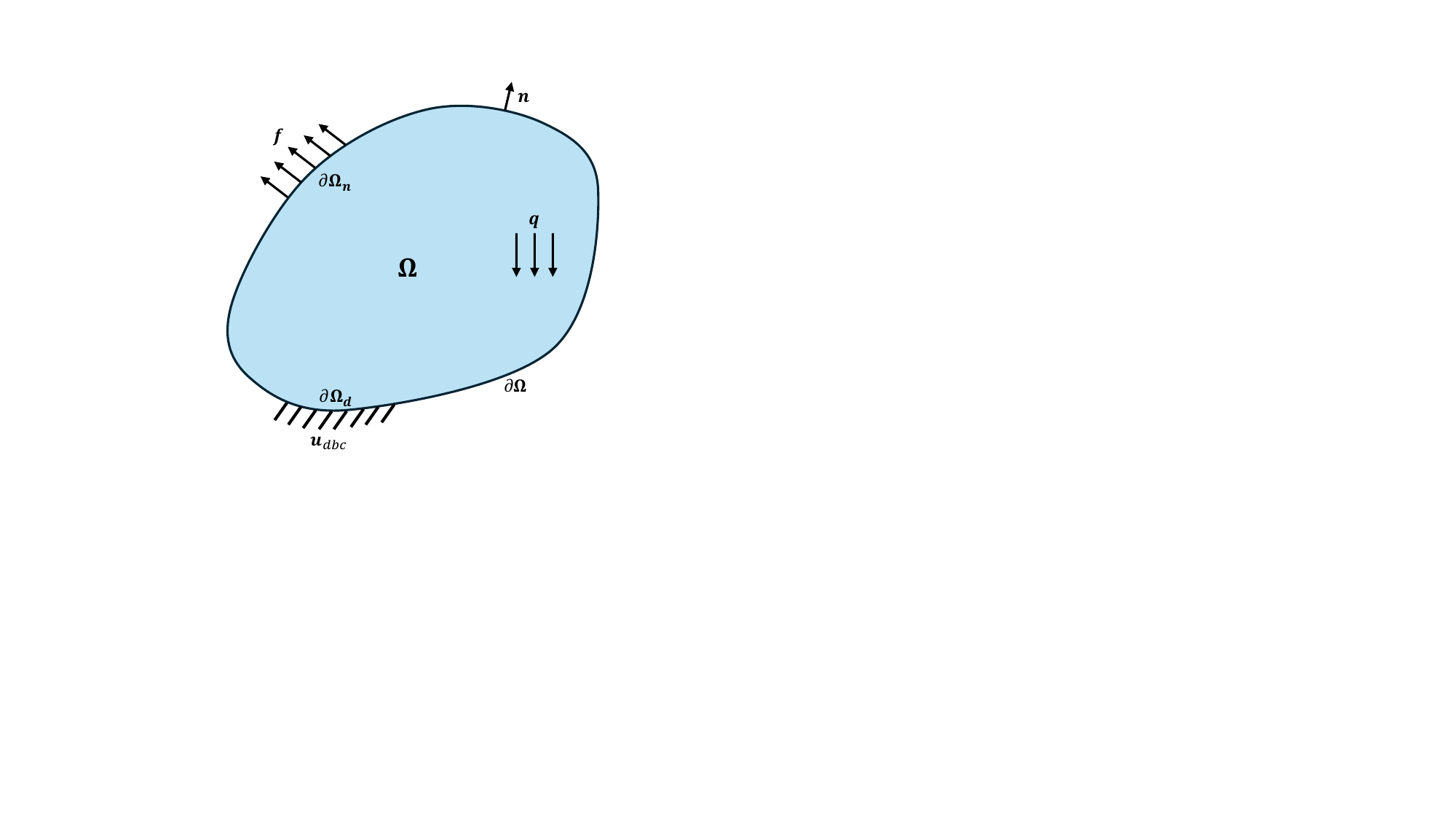}
    \caption{A static solid mechanics problem. The domain is denoted by $\boldsymbol{\Omega}$, with its boundary represented by $\partial \boldsymbol{\Omega}$. The boundary segments where Dirichlet and Neumann boundary conditions are applied are denoted by $\partial \boldsymbol{\Omega}_d$ and $\partial \boldsymbol{\Omega}_n$, respectively.}
    \label{fig:p0}
\end{figure}

The governing equation of a structural dynamic problem with zero external body force is stated as:
\begin{equation}
m\nabla_t^2 \boldsymbol{u} + c \nabla_t \boldsymbol{u} + \nabla_{\boldsymbol{x}} \cdot \boldsymbol{\sigma} = \boldsymbol{0}
\label{dsmp}
\end{equation}
where $\boldsymbol{u}$ and $\boldsymbol{\sigma}$ denote the displacement and stress, respectively, $m$ is the mass and $c$ is the damping coefficient. 
The terms with derivatives in the time dimension vanish when the operating time is sufficiently long, such that the first and second-order derivatives of displacement with respect to time become negligibly small. The problem then reduces to a static or quasi-static solid continuum problem, whose full statement of the partial differential equations is expressed as:
\begin{equation}
\nabla \cdot \boldsymbol{\sigma} = \boldsymbol{0}
\label{dsmp2}
\end{equation}
This work focuses on addressing the challenge of non-uniform spatial space induced by complex solid geometries or structures. A schematic of a solid mechanics problem is presented in Fig.\ref{fig:p0}, which includes the demonstration of defined Dirichlet and Neumann boundary conditions. 

\subsection{PINN and FEM in solving solid mechanics problems}
\begin{figure}[htbp]
    \centering
    \includegraphics[width=0.85\linewidth]{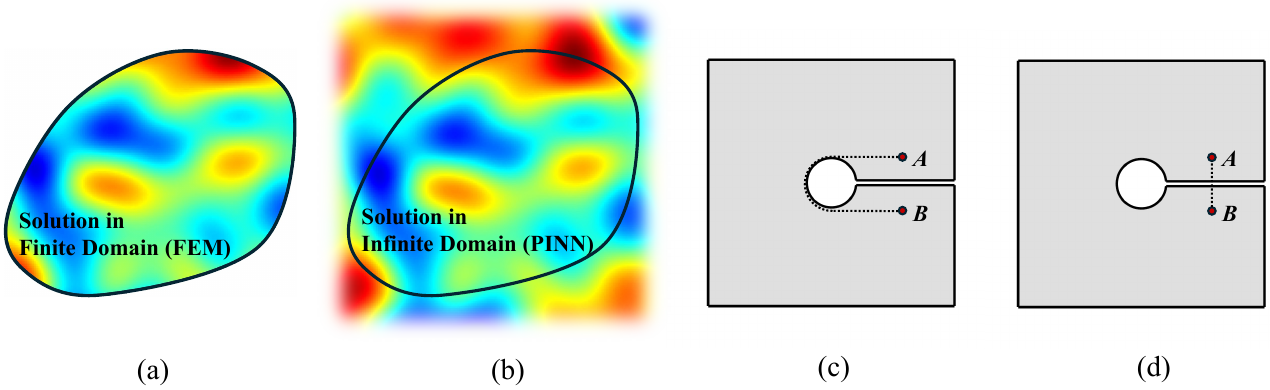}
    \caption{(a) Solution domain defined in FEM. (b) Solution domain defined in PINN. (c) Geodesic distance between points A and B. (d) Euclidean distance between points A and B.}
    \label{fig:p1}
\end{figure}

Solid structures usually possess discontinuous and non-smooth characteristics, leading to discontinuous and non-smooth solution fields in solid mechanics problems. Consequently, the inherently continuous nature of PINN limits its effectiveness when applied to such problems. It is well established that the FEM is the most widely adopted numerical technique for solving solid mechanics problems. A comparison between PINN and FEM can yield valuable insights into the computational approaches of these two methods. The two most significant differences between these methods are:
\begin{difference}
PINNs perform computations over an infinite domain, yielding an infinitely continuous representation of the solution, whereas FEM operates on a discretised, finite domain;
\label{p1}
\end{difference}
\begin{difference}
The solution space of PINN is defined in the Euclidean space, whereas FEM's solution space is represented in a metric (topological) space defined by the finite element mesh.
\label{p2}
\end{difference}
A schematic is presented in Fig.\ref{fig:p1}, illustrating the two core differences between the finite element method and the physics-informed neural network. The explanations are as follows.

\begin{figure}[htbp]
    \centering
    \includegraphics[width=0.65\linewidth]{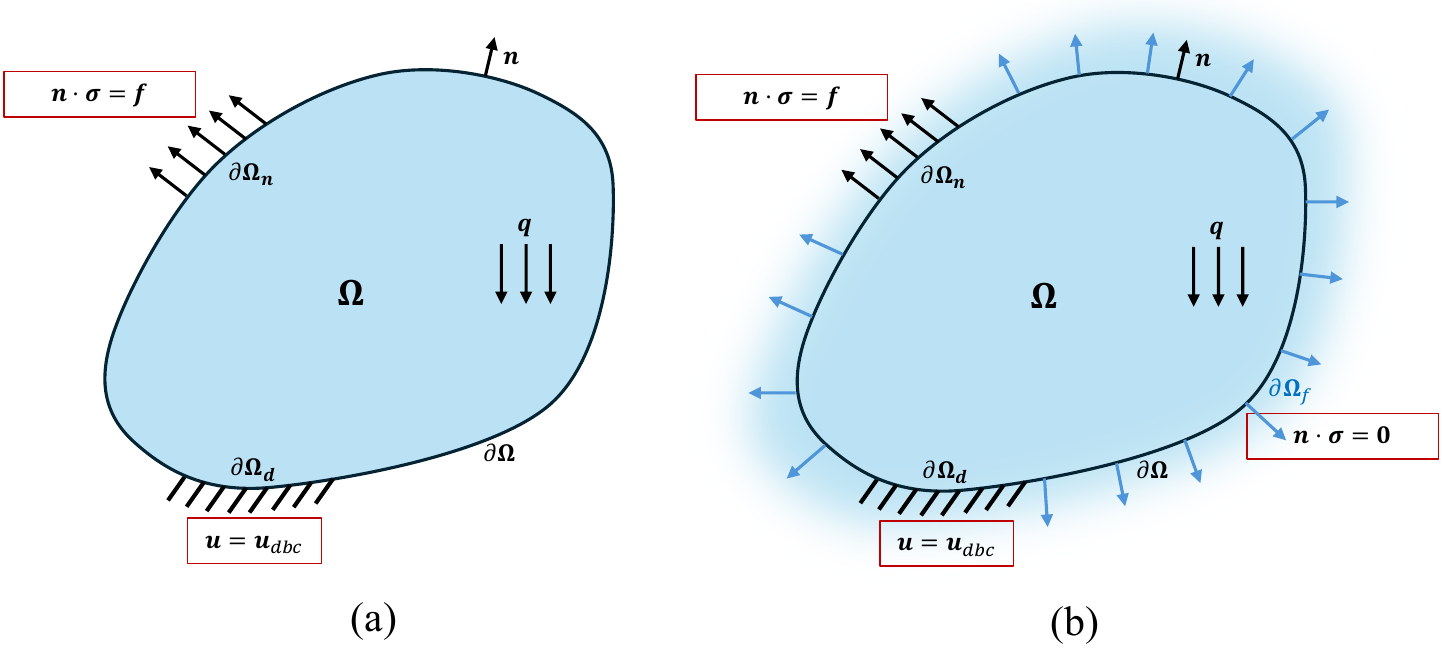}
    \caption{Schematic of general solid mechanics problems defined in a finite domain (a) and an infinite domain (b). Extra Neumann boundary conditions, $\boldsymbol{n} \cdot \boldsymbol{\sigma} = \boldsymbol{0}$, are required to be applied on free boundaries, denoted by $\partial \boldsymbol{\Omega}_f$.}
    \label{fig:p3}
\end{figure}

\begin{figure}[htbp]
    \centering
    \includegraphics[width=0.85\linewidth]{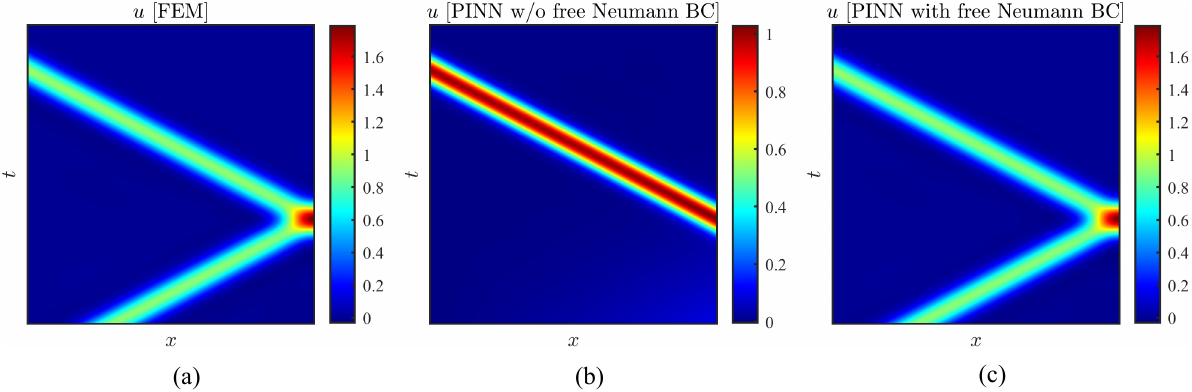}
    \caption{Response distribution of displacements over the spatial (horizontal) and temporal (vertical) domains calculated by FEM (a) and PINN models without (b) and with (c) free-surface boundary conditions. A simple displacement impulse is applied to the left end of a 1D segment, with the right end remaining free. The results indicate that the PINN model without free-surface boundary conditions shows no reflection at the right end, whereas the model with free-surface boundary conditions exhibits reflection at the right end.}
    \label{fig:pk}
\end{figure}

For the first point, PINNs represent the solution as a continuous and differentiable function defined over an infinite domain (Fig.\ref{fig:p1}(b)), although only the solution within the effective domain is considered. FEM, on the other hand, provides discrete values at specific points (nodes) and solves the problem over a finite domain throughout (Fig.\ref{fig:p1}(a)). From a mathematical perspective, the finite domain of FEM leads to an important characteristic: FEM’s numerical model implicitly satisfies the free Neumann boundary condition $\boldsymbol{n} \cdot \boldsymbol{\sigma} = 0$ on its free boundaries. In contrast, PINN models require the explicit application of the constraint $\boldsymbol{n} \cdot \boldsymbol{\sigma} = 0$ on all free boundaries (Fig.\ref{fig:p3}), or otherwise assume an initially infinite boundary. This difference is more apparent in dynamic problems. Fig.\ref{fig:pk} presents an example showing the dynamic response of a 1D finite line subjected to an impulse actuation on its left side, solved by FEM and PINN models both with and without consideration of the free Neumann boundary conditions, respectively. It can be observed that FEM returns a response featuring a reflection at the right end of the line due to the finite domain definition implicitly embedded within FEM (Fig.\ref{fig:pk}(a)). However, the PINN model performed with the strong form PDE loss produces a response with no reflection if no additional constraints are imposed for the free boundary at the right end (Fig.\ref{fig:pk}(b)). The challenge of defining a finite domain has also been identified in \cite{rasht2022physics}, where the author used PINN to solve the wave propagation equation in a semi-infinite domain. In that case, an additional free-boundary constraint was applied on the top surface (finite boundary) during PINN training. This implies that, to solve a typical solid mechanics problem using PINN, if the strong form PDE loss is used, all free boundaries must be explicitly assigned the free Neumann boundary condition $\boldsymbol{n} \cdot \sigma = 0$. This is also the only way that enables a traditional PINN model to understand the geometry of specific structures. In summary, PINN and FEM handle free boundaries in fundamentally opposite ways: PINN is defined over an infinite solution domain, necessitating additional considerations when a finite domain is required, whereas FEM naturally operates within a finite solution domain. This characteristic is intrinsic to the name “Finite Element Method”, where the term "finite" refers to the finite solution domain. Interestingly, for certain problems involving infinite domains, such as wave propagation or specific structural dynamics problems, researchers have adapted FEM into the Infinite Element Method (IFEM) to handle infinite boundaries as needed \cite{zienkiewicz1985mapped}.

\begin{figure}[htbp]
    \centering
    \includegraphics[width=0.55\linewidth]{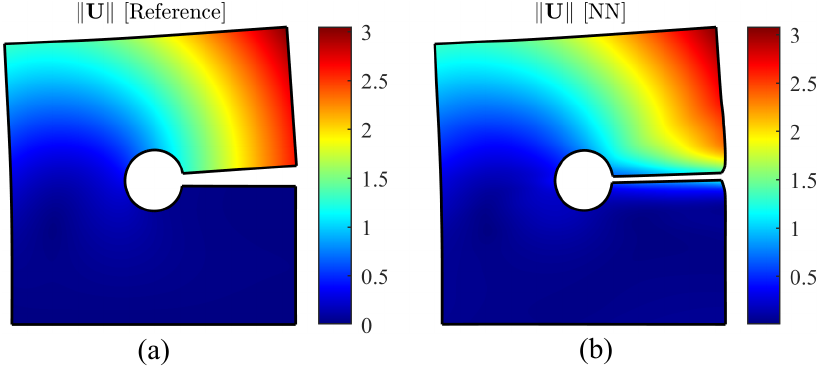}
    \caption{Approximation results of using a neural network to approximate the displacement field of an open-notch structure. (a) Reference displacement field. (b) Learned displacement field by a neural network after training.}
    \label{fig:pi1}
\end{figure}

For the second point, since traditional PINNs' solution is given in the form $\boldsymbol{u}=f\left(\boldsymbol{x}\right)$, which is a function in Euclidean space, the distance between points is measured using the Euclidean distance, as it takes the Euclidean coordinates $\boldsymbol{x}$ as inputs. FEM, on the other hand, defines its solution space within the metric space embedded in the finite element mesh, where distances between locations are represented by geodesic distances. Fig.\ref{fig:p1}(c) and (d) illustrate the Euclidean distance and the geodesic distance, respectively. This distinction can also be interpreted using graph theory, as FEM operates based on a graph (the finite element mesh), which represents the solution space via the connectivity matrix (commonly the stiffness matrix in solid mechanics). 
This fundamental difference provides FEM with an advantage in performing approximations within complex geometries but poses significant challenges for PINNs. An example is shown in Fig.\ref{fig:pi1}, where a neural network model is used to approximate the resulting field of a tensile simulation of an open-notch structure. This simulation is performed solely using data-driven learning, i.e., employing an MLP to approximate the solution distribution without integrating any physical information. It is evident that the presence of the narrow notch leads to significant discrepancies at the notch location. This phenomenon is noticeable in this pure learning-based approximation, and when the differentiation-based physical information is incorporated in a PINN, the derivatives obtained near the notch can become extremely ill-conditioned, preventing the PINN from converging to reasonable results. A case can be considered where the notch width approaches zero, resulting in a crack; in such a scenario, the PINN model would fail to approximate the resulting field, as the function becomes discontinuous and non-differentiable at the notch (now a crack).

In summary, based on the two differences observed between FEM and PINN, two significant challenges can be identified when using PINN to solve solid mechanics problems:
\begin{problem}
PINN generates solutions defined over an infinite domain, which conflicts with the finite domain characteristic of most solid structures;
\label{prob3}
\end{problem}
\begin{problem}
The Euclidean space is not an appropriate space for PINN to effectively learn or solve the solution field of a solid mechanics problem.
\label{prob4}
\end{problem}
These two challenges are the primary reasons why PINN often fails to solve general solid mechanics problems, particularly those involving complex geometries or domains. This work aims to address these challenges within a PINN framework by proposing a new PINN model capable of effectively handling general solid mechanics problems.

\subsection{Possible solutions of using PINN in solid problems}
To address the two challenges and enable the use of PINN in general solid mechanics problems, several solutions have been proposed by researchers. 

Problem \ref{prob3} has been a longstanding challenge, not just for PINN but for nearly all mesh-free methods used in solving partial differential equations. {\color{black} If the problem adopts the weak form of the PDE, as in the Ritz method or the Galerkin method—which are commonly used to solve certain solid mechanics problems—the homogeneous Neumann boundary conditions can be automatically satisfied during the solution process, since the integral in the weak form inherently accounts for the work done by the zero Neumann boundary condition. Additionally, the non-homogeneous Neumann boundary conditions are automatically considered as the external work introduced by applied loads. In this case, the PDE problem is converted into a variational problem, which is typically easier to address using certain deep learning methods. Examples of methods developed for this purpose include the Deep Ritz method~\cite{yu2018deep} and the Deep Energy approach~\cite{samaniego2020energy}.

While these methods have demonstrated good performance in solving forward solid mechanics problems, as reported in the literature, and typically outperform approaches based on strong form losses, their variational weak form exhibits poor performance when dealing with inverse problems. These include load identification and deformation sensing, where the boundary conditions, PDE parameters, or other associated data are either unknown or uncertain. Additionally, the integration involved in the weak form enforces a global solution approach, making it challenging to apply this formulation to localised problems, such as sparse data reconstruction.

To apply the strong form in solving solid mechanics problems, zero Neumann boundary conditions, as illustrated in Problem~\ref{prob3}, must be explicitly assigned to the neural networks.
}
The most common method for implementing this in PINN is to include an additional loss term, $loss = \boldsymbol{n} \cdot \boldsymbol{\sigma}$, in the loss function, which is then minimised to enforce weak imposition of free traction on free surfaces \cite{haghighat2021physics}. However, incorporating such a loss term presents significant challenges. Firstly, assembling and including this loss term is highly complex, especially for intricate structures, as it requires information about all the normal vectors across all free surfaces. Secondly, optimising a hybrid loss function in a PINN model is difficult \cite{bai2023physics,wang2021understanding,wang2022and}, as the optimiser often struggles to focus effectively on which component of the loss to minimise.

Rather than using loss function methods, some researchers have proposed automatically imposing exact boundary conditions within a PINN model \cite{sukumar2022exact}. This idea also builds on earlier work involving mesh-free models that were studied in the previous century \cite{rvachev1995r}. \cite{sukumar2022exact} introduces a comprehensive method that imposes both Dirichlet and Neumann boundary conditions on a PINN model by utilising distance functions. However, implementing this approach is challenging in solid mechanics problems, particularly because the Neumann boundary conditions in solid mechanics problems are defined in vector fields and are inhomogeneous with respect to the normal vector direction at specific locations. Although paper \cite{sukumar2022exact} presents methods for imposing inhomogeneous Neumann boundary conditions, the resulting inhomogeneous distance function becomes very large, as nearly all surfaces in a typical solid mechanics problem are free (Fig.\ref{fig:p3}(b)), leading to expensive computational and memory requirements.

Problem \ref{prob4} can be addressed by modifying the input space of PINN. Inspired by the finite element method, one direct approach to achieving this is to replace the traditional neural network with graph neural networks. The so-called physics-informed graph neural network (PIGNN) method was proposed in \cite{gao2022physics}. Instead of using a neural network to directly approximate the resulting field, a graph neural network performs computations by calculating the convolution between nodes, which are connected by edges. This approach is very similar to the finite element method, where both input and output data are in discrete form, and the graph can be applied to specific structures to incorporate geometrical information into the neural network model. While this method maintains a useful structure with a mathematical formulation similar to the finite element method, it also loses many of the traditional advantages of PINN, such as efficient memory usage, continuous solution representation, and suitability for inverse problems. 

\begin{figure}[htbp]
    \centering
    \includegraphics[width=0.20\linewidth]{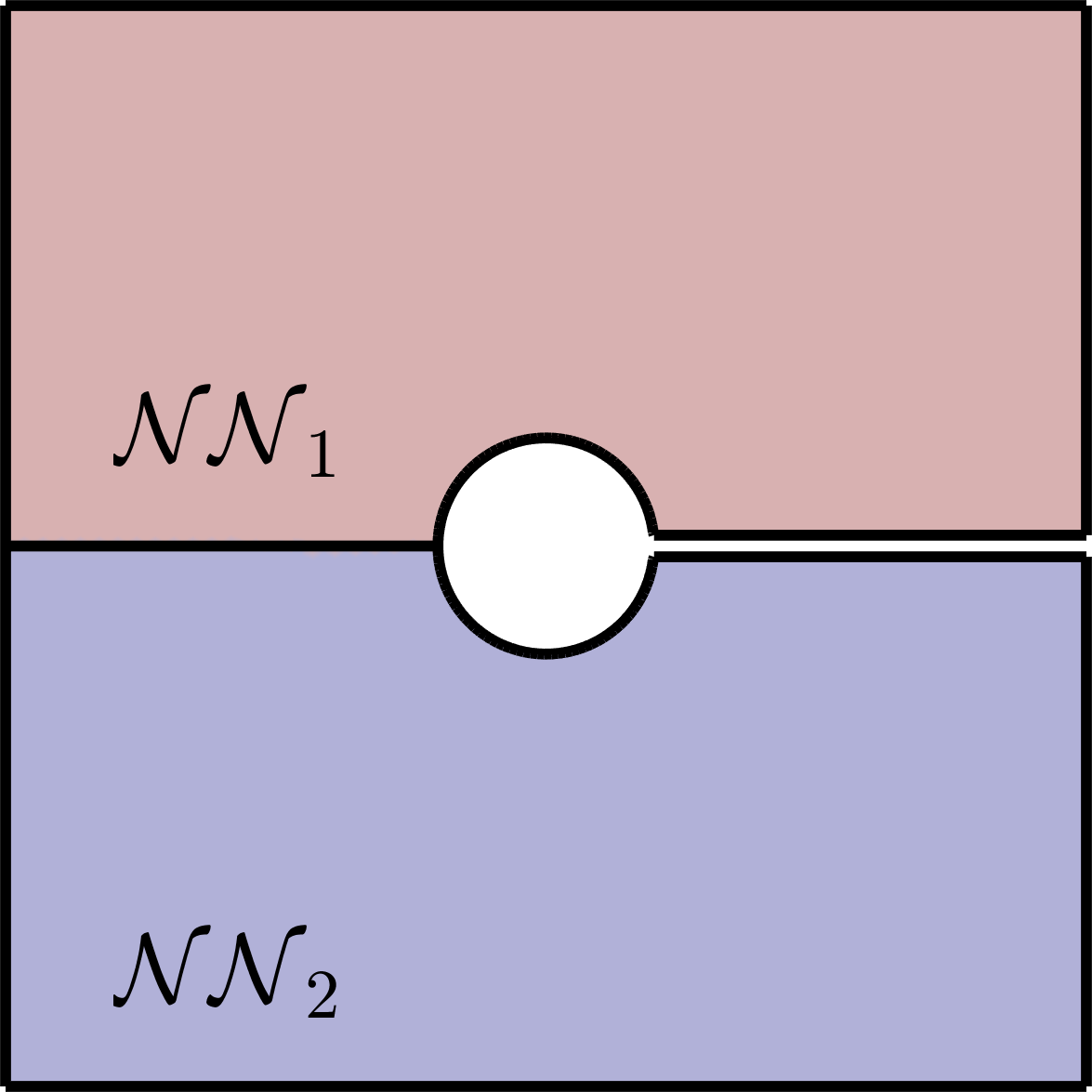}
    \caption{A possible domain decomposition of the open-notch structure using the XPINN method.}
    \label{fig:pnn}
\end{figure}

In addition, another method called the XPINN has been proposed \cite{jagtap2020extended}. XPINN utilises multiple sub-neural-networks instead of a single network model for the computation. The main concept is to decompose the entire effective domain into smaller sub-domains, thereby representing the solution space with several individual sub-neural network functions. This approach offers a more desirable input space than the traditional Euclidean space by decomposing areas with large physical distances into distinct sub-regions instead of treating them as a unified whole. For example, as shown in Fig.\ref{fig:pnn}, a possible solution using XPINN is to divide the entire domain into two sub-domains by a horizontal middle line, allowing the top and bottom parts to be approximated by two separate neural networks. The XPINN method can handle general partial differential equations with complex domains. However, the use of sub-networks makes training challenging, as the sub-networks are joined through loss functions applied at their boundaries rather than being trained as a unified system. Additionally, the method has been observed to be sensitive to the quality of domain decomposition, particularly for complex geometries.

\color{black}
To improve the handling of geometric complexity in weak-form-based methods, some researchers have proposed the parametric Deep Energy Method (p-DEM). This approach extends the traditional Deep Energy Method by transforming the physical domain into a parametric domain, where neural networks can more effectively approximate the solution fields. In p-DEM, NURBS curves are used as the fundamental geometric representation to construct the parametric domain, facilitating the solution process in geometrically complex cases. Although p-DEM shows promise in solving Problem~\ref{prob4}, its application remains limited when dealing with non-simply connected domains, which are common in solid mechanics problems (e.g., open-hole structures). In such cases, the domain must be partitioned into multiple subdomains to define the parametric mappings properly. This approach mirrors the strategies employed in the XPINN method~\cite{jagtap2020extended} in such contexts, while also aligning with traditional numerical techniques such as Isogeometric Analysis \cite{hughes2005isogeometric}.
\color{black}

Apart from the above methods, an alternative PINN model called the $\Delta$-PINN model is proposed in \cite{costabal2024delta}, which replaces the Cartesian coordinate inputs with Laplace eigenfunctions. In this approach, the input space of the PINN model is transformed from Euclidean space to Riemannian manifolds represented by the Laplace eigenfunctions. The Laplace eigenfunctions are orthogonal vectors in a functional space and are computed based on the geometry of the structures under study. This Riemannian space defined by Laplace eigenfunctions is embedded with comprehensive topological information of the structures, making the spatial distances between locations in this input space equivalent to the desired geodesic distances.

\cite{costabal2024delta} demonstrates the application of the $\Delta$-PINN model in various PDE problems, highlighting that due to its ability to handle problems defined over complex geometries or domains, applying $\Delta$-PINN to solid mechanics problems could be a promising direction. Moreover, Laplace eigenfunctions possess additional properties that are beneficial for solving solid mechanics problems. A more in-depth mathematical exploration of these properties will be discussed in the next section.

\section{Methodology}
\label{sec3}
This section presents the proposed Finite-PINN method in a structured sequence: the Laplace-Beltrami Operator eigenfunctions, the Finite-PINN architecture, and the model's properties.
\subsection{Laplace-Beltrami Operator (LBO) eigenfunctions}
The Laplace eigenfunctions used in this study are calculated by solving the following Laplace-Beltrami operator eigenvalue problem:
\begin{equation}
\left\{
\begin{aligned}
-\Delta u\left(\boldsymbol{x}\right)=\lambda u\left(\boldsymbol{x}\right) & \quad \forall \quad \boldsymbol{x} \in \mathbf{\Omega} \\ 
-\nabla u \cdot \boldsymbol{n}=0  & \quad \forall \quad  \boldsymbol{x} \in \partial \mathbf{\Omega}
\end{aligned}
\right.
\label{LBO}
\end{equation}
where $u$ is the output, $x$ represents the input coordinates, $\boldsymbol{n}$ denotes the outward normal vector on the boundary, $\Omega$ is the effective domain, and $\lambda$ represents the eigenvalue. Eq.\ref{LBO} defines a Laplace-Beltrami problem with homogeneous Neumann boundary conditions. The Laplace-Beltrami operator (LBO) eigenfunctions can be obtained by solving Eq.\ref{LBO}:
\begin{equation}
\boldsymbol{\phi} = \phi_1\left(x\right),\phi_2\left(x\right), ... \phi_k\left(x\right), ... 
\end{equation}
where $\phi_k\left(x\right)$ represents the $k$th LBO function, with the eigenfunctions sorted by their eigenvalues in ascending order. The obtained eigenfunctions serve as basis functions in the Hilbert space $L^2\left(\mathbf{\Omega}\right)$ and are orthogonal to each other within this functional space. Any functions in $L^2\left(\mathbf{\Omega}\right)$ can be represented as a linear combination of these LBO functions.

LBO functions are useful in functional analysis and have been employed as powerful mathematical tools in principal component analysis, basis computation, and more. Unlike Fourier bases, which are derived from Fourier transformations and require periodicity of the operating domain, LBO eigenfunctions can handle arbitrarily defined domains. More importantly, they provide information about specific domains across different bases. For instance, the second LBO function is known as the Fiedler vector in graph theory \cite{chen2017fiedler}, which conveys comprehensive information about the topology under study and is sometimes used to estimate the algebraic connectivity of a graph.

The properties of LBO functions make them particularly useful for addressing problems involving complex structures or geometries. As mentioned earlier, the $\Delta$-PINN model employs LBO functions for solving PDEs. Additionally, \cite{chen2023learning} proposed the Laplace Neural Operator, which utilises LBO functions in operator learning. The common feature of these two works is the use of LBO functions as inputs to incorporate the geometries of the studied structures, allowing their models to effectively handle complex geometries that other models struggle with. This work also focuses on using LBO eigenfunctions to solve physics-informed neural network problems in solid mechanics.

\subsection{Finite-PINN}
\label{sec32}
To develop a model capable of solving general solid mechanics problems using physics-informed neural networks, specifically, to address the two challenges discussed in Section \ref{sec2}, we propose the following function architecture, named Finite-PINN, which aims to solve solid mechanics problems defined within general geometries:
\begin{equation}
\color{black}
u_{i}\left(\boldsymbol{x} \right) = f_{i}\left(\boldsymbol{x}, \boldsymbol{\phi}^{n_u}\left(\boldsymbol{x} \right)\right),
\label{fpinnl1}
\end{equation}
where $i= 1, 2$ for 2D problems and $i= 1, 2, 3$ for 3D problems, $\boldsymbol{\phi}^{n_u}$ represents the first $n_u$ LBO basis functions. In this work, the function $\boldsymbol{f}$ is approximated by neural networks:
\begin{equation}
\boldsymbol{f}\left(\boldsymbol{x} \right) = \mathcal{NN}_u\left(\boldsymbol{x},\boldsymbol{\phi}^{n_u};\boldsymbol{\theta}\right),
\label{NNl1}
\end{equation}
\color{black}
The solution process can be carried out using either the weak form or the strong form of the governing PDE. Both approaches are introduced below.

\subsubsection{Weak form Finite-PINN}
The weak form Finite-PINN is designed to solve direct forward problems where all boundary conditions and PDE parameters are exactly known. In this approach, the neural network is trained by minimising the energy functional of the PDE Eq.\ref{dsmp2}:
\begin{equation}
I = \int_{\Omega} \boldsymbol{\sigma}\left(\boldsymbol{x}\right) \cdot \boldsymbol{\varepsilon}\left(\boldsymbol{x}\right) d V -\int_{\partial \Omega_n} \overline{\boldsymbol{t}} \cdot \boldsymbol{u}(\boldsymbol{x}) d A
\label{fnc}
\end{equation}
where $\boldsymbol{\sigma}\left(\boldsymbol{x}\right)$ and $\boldsymbol{\varepsilon}\left(\boldsymbol{x}\right)$ denote the stress and strain, respectively, $\overline{\boldsymbol{t}}$ denotes the traction loads. 

The problem can then be solved by incorporating this functional and the boundary conditions into the loss function used to train $\mathcal{NN}_u$:
\begin{equation}
\begin{aligned}
\mathcal{L} & = \mathcal{L}_{dbc} + \mathcal{L}_{pde}, \\
\mathcal{L}_{dbc} & = \frac{1}{N_{dbc}}\sum^{N_{dbc}}_{i=1}\left(\boldsymbol{u}\left(\boldsymbol{x}^i\right)-\boldsymbol{u}_{bc}\left(\boldsymbol{x}^i\right)\right)^2, & \boldsymbol{x}^i & \in \partial\boldsymbol{\Omega}_d, \\
\mathcal{L}_{pde} & = \sum^{N_{pde}}_{i=1}v^i \cdot \boldsymbol{\varepsilon}\left(\boldsymbol{x}^i\right) \cdot \left(\mathbb{C} : \boldsymbol{\varepsilon}\left(\boldsymbol{x}^i\right)\right) - \sum^{N_{nbc}}_{i=1} a^i \cdot \overline{\boldsymbol{t}} \cdot \boldsymbol{u}\left(\boldsymbol{x}^i\right), & \boldsymbol{x}^i & \in \boldsymbol{\Omega}
\end{aligned}
\label{loss1}
\end{equation}
where 
$\boldsymbol{\varepsilon}\left(\boldsymbol{x}\right) = \frac{1}{2}\left(\nabla \boldsymbol{u}\left(\boldsymbol{x}\right) + \nabla \boldsymbol{u}^{T}\left(\boldsymbol{x}\right)\right)$. $v^i$ and $a^i$ denote the representative volume and area associated with each collocation point. $N_{dbc}$, $N_{nbc}$ and $N_{pde}$ denote the number of collocation points for the Dirichlet boundary conditions, Neumann boundary conditions and integration points, respectively. It is seen that the boundary condition loss only considers the Dirichlet boundary conditions, since the Neumann boundary conditions are automatically included in Eq.\ref{fnc}. 

\subsubsection{Strong form Finite-PINN}
For the strong form Finite-PINN, an additional neural network is introduced to approximate the stress field alongside the displacement field. Combining these two approximators, the strong form Finite-PINN is formulated as follows:
\color{black}
\begin{equation}
\begin{aligned}
\sigma_{ij}\left(\boldsymbol{x} \right) & = p_{ik}\left(\boldsymbol{x} \right)\phi_{k,j}\left(\boldsymbol{x} \right),  \\
u_{i}\left(\boldsymbol{x} \right) & = f_{i}\left(\boldsymbol{x}, \boldsymbol{\phi}^{n_u}\left(\boldsymbol{x} \right)\right)
\label{fpinn}
\end{aligned}
\end{equation}
where $i, j = 1, 2$ for 2D problems and $i, j = 1, 2, 3$ for 3D problems. The index $k = 1, 2, 3, \dots, n_\sigma$ denotes the labels of the LBO eigenfunctions, where $n_\sigma$ is the number of eigenfunctions used in the stress approximation. The equation follows the Einstein summation convention for tensors. $\sigma_{ij}$ and $u_i$ represent the stress and displacement fields to be solved, respectively. $\phi_{k,j}$ denotes the partial derivative of the $k$th LBO eigenfunction with respect to $x_j$, i.e., $\phi_{k,j} = \frac{\partial \phi_k}{\partial x_j}$. $\boldsymbol{p}$ and $\boldsymbol{f}$ are direct outputs of the neural network, used to approximate the stress field $\boldsymbol{\sigma}$ and displacement field $\boldsymbol{u}$, respectively. In this work, $\boldsymbol{p}$ and $\boldsymbol{f}$ are approximated by neural networks, presented as follows:
\begin{equation}
\begin{aligned}
\boldsymbol{p}\left(\boldsymbol{x} \right) & = \mathcal{NN}_\sigma\left(\boldsymbol{x};\boldsymbol{\theta}\right), \\
\boldsymbol{f}\left(\boldsymbol{x} \right) & = \mathcal{NN}_u\left(\boldsymbol{x},\boldsymbol{\phi}^{n_u};\boldsymbol{\theta}\right),
\end{aligned}
\label{NN}
\end{equation}
where $\mathcal{NN}$ represents a given neural network, $\boldsymbol{x}$ denotes the Cartesian coordinate input, and $\boldsymbol{\phi}^{n_u}$ represents the first $n_u$ LBO eigenfunctions used in the approximation of the displacement field. $\boldsymbol{\theta}$ represents the trainable parameters of the neural networks, which are to be optimised. $\mathcal{NN}_\sigma$ and $\mathcal{NN}_u$ are neural networks used to approximate the stress and displacement fields, respectively. As described by Eqs. \ref{fpinn} and \ref{NN}, $\mathcal{NN}_\sigma$ takes inputs of dimension $dim\left(\boldsymbol{x}\right)$ and returns an output with dimension $dim\left(\boldsymbol{x}\right)\times n_\sigma$, corresponding to the number of elements in the stress tensor. Similarly, $\mathcal{NN}_u$ takes inputs of dimension $dim\left(\boldsymbol{x}\right) + n_u$ and produces outputs of dimension $dim\left(\boldsymbol{x}\right)$. For example, a 2D problem using 8 LBO bases for both the stress and displacement approximations requires an $\mathcal{NN}_\sigma$ with 2 inputs and 16 outputs for the stress fields, and an $\mathcal{NN}_u$ with 10 inputs and 2 outputs for the two-dimensional displacement fields.

In the model, note: a) the numbers of LBO bases used in $\mathcal{NN}_\sigma$ and $\mathcal{NN}_u$ serve as hyperparameters for each of the neural networks, and they may differ from each other; b) $\mathcal{NN}_\sigma$ and $\mathcal{NN}_u$ are general representations of neural networks for approximating stress and displacement, and can be divided into different sub-networks for each output or combined in various ways.

The loss function for the Finite-PINN model is defined as:
\begin{equation}
\begin{aligned}
\mathcal{L} & = \mathcal{L}_{data} + \mathcal{L}_{bc} + \mathcal{L}_{pde} + \mathcal{L}_{C} , \\
\mathcal{L}_{data} & = \frac{1}{N_{\sigma}}\sum^{N_{\sigma}}_{i=1}\left(\boldsymbol{\sigma}\left(\boldsymbol{x}^i\right)-\boldsymbol{\sigma}_0\left(\boldsymbol{x}^i\right)\right)^2 + \frac{1}{N_{u}}\sum^{N_{u}}_{i=1}\left(\boldsymbol{u}\left(\boldsymbol{x}^i\right)-\boldsymbol{u}_0\left(\boldsymbol{x}^i\right)\right)^2, & \boldsymbol{x} & \in \boldsymbol{\Omega}, \\
\mathcal{L}_{bc} & = \frac{1}{N_{n}}\sum^{N_{n}}_{i=1}\left(\boldsymbol{\sigma}\left(\boldsymbol{x}^i\right)-\boldsymbol{f}_{bc}\left(\boldsymbol{x}^i\right)\right)^2 + \frac{1}{N_{d}}\sum^{N_{d}}_{i=1}\left(\boldsymbol{u}\left(\boldsymbol{x}^i\right)-\boldsymbol{u}_{bc}\left(\boldsymbol{x}^i\right)\right)^2, & \boldsymbol{x} & \in \partial\boldsymbol{\Omega}_{d/n}, \\
\mathcal{L}_{pde} & = \frac{1}{N_{pde}}\sum^{N_{pde}}_{i=1}\left(\nabla \cdot\boldsymbol{\sigma}\left(\boldsymbol{x}^i\right)-\boldsymbol{q}\right)^2, & \boldsymbol{x} & \in \boldsymbol{\Omega}, \\
\mathcal{L}_{C} & = \frac{1}{N_{C}}\sum^{N_{C}}_{i=1}\left(\boldsymbol{\sigma}\left(\boldsymbol{x}^i\right) - \mathbb{C}:\frac{1}{2}\left(\nabla\boldsymbol{u}\left(\boldsymbol{x}^i\right) + \left(\nabla\boldsymbol{u}\left(\boldsymbol{x}^i\right)\right)^T\right)\right)^2, & \boldsymbol{x} & \in \boldsymbol{\Omega} , 
\end{aligned}
\label{loss2}
\end{equation}
where $\mathcal{L}$ is the total loss function, composed of four components:

a) $\mathcal{L}_{data}$ denotes the data loss, which uses stress or displacement as supervised data. Here, $N_{\sigma}$ and $N_{u}$ are the numbers of collocation points for the stress and displacement labels, respectively.

b) $\mathcal{L}_{bc}$ represents the boundary condition loss, accounting for the two most common boundary conditions applied in general solid mechanics problems, i.e., Dirichlet and Neumann boundary conditions. Similarly, $N_{d}$ and $N_{n}$ denote the number of collocation points for the Dirichlet and Neumann boundary conditions, respectively. $f_{bc}$ represents the applied traction, and $u_{bc}$ represents the boundary displacement.

c) $\mathcal{L}_{pde}$ denotes the PDE loss for the linear elasticity partial differential equation, as stated by Eq.\ref{dsmp2}.

d) $\mathcal{L}_{C}$ is the constitutive loss that links the displacement field and the stress field through the constitutive formulation in solid mechanics, where $\mathbb{C}$ is the fourth-order constitutive tensor dependent on the constitutive behaviours of so-defined problems. This loss term $\mathcal{L}_{C}$ serves a similar function as in the separate PINN model proposed in \cite{haghighat2021physics}, connecting the stress and displacement fields when they are approximated by independent networks.

\begin{figure}[htbp]
    \centering
    \includegraphics[width=0.99\linewidth]{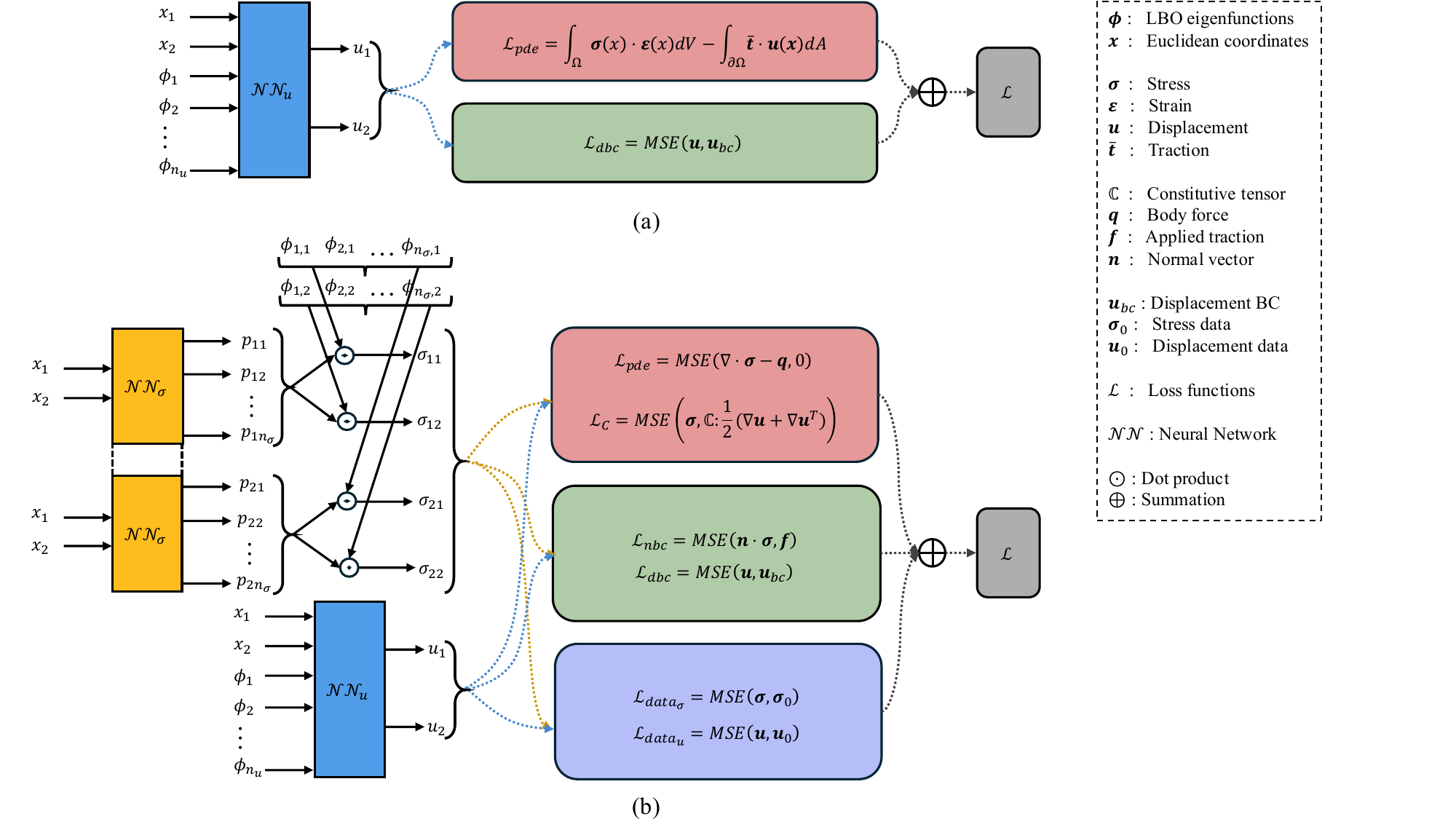}
    \caption{\color{black} Schematic of the Finite-PINN architecture for 2D cases: (a) Weak form Finite-PINN; (b) Strong form Finite-PINN. The neural network architectures and the corresponding data and physics loss functions are shown on the left. The symbols and parameters are presented on the right within the dotted box.}
    \label{fig:p4}
\end{figure}

A detailed schematic of both the weak form and strong form Finite-PINN architectures for a general 2D solid mechanics problem is shown in Fig.\ref{fig:p4}.

\subsection{Properties of Finite-PINN}
The proposed Finite-PINN model exhibits several remarkable properties that are advantageous for modelling solid mechanics problems. The function form of the Finite-PINN model is provided by Eq.\ref{fpinnl1}, which utilises the LBO eigenfunctions obtained from the problem defined in Eq.\ref{LBO}. The following are some significant features that make the Finite-PINN model particularly well-suited for solving general solid mechanics problems in a clear and efficient manner:

\begin{remark}
The Finite-PINN model inherently satisfies free-traction boundary conditions on free boundaries;
\label{r1}
\end{remark}
The mathematical statement of Remark \ref{r1} is
\begin{equation}
-\boldsymbol{n}\left(\boldsymbol{x}\right) \cdot \boldsymbol{\sigma}\left(\boldsymbol{x}\right) = \boldsymbol{0} \quad \forall \quad \boldsymbol{x} \in \partial \mathbf{\Omega}.
\label{mathr1}
\end{equation}
\color{black}
The weak form Finite-PINN automatically accounts for free-traction boundary conditions through its variational formulation in the weak form model. For the strong form Finite-PINN, the following proof is provided to demonstrate Remark \ref{r1}.
\color{black}
\begin{proof}
The second equation of Eq.\ref{LBO} can be rewritten using Einstein summation convention as:
\begin{equation}
- u_{i,j}\left(\boldsymbol{x}\right)n_j\left(\boldsymbol{x}\right) = 0 \quad \forall \quad x \in \partial \mathbf{\Omega},
\label{pc1}
\end{equation}
which determines that the used LBO eigenfunctions also follow that:
\begin{equation}
- \phi_{i,j}\left(\boldsymbol{x}\right)n_j\left(\boldsymbol{x}\right) = 0 \quad \forall \quad x \in \partial \mathbf{\Omega},
\label{pc2}
\end{equation}
and since $\sigma_{ij}\left(\boldsymbol{x}\right) = p_{ik}\left(\boldsymbol{x}\right)\phi_{k,j}\left(\boldsymbol{x}\right)$ as stated by Eq.\ref{fpinn}, 
\begin{equation}
- \sigma_{ij}\left(\boldsymbol{x}\right) n_j\left(\boldsymbol{x}\right) = - p_{ik}\left(\boldsymbol{x}\right) \phi_{k,j}\left(\boldsymbol{x}\right) n_j\left(\boldsymbol{x}\right) = - p_{ik}\left(\boldsymbol{x}\right) \left(\phi_{k,j}\left(\boldsymbol{x}\right) n_j\left(\boldsymbol{x}\right)\right) = 0  \quad \forall \quad x \in \partial \mathbf{\Omega},
\label{pc3}
\end{equation}
which is equivalent to Eq.\ref{mathr1} that is expressed in vector form.
\end{proof}

\begin{figure}[htbp]
    \centering
    \includegraphics[width=0.42\linewidth]{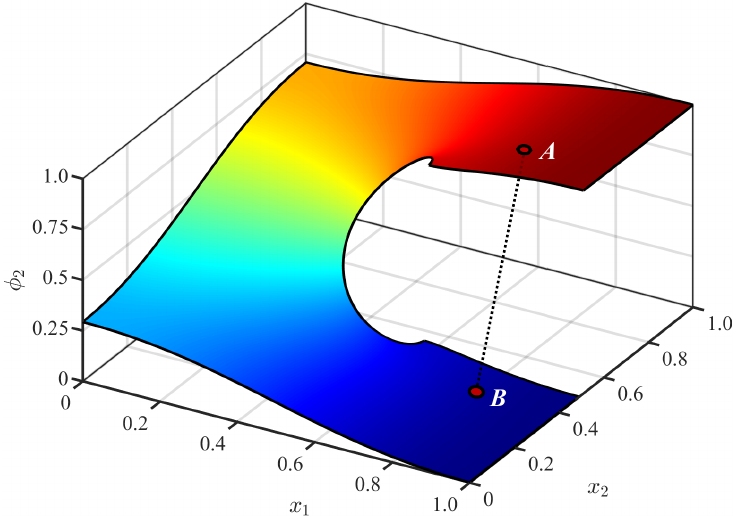}
    \caption{The topology of the open-notch structure in a Euclidean-Riemannian hybrid space, constructed using two Euclidean dimensions and a Riemannian manifold defined by the second LBO basis. The spatial distance between points A and B in this space is represented by the dotted line.}
    \label{fig:p5}
\end{figure}

\begin{remark}
The Finite-PINN model approximates solutions within a Euclidean-Topological hybrid space that includes the topological information of specific structures.
\end{remark}
As shown in Eq.\ref{fpinnl1}, the displacement field to be solved is represented by the neural network $\mathcal{NN}_u$, which takes as input the concatenation of the Cartesian coordinate $\boldsymbol{x}$ and the LBO basis functions $\boldsymbol{\phi}$. Since the Cartesian coordinate is defined in Euclidean space and the LBO basis functions are typical Riemannian manifolds, this combination of inputs forms a Euclidean-Riemannian hybrid space for the PINN model to approximate the solution. The distance between locations in this hybrid space is a combination of Euclidean and geodesic distances. Fig.\ref{fig:p5} shows the input space of the 2D open-notch model presented in Fig.\ref{fig:p5}, which combines the two Cartesian dimensions with the second LBO basis (the Fiedler vector) to form a 3D joint input space. It is evident that the new input space provides a distance representation more similar to the real geodesic distance (Fig.\ref{fig:p1}(c)) than the pure 2D input space Fig.\ref{fig:p1}(d).

{\color{black} Such a network architecture, which incorporates the basis functions $\phi$ as additional input dimensions, can be considered a geometric encoding technique for neural networks. Similar techniques have been used in~\cite{chen2023learning}, where Laplace–Beltrami operator (LBO) eigenfunctions serve as inputs to operator learning networks. Their approach is referred to as *Learning Neural Operators on Riemannian Manifolds*, specifically employing LBO manifolds, which are typical examples of Riemannian manifolds. On the other hand, our method incorporates Riemannian manifold information by augmenting the original Euclidean coordinate inputs with additional dimensions, thereby constructing a combined Euclidean–Riemannian space in which the neural network can more effectively learn the underlying behaviours.
}

\begin{remark}
The strong form Finite-PINN model features a separate architecture for approximating the stress and displacement fields independently.
\end{remark}
This separate architecture in the strong form Finite-PINN model simplifies the assignment of both Dirichlet and Neumann boundary conditions. Specifically, Dirichlet boundary conditions can be directly assigned to the neural network approximating the displacement field, while Neumann boundary conditions can be directly assigned to the neural network approximating the stress field.

From a computational method standpoint, this architecture can also be compared to some mathematical approaches that independently focus on the stress and displacement fields for solving specific solid mechanics problems. For example, the Airy stress function \cite{vallee2010airy} is used to approximate the stress field through a scalar potential function, which significantly aids in deriving analytical or semi-analytical solutions for 2D solid mechanics problems. Another example is the dual boundary element method, which separates the displacement and load computing in solid mechanics problems, making it effective for solving fracture and crack problems \cite{portela1992dual}.

It can be observed that the first two remarks directly address the two key challenges faced by PINN in solving general solid mechanics problems. The first remark provides a weak solution to Problem \ref{p1} by automatically applying the free traction constraints on free boundaries. The second feature modifies the input space of PINN from a uniform Euclidean space to a Euclidean-Topological hybrid space that is embedded with geometric information of the operating domain.

These remarks make the Finite-PINN model similar to the operation of finite element methods while retaining the use of neural networks to solve the solid mechanics PDEs. The connection between finite element concepts and neural networks is also the basis for the name 'Finite-PINN'. The implementation details of Finite-PINN are introduced in the next section.

\section{Implementations}
Since this method is geometry-specific, with a prepared geometric encoder focused on a fixed geometry for solving all related solid mechanics problems, the implementation is divided into two stages: an offline stage for preparing the geometric encoder and an online stage for problem-solving within the given geometry.
\label{sec4}
\subsection{Acquirement of partial derivatives}
The PDE loss term $\mathcal{L}_{pde}$ and the constitutive loss term $\mathcal{L}_{C}$ require certain partial derivatives of stress and displacement with respect to the spatial coordinates $\boldsymbol{x}$. Specifically, the terms that require partial derivatives with respect to the input are $\nabla \cdot \boldsymbol{\sigma}$ and $\nabla \boldsymbol{u}$, i.e., $\sigma_{ij,j}$ and $u_{i,j}$ in Einstein notation for tensors. By applying the chain rule of partial derivatives to Eq.\ref{fpinnl1} combining Eq.\ref{NNl1}, it is obtained that:
\begin{equation}
\begin{aligned}
u_{i,j} = & \ f_{i,j}.
\end{aligned}
\label{pd1}
\end{equation}
And since the LBO eigenfunctions $\boldsymbol{\phi}$ is also a function of $\boldsymbol{x}$, a further step is required to calculate the partial derivatives of $f_{i,j}$
\begin{equation}
\begin{aligned}
f_{i,j} = \frac{\partial f_{i}}{\partial x_{j}} = \frac{\partial f_{i}}{\partial x^*_{j}} + \sum^{n_u}_{k=1} \left(\frac{\partial f_{i}}{\partial \phi_{k}} \cdot \frac{\partial \phi_{k}}{\partial x_{j}}\right). \\
\end{aligned}
\label{pd2}
\end{equation}
Note that $x^*_{j}$ specifically represents the Cartesian input to the neural network, rather than the general Cartesian coordinate used in mathematics. 

\color{black}
The strong form Finite-PINN also requires the derivatives of the stress field $\sigma_{ij}$ with respective to the coordinates:
\begin{equation}
\begin{aligned}
\sigma_{ij,j} = & \ \left(p_{ik}\phi_{k,j}\right)_{,j} \\
              = & \ p_{ik,j}\phi_{k,j} + p_{ik}\phi_{k,jj}, \\
\end{aligned}
\label{pdkc}
\end{equation}
\color{black}

{\color{black}In this work, partial derivatives are computed using the above chain rule. In Eqs.\ref{pd2} and \ref{pdkc}, the terms $p_{ik,j}$, $\frac{\partial f_{i}}{\partial x_{j}}$, and $\frac{\partial f_i}{\partial \phi_{k}}$ are obtained from gradient of the neural networks through the computational graph during training. The terms $\phi_{k,j}$ and $\phi_{k,jj}$ represent the first and second derivatives of the LBO eigenfunctions, which are calculated using numerical methods.} This approach differs from the approach used in \cite{costabal2024delta}, where numerical methods (specifically finite element methods) were used to acquire all derivatives during the training process. The reason for using FEM in \cite{costabal2024delta} is that they require high-order derivatives in certain PDE problems, and applying the chain rule for partial derivatives becomes extremely complex when the derivative order is equal to or higher than 2. In contrast, even though the linear elasticity equation in this work involves partial derivatives of order 2, the implementation only requires derivatives up to the first order. This is because the Finite-PINN model employs two neural networks to approximate the stress and displacement fields separately, which reduces the governing PDE to a first-order PDE with respect to the stress $\sigma$, while the stress and displacement fields also maintain a first-order partial derivative relationship. By calculating the partial derivatives in this way, the Finite-PINN largely preserves the original implementation of PINN, thereby retaining the core advantages of PINN to the greatest extent possible.

\subsection{LBO eigenfunctions and their derivatives}
The Finite-PINN model requires the preparation of LBO eigenfunctions $\phi_k$ and their derivatives $\phi_{k,j}$ and $\phi_{k,jj}$, as demonstrated in Eqs.\ref{pd2} and \ref{pdkc}. The strong form of the PDE for the employed Laplace-Beltrami operator eigenvalue problem is provided in Eq.\ref{LBO}. For a 1D problem, Eq.\ref{LBO} reduces to an ordinary differential equation given by:
\begin{equation}
\left\{
\begin{aligned}
- u_{,xx}\left(x\right) = \lambda u\left(x\right) & \quad \forall \quad x \in \mathbf{\Omega} \\ 
- u_{,x}\left(x\right)=0  & \quad \forall \quad  x \in \partial \mathbf{\Omega}
\end{aligned}
\right. ,
\label{ef1}
\end{equation}
where $u_{,xx}$ and $u_{,x}$ denotes the second and first order derivatives of $u$ with respective to $x$. For a simple example of an 1D homogeneous segment, a possible solution of Eq.\ref{ef1} can be easily obtained as:
\begin{equation}
\left\{
\begin{aligned}
\lambda_k = & \left(\frac{k\pi}{L}\right)^2 \\ 
\phi_k \left(x\right)=  & \cos{\left(\frac{k\pi x}{L}\right)} 
\end{aligned}
\right. ,
\label{ef2}
\end{equation}
where $\lambda_k$ are the eigenvalues, $\phi_k$ are the eigenfunctions, and $L$ denotes the length of the domain. The solution is obtained by assuming a general trigonometric form and applying free Neumann boundary conditions at both ends of the segment: $ - u_{,x}\left(0\right) = 0$ and $ - u_{,x}\left(L\right) = 0$.

For general 2D or 3D problems, it is typically challenging to obtain analytical solutions from the strong form of the PDE Eq.\ref{LBO}. A common approach to address this challenge is to use numerical methods to derive discrete solutions based on a weak formulation of the PDE. This work introduces the finite element method and the meshfree method based on the radial basis functions to obtain LBO eigenfunctions for general 2D or 3D geometries. {\color{black} The finite element formulation and the RBF-based meshfree formulation of the eigenproblem are presented in the appendix. The codes for implementing both methods to calculate the eigenfunctions are provided in the next section.}

\subsection{Hybrid LBO eigenfunctions}
The proposed Finite-PINN model inherently satisfies the free-traction constraints $-\boldsymbol{n} \cdot \boldsymbol{\sigma} = \boldsymbol{0}$ on all boundaries, as stated in Remark \ref{r1} and Eq.\ref{mathr1}. However, a general solid mechanics problem (Fig.\ref{fig:p0}) is typically defined with specific Dirichlet and Neumann boundary conditions, for which the regions defined by those conditions often have $-\boldsymbol{n} \cdot \boldsymbol{\sigma} \neq \boldsymbol{0}$. To address this issue, specifically for the strong form model, we propose using hybrid LBO eigenfunctions as basis inputs for those solid mechanics problems. The details of getting the hybrid LBO eigenfunctions are demonstrated in Appendix \ref{apc}.

\color{black}
\subsection{Boundary conditions}
As introduced in Section~\ref{sec32}, Dirichlet and Neumann boundary conditions, other than free-traction boundary conditions, must be incorporated into the loss functions. For non-zero Neumann boundary conditions specifically, if they are explicitly defined in a forward problem, and by using the weak form Finite-PINN model, they can be automatically imposed through the loss term introduced in Eq.~\ref{loss1}. For the Finite-PINN model using the strong form loss, the boundary conditions are assigned using the collocation method, which minimises the difference between the predicted stress values and the prescribed traction values at the collocation points. This process is exactly analogous to assigning Dirichlet boundary conditions on the displacement field.

As for non-homogeneous Neumann boundary conditions, the obtained LBO eigenfunctions can be used to compute the surface normal, which defines the direction of the applied surface pressure. The approach involves calculating the gradient of the LBO basis functions at the boundaries to determine the surface normal direction. This is valid because the LBO eigenfunctions are obtained under traction-free boundary conditions, ensuring that their gradients at the boundaries naturally align with the outward surface normal.

From the loss function terms of the strong form Finite-PINN model, it is observed that Dirichlet and non-zero Neumann boundary conditions are still imposed through soft constraints, that is, by including corresponding loss terms. This is consistent with the original PINN framework. However, the key distinction is that homogeneous Neumann boundary conditions no longer need to be explicitly assigned, thanks to the finite geometric encoding used in stress field approximation. This avoids one of the most computationally expensive loss terms in training, especially for complex geometries.

Readers are also referred to \cite{sukumar2022exact} for techniques involving the exact imposition of boundary conditions to improve forward problem solutions. By imposing hard boundary conditions through direct modification of the network output, the boundary condition loss terms presented in Eqs.\ref{loss1} and \ref{loss2} can be eliminated, resulting in a simpler implementation. For example, exact Dirichlet boundary conditions can be imposed as follows:
\begin{equation}
\boldsymbol{u}_{bc}(\boldsymbol{x}) = g(\boldsymbol{x}) + R(\boldsymbol{x}) \boldsymbol{u}_{NN}(\boldsymbol{x}),
\label{ebc}
\end{equation}
where $g(\boldsymbol{x})$ is a smooth field extended from the boundary data, and $R(\boldsymbol{x})$ is a distance function that equals zero on the domain boundaries. While these methods are fully compatible with the Finite-PINN framework, a detailed exploration of their application is beyond the scope of this work.

\subsection{Training}
The training of PINNs is essentially the process of solving the PDE. In this work, we use multilayer perceptron to approximate all problems, which is a well-established approach in modern deep learning model development. We consistently apply best practices while keeping the process simple by using the \textit{ADAM} optimiser for all training tasks. A constant learning rate, ranging from $1 \times 10^{-4}$ to $1 \times 10^{-3}$, is used throughout the entire training process without employing any learning rate scheduling.

Sampling points are also a critical factor that may influence the training process and, consequently, the results. Since the Finite-PINN model requires LBO eigenfunctions or their derivatives as geometric encodings of certain geometries, the sampling points are usually determined by the numerical methods used to compute these eigenfunctions. For example, if the finite element method is used to solve the LBO eigenproblem, a convenient way to select the collocation points is to use the integration points (such as Gaussian integration points when Gaussian quadrature is applied) from FEM. This simplifies obtaining the values and derivatives of the resulting fields at those points. Additionally, the areas or volumes required by the integrals in the weak form loss function (Eq.~\ref{loss1}) can be easily computed using the Jacobian matrix in FEM. Similarly, in meshfree methods that employ radial basis functions (RBFs), as introduced in Appendix \ref{apb}, the integration points used in RBF methods are also suitable as sampling points during training.

In this context, the distribution of sampling points depends on the numerical methods adopted. The advantage is that many well-established and mature discretisation techniques exist in traditional numerical methods. For example, there are numerous robust tools available for finite element mesh generation. These methods, which facilitate finite element analysis, also significantly support the training of PINN models when their selected integration points are used as sampling or collocation points. For instance, in the finite element method (FEM), a finer mesh is typically applied near boundaries with significant geometric variations and stress concentrations. This refinement strategy can also benefit the PINN model by enabling more accurate approximations in those critical regions. Besides, if a uniform structured mesh is used in FEM to obtain the LBO eigenfunctions, there is no need to calculate the representative areas or volumes of each collocation point for the weak form formulation, since all points represent equal areas or volumes in the uniform mesh. In this case, a simple averaging of all collocation point values can replace the numerical integration in Eq.\ref{loss1}, further simplifying the process.
\color{black}

\subsection{Workflow}
Based on the above introduction, the details of the two stages are: a) an offline stage to prepare the required LBO eigenfunctions for a given structure, and b) an online stage to solve specific solid mechanics problems. The offline stage prepares the data $\phi_k$, $\phi_{k,j}$, and $\phi_{k,jj}$.

The Finite-PINN model introduces additional hyperparameters beyond those in common neural networks due to its specific architecture. The two fundamental ones are $n_u$ and $n_\sigma$, which represent the number of LBO eigenfunctions used to approximate the displacement and stress fields, respectively.

The parameter $n_u$ is focused on incorporating manifold dimensions into the input space and can be set to zero for problems with a uniform domain, such as square or cubic domains. On the other hand, for the Finite-PINN using strong from loss, $n_\sigma$ must be large enough to encompass all the basis functions needed to approximate a random stress field accurately. Thus, the selection of $n_u$ and $n_\sigma$ depends on the specifics of the problem at hand.

\begin{figure}[t]
\color{black}
    \centering
    \includegraphics[width=0.999\linewidth]{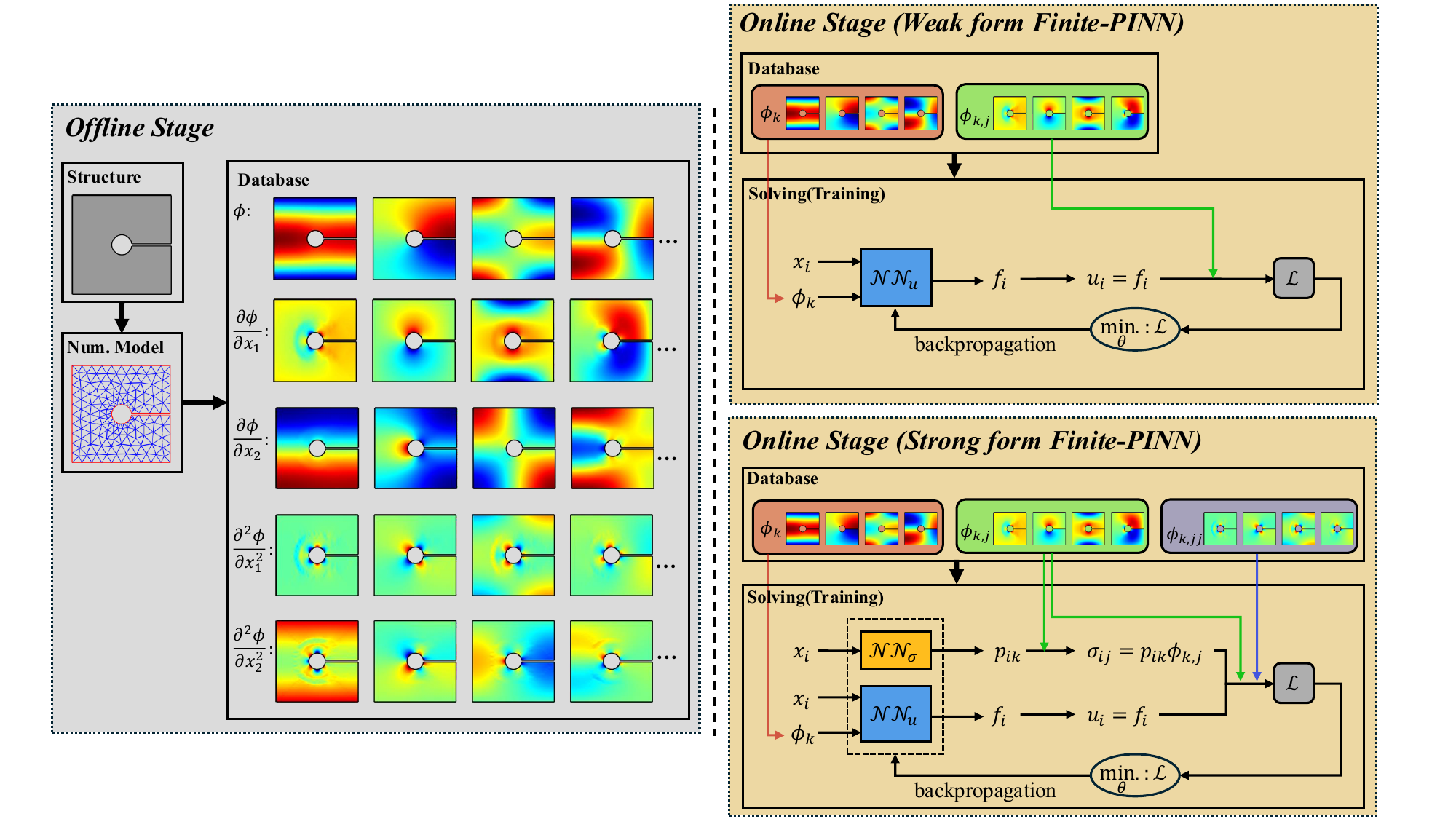}
    \caption{Two-stage implementation of the Finite-PINN method. The offline stage (left) focuses on preparing the database, while the online stage (right) is used to solve specific solid mechanics problems. }
    \label{fig:p6}
\end{figure}

A further and more detailed discussion on how to select appropriate values for $n_u$ and $n_\sigma$ is provided in Section \ref{sec6}. An illustration of the complete implementation of the Finite-PINN model for solving a solid mechanics problem is presented in Fig.\ref{fig:p6}.

\section{Benchmarks and examples}
\label{sec5}
This section presents several cases where the Finite-PINN model is used to solve different solid mechanics problems. The code is written in PyTorch and available at \href{https://github.com/hl4220/Finite_PINN.git}{GitHub}, including the offline stage preparation of the LBO eigenfunctions by FEM (Appendix \ref{apa}) and RBF-basedmeshfree method (Appendix \ref{apb}). In all cases, unless otherwise stated, the test loss is calculated by comparing the predictions with reference solutions obtained from FEM with a fine mesh at all FEM nodal locations. \color{black}Illustrations of whether the examples are solved using the strong form or the weak form Finite-PINN are provided in each case. Generally, forward problems with all known boundary conditions and parameters are solved using the weak form model, while inverse problems are addressed using the strong form model. Further discussion comparing the weak and strong form models is presented in Section \ref{sec6}.
\color{black}

\subsection{Example 1: a 1D problem}
A solid mechanics problem in the 1D case reduces the partial differential equation, Eq.\ref{dsmp2}, to an ordinary differential equation:
\begin{equation}
\left\{
\begin{aligned}
\sigma_{,x} = & 0 \\
\sigma = & C \cdot u_{,x} \\
\end{aligned}
\right. ,
\label{Ex1}
\end{equation}
where $C$ is the constitutive constant. The Finite-PINN model presented in Eq.\ref{fpinn} could thus be simplified as:
\begin{equation}
\left\{
\begin{aligned}
\sigma\left(x\right) = & p_k\left(x\right) \phi_{k,}\left(x\right) \\
u\left(x\right) = & f\left(x\right) \\
\end{aligned}
\right. .
\label{Ex2}
\end{equation}
\begin{figure}
    \centering
    \includegraphics[width=0.95\linewidth]{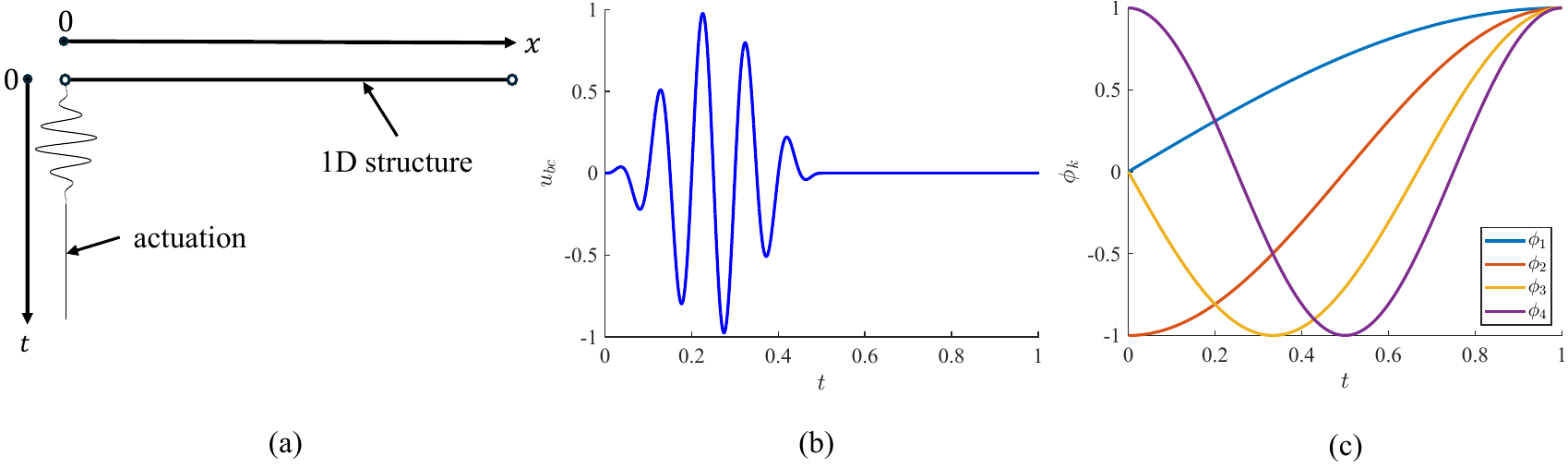}
    \caption{Definition of Example 1. (a) The problem is to actuate a wave package on the left end of the finite rod to find its response over the spatial and temporal domain. (b) The excitation wave package. (c) The four eigenfunctions use in Example 1.}
    \label{fig:p7}
\end{figure}

Furthermore, since a 1D problem involves a scalar field for both displacement $u$ and stress $\sigma$, the connection between the stress field $\boldsymbol{\sigma}$ and the displacement field $u$ is straightforward, represented by a simple first-order derivative: $\boldsymbol{\sigma} = C \cdot \frac{\partial u}{\partial x}$. In this case, there is no need to separate the learning of stress and displacement. The Finite-PINN model in Eq.\ref{Ex2} can be simplified to an alternative model, as follows:
\begin{equation}
\begin{aligned}
u = \mathcal{NN}\left(\phi^{n_u};\boldsymbol{\theta}\right) .
\end{aligned}
\label{Ex3}
\end{equation}
This simplified Finite-PINN model still inherently satisfies the free boundary constraints, as the chain rule implies that:
\begin{equation}
\begin{aligned}
u_{,x} = \frac{\partial u}{\partial x} =  \sum_k \left( \frac{\partial u}{\partial \phi_k} \cdot \frac{\partial \phi_k}{\partial x} \right)
\end{aligned} .
\label{Ex4}
\end{equation}
and all LBO basis functions satisfy the free Neumann boundary conditions: $\frac{\partial \phi_k}{\partial x} = 0$ at both ends (boundaries of a 1D structure).

In this example, we investigate the dynamic response of a homogeneous solid rod, which can be treated as a 1D problem. The structural dynamic equation for a 1D solid problem is given by:
\begin{equation}
\rho  u_{,tt} + c u_{,t} - C u_{,xx} = 0 \\
\label{dy2}
\end{equation}
where $\rho$ is the density of the rod material, and $c$ is the damping coefficient. Therefore, the neural network architecture used for this 1D solid problem is defined as:
\begin{equation}
u = \mathcal{NN}\left(\boldsymbol{\phi}^{n_u}\left(x\right),t;\boldsymbol{\theta}\right)
\label{Exp3}
\end{equation}
where $t$ is the time dimension. The loss function for learning is stated as:
\begin{equation}
\begin{aligned}
\mathcal{L} & = \mathcal{L}_{bc} + \mathcal{L}_{pde} , \\
\mathcal{L}_{bc} & = \frac{1}{N_{d}}\sum^{N_{d}}_{i=1}\left(u\left(x,t\right)-u_{bc}\left(x,t\right)\right)^2, & x & \in \partial \Omega, & t & \in [0,T] , \\
\mathcal{L}_{pde} & = \rho u_{,tt}\left(x,t\right) + c u_{,t}\left(x,t\right) - C u_{,xx}\left(x,t\right), & x & \in \Omega, & t & \in [0,T] ,
\end{aligned}
\label{lossk}
\end{equation}
where $T$ represents the time limit. Note that there is no labelled data involved in solving this case, so there is no data loss term in the loss function as can be observed.
\begin{figure}[htbp!]
    \centering
    \includegraphics[width=0.999\linewidth]{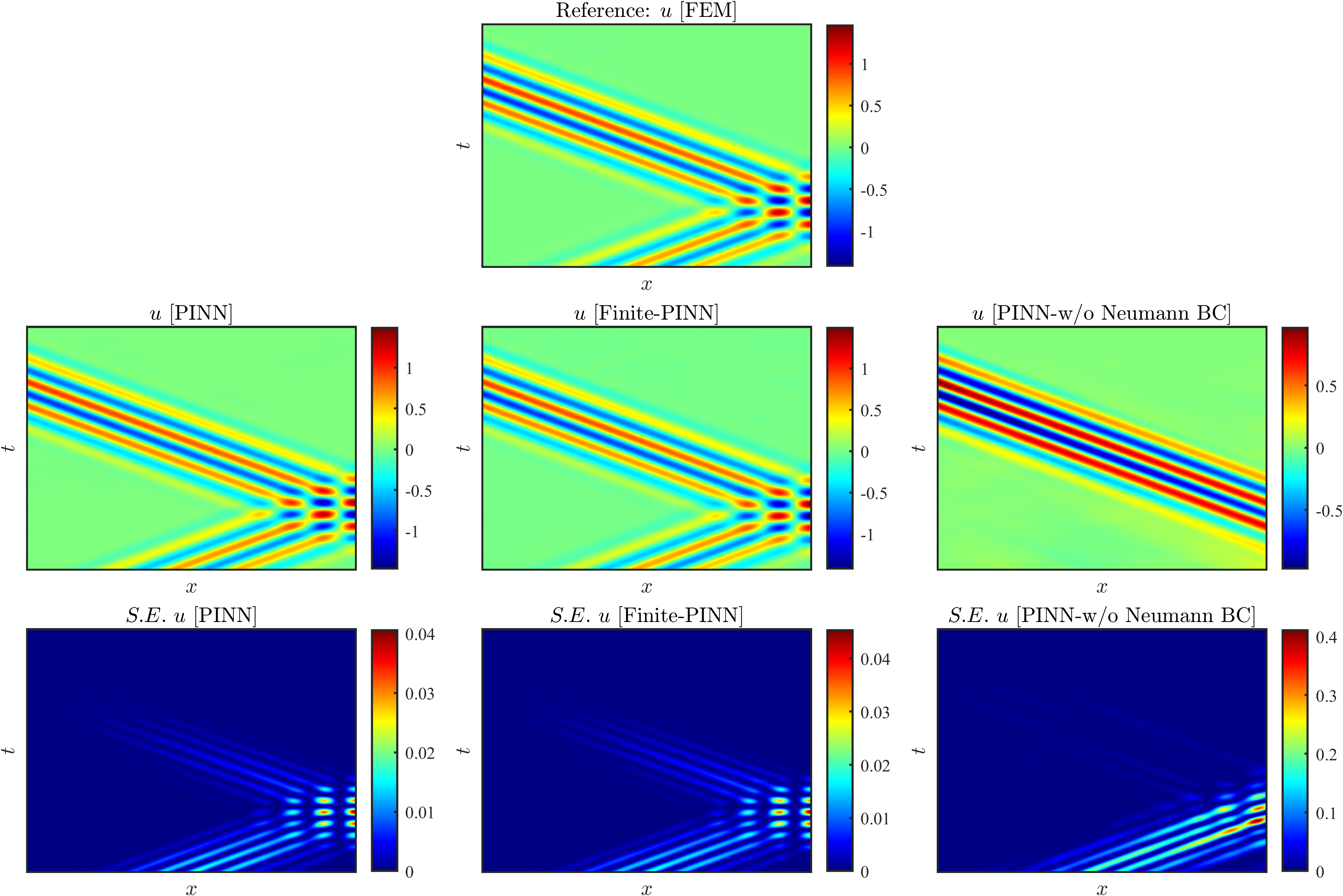}
    \caption{Results of Example 1. The top figure shows the reference result calculated by FEM with a fine mesh and time steps. The second row presents the results obtained by PINN, Finite-PINN, and PINN without the free-surface Neumann boundary condition, respectively, from left to right. The third row shows the corresponding squared errors of these models compared to the reference result.}
    \label{fig:DY1}
\end{figure}

In this example, the problem is defined to determine the dynamic response of a homogeneous solid rod with an input signal (displacement) applied at the left end, while the right end remains free. The problem setup and the input actuation signal are illustrated in Fig.\ref{fig:p7}. The problem is solved using both FEM and the Finite-PINN model. Additionally, the traditional PINN model is also employed for comparison.

In this 1D example, the LBO eigenfunctions can be explicitly obtained by directly solving the problem defined in Eq.\ref{LBO} in one dimension. One of the analytical solution can be obtained as: 
\begin{equation}
\phi\left(x\right) = cos\left(\frac{k \pi \left(x-L\right)}{2L}\right), \quad k = 1,2,3, \dots
\label{anc}
\end{equation}
In such a case, the utilised neural network becomes:
\begin{equation}
u = \mathcal{NN}\left(cos\left(\frac{k \pi \left(x-L\right)}{2L}\right),t;\boldsymbol{\theta}\right), \quad k = 1,2,3, \dots
\label{Exc3}
\end{equation}

The neural network $\mathcal{NN}$ used here consists of 4 hidden layers with 64 nodes each. We use the \textit{SIREN} neural network for the approximation, in which the activation functions are sine functions. \textit{SIREN} is an implicit neural network architecture \cite{sitzmann2020implicit} that has been previously used to solve point-source wave propagation problems with the PINN method in \cite{huang2021solving}. The number of trainable parameters is 35,409. 

\begin{figure}[htbp!]
    \centering
    \includegraphics[width=0.999\linewidth]{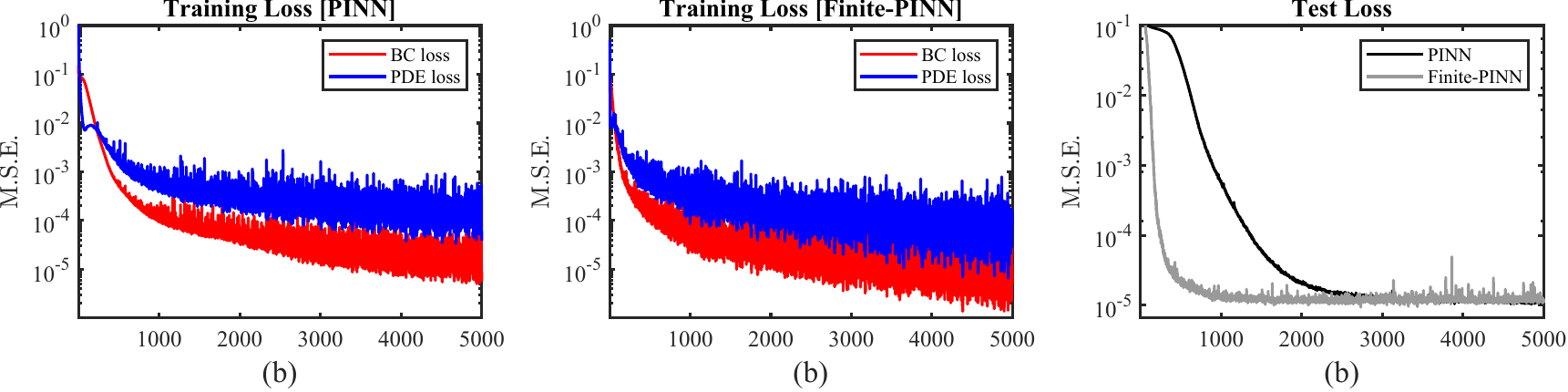}
    \caption{(a) Evolution of training loss over epochs for Example 1 using the traditional PINN model. (b) Evolution of training loss over epochs for Example 1 using the Finite-PINN model. (c) Evolution of test loss for both models.}
    \label{fig:DY2}
\end{figure}

The first four LBO eigenfunctions are used as inputs to the neural network, i.e., $k=1,2,3,4$ in Eq.\ref{Exc3}. These four eigenfunctions are shown in Fig.\ref{fig:p7}(c). We used 500 spatial collocation points in the domain for the PDE loss and one spatial collocation point (at the left end) for the boundary condition loss. The time domain is divided into 2,000 time steps for training. The neural network is trained for 5,000 epochs with a batch size of 32. The training results are shown in Fig.\ref{fig:DY1}. The evolution of the training loss and test loss is illustrated in Fig.\ref{fig:DY2}.

Fig.\ref{fig:DY1} depicts the solutions of the dynamic equation solved by FEM, PINN, Finite-PINN, and PINN without considering the free boundary condition. The error distributions show that both the PINN and Finite-PINN models achieve results comparable to the reference FEM solutions after training. However, as can be seen in Fig.\ref{fig:DY2}, which presents the evolution of the training and test losses during the learning process, the Finite-PINN model demonstrates faster convergence compared to the traditional PINN. This is because the Finite-PINN model inherently satisfies the free-surface boundary condition, eliminating the need for an additional term to account for the free boundary in the total physics-informed loss, thus simplifying the training process by reducing the number of objectives to optimise.

\begin{figure}[htbp!]
    \centering
    \includegraphics[width=0.75\linewidth]{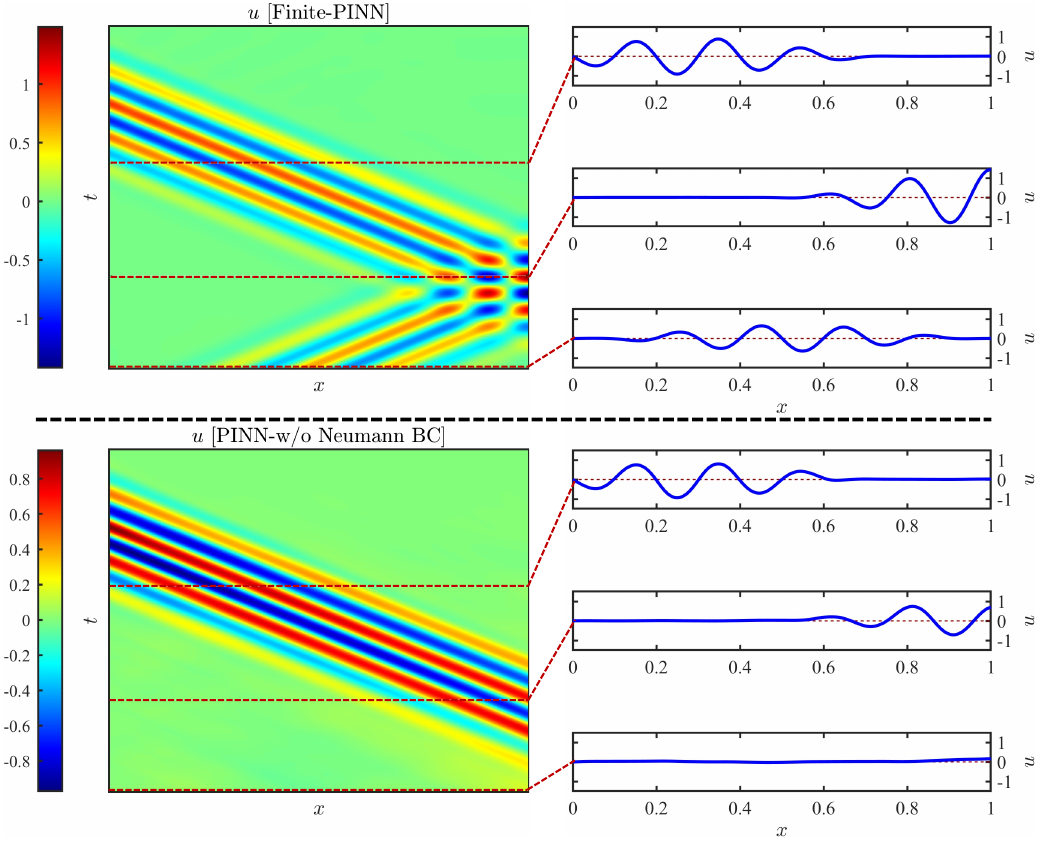}
    \caption{Displacement response at specific times ($t=0.4s,0.7s,1.0)$ for the problem defined in Example 1, solved by the Finite-PINN model (top) and the traditional PINN model without applying the free boundary condition (bottom).}
    \label{fig:p9}
\end{figure}

Fig.\ref{fig:DY1} also shows the results obtained by the PINN model without free-surface loss terms. A detailed comparison is provided in Fig.\ref{fig:p9}, illustrating the displacement response at different times. It is observed that the actuation moves from the left end to the right end but vanishes at the right end without any reflections for the traditional PINN without considering the free boundary loss. This behaviour corresponds to the fact that the mesh-free representation of PINN models is defined over an infinite domain. 

\subsection{Example 2: a benchmark for a 2D square structure including both forward and inverse problems}
This example aims to solve a solid mechanics problem for a 2D square structure, including both a forward problem and an inverse problem. The example is used to validate the proposed method and verify the implementation.
\begin{figure}[ht]
    \centering
    \includegraphics[width=0.61\linewidth]{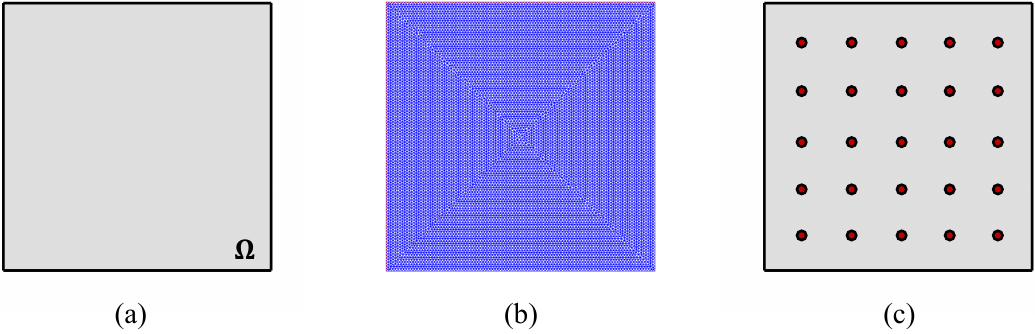}
    \caption{(a) Square domain defined in Example 2. (b) Finite element mesh of the square. (c) Locations where data are acquired for the inverse problem.}
    \label{fig:p10}
\end{figure}

\begin{figure}[ht]
    \centering
    \includegraphics[width=0.59\linewidth]{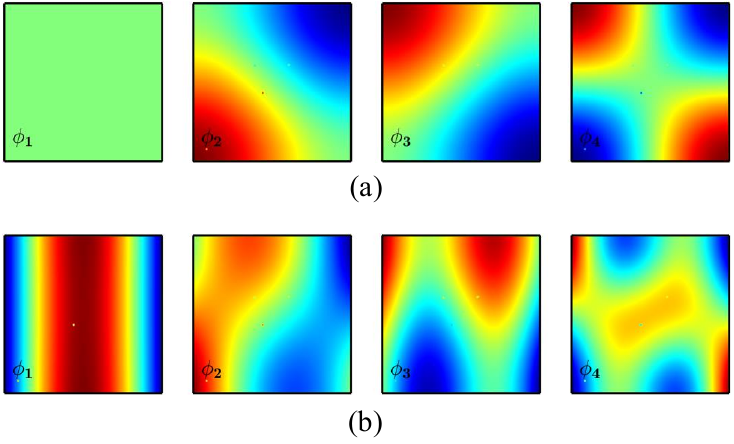}
    \caption{The first 4 Homogeneous LBO eigenfunctions (a) and the first 4 hybrid LBO eigenfunctions (b) of the square structure used in Example 2. (a) is used in the weak form model and (b) is employed by the strong form model. }
    \label{fig:sq3}
\end{figure}

The structure is a 2D square with a side length of 1. The bottom boundary is fully fixed, and various boundary conditions are applied to the top surface of the square. An illustration of the structure is provided in Fig.\ref{fig:p10}(a). The FEM formulation introduced in Section \ref{sec4} is used to calculate the LBO eigenfunctions. The structure is meshed using first-order triangular finite elements, and the mesh is shown in Fig.\ref{fig:p10}(b). The LBO eigenfunctions used in this model are presented in Fig.\ref{fig:sq3}. 

\subsubsection{Forward problem}
The simple 2D square is assumed to be fixed on the ground by its bottom side and a load or displacement can be applied or assigned on its top surface. The forward problems in this case are to solve the displacement of this structure field under different boundary conditions with no labelled data. 

\color{black}
In this case, the hard Dirichlet boundary conditions are imposed by Eq.\ref{ebc}. The weak form loss function is then stated as:
\begin{equation}
\begin{aligned}
\mathcal{L} & = \mathcal{L}_{pde} , \\
\mathcal{L}_{pde} & = \sum^{N_{pde}}_{i=1}v^i \cdot \boldsymbol{\varepsilon}\left(\boldsymbol{x}^i\right) \cdot \left(\mathbb{C} : \boldsymbol{\varepsilon}\left(\boldsymbol{x}^i\right)\right) - \sum^{N_{nbc}}_{i=1} a^i \cdot \overline{\boldsymbol{t}} \cdot \boldsymbol{u}\left(\boldsymbol{x}^i\right), & \boldsymbol{x}^i & \in \boldsymbol{\Omega} .
\end{aligned}
\label{losskf}
\end{equation}
\color{black}
The strong form loss function is stated as:
\begin{equation}
\begin{aligned}
\mathcal{L} & = \mathcal{L}_{nbc} + \mathcal{L}_{pde} +\mathcal{L}_{C} , \\
\mathcal{L}_{nbc} & = \frac{1}{N_{nbc}}\sum^{N_{nbc}}_{i=1}\left(\boldsymbol{n}\cdot \boldsymbol{\sigma}\left(\boldsymbol{x}^i\right)-f_{bc}\left(\boldsymbol{x}^i\right)\right)^2,  & \boldsymbol{x} & \in \partial \boldsymbol{\Omega}_n ,\\
\mathcal{L}_{pde} & = \frac{1}{N_{pde}}\sum^{N_{pde}}_{i=1}\left(\nabla \cdot \boldsymbol{\sigma}\left(\boldsymbol{x}^i\right)\right)^2, & \boldsymbol{x} & \in \boldsymbol{\Omega}, \\
\mathcal{L}_{C} & = \frac{1}{N_{C}}\sum^{N_{C}}_{i=1}\left(\boldsymbol{\sigma}\left(\boldsymbol{x}^i\right) - \mathbb{C}:\frac{1}{2}\left(\nabla\boldsymbol{u}\left(\boldsymbol{x}^i\right) + \left(\nabla\boldsymbol{u}\left(\boldsymbol{x}^i\right)\right)^T\right) \right) ^2, & \boldsymbol{x} & \in \boldsymbol{\Omega}, 
\end{aligned}
\label{loss3k}
\end{equation}
where$\mathcal{L}_{nbc}$ denote the loss terms for the Neumann boundary condition, respectively, as defined on $\partial \boldsymbol{\Omega}_n$. The number of collocation points $N_nbc = 101$, $N_{pde} = 10285$. No data loss term is included since there is no labelled data used for training in this forward problem. 

The Finite-PINN model used is defined in Eqs.\ref{fpinnl1} and \ref{fpinn}. The first 4 eigenfunctions are used in both $\mathcal{NN}_\sigma$ and $\mathcal{NN}_u$, which means that $\mathcal{NN}_\sigma$ takes an input of dimension 2 and outputs in a dimension of 8, while $\mathcal{NN}_u$ takes an input of dimension 6 and outputs in a dimension of 2. In this forward problem, the homogeneous eigenfunctions as shown in Fig.\ref{fig:sq3}(a) are utilised in the model training with the weak form loss, and the hybrid eigenfunctions in Fig.\ref{fig:sq3}(b) are used by the strong form model. Both networks have 4 hidden layers with 64 nodes each, using the \textit{GELU} activation function. The neural network is trained for 1,000 epochs with a batch size of 200. The material of the structure has an elastic modulus of $1.0$ and a Poisson's ratio of $0.3$.

Three cases are investigated in this problem, corresponding to three different boundary conditions assigned to the top surface of the square while the bottom surface is fully fixed:
\begin{itemize}
\item Case 1: $\boldsymbol{u}\left(\boldsymbol{x}_{b}\right) = \boldsymbol{0}$, $f_{bc}\left(\boldsymbol{x}_{t}\right) = x_1$,
\item Case 2: $\boldsymbol{u}\left(\boldsymbol{x}_{b}\right) = \boldsymbol{0}$, $f_{bc}\left(\boldsymbol{x}_{t}\right) = 0.5 + sin\left(2\pi \cdot x_1\right)$,
\item Case 3: $\boldsymbol{u}\left(\boldsymbol{x}_{b}\right) = \boldsymbol{0}$, $u_1\left(\boldsymbol{x}_{t}\right) = 1, u_2\left(\boldsymbol{x}_{t}\right) = 1$,
\end{itemize}
in which $\boldsymbol{x}_{b}$ denotes the coordinates of the bottom surface and $\boldsymbol{x}_{t}$ represents the top surface, the dimension number is defined as that $1$ represents the horizontal direction and $2$ denotes the vertical direction. 
\begin{figure}[htbp]
    \centering
    \includegraphics[width=0.99\linewidth]{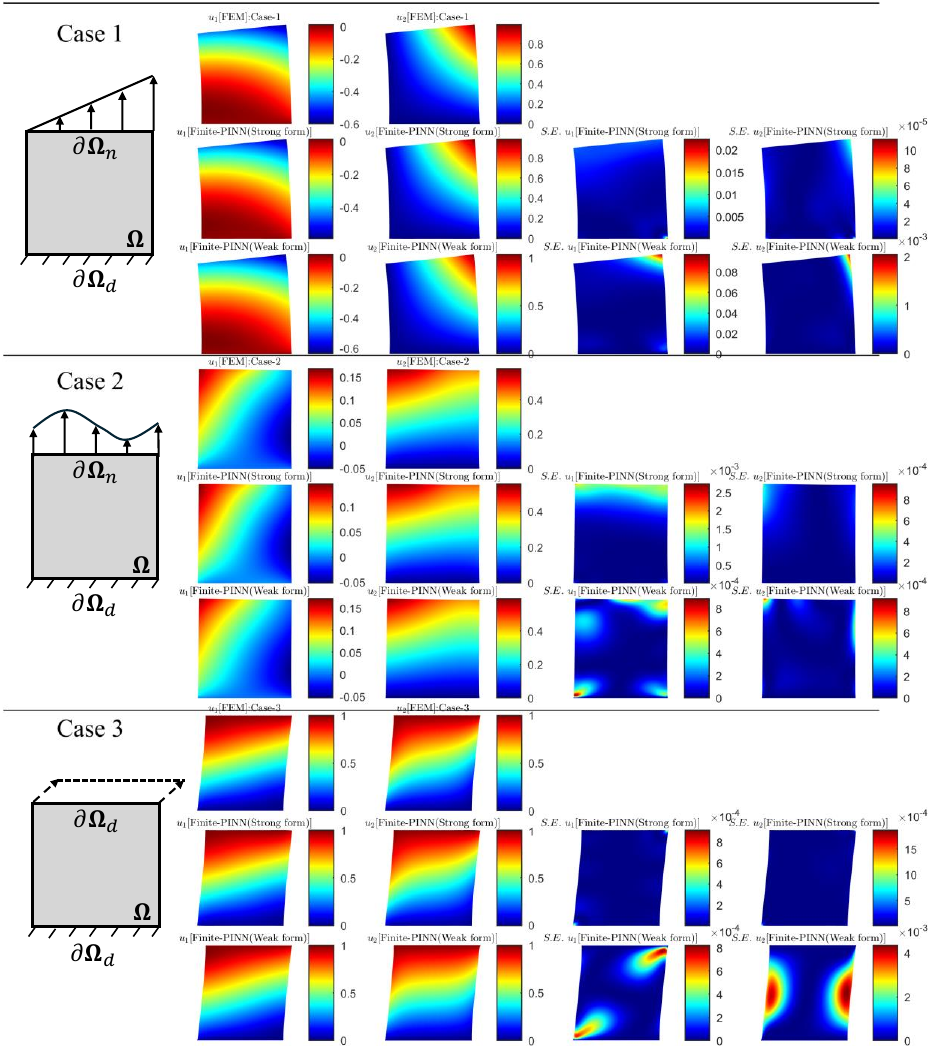}
    \caption{\color{black} Results of the three forward problems in Example 2. The left side shows the schematic representation of the three cases. The right side presents the reference field calculated by FEM, the predicted field by the Finite-PINN models, and their corresponding errors.}
    \label{fig:p8}
\end{figure}

\begin{figure}[ht]
    \centering
    \color{black}
    \includegraphics[width=0.99\linewidth]{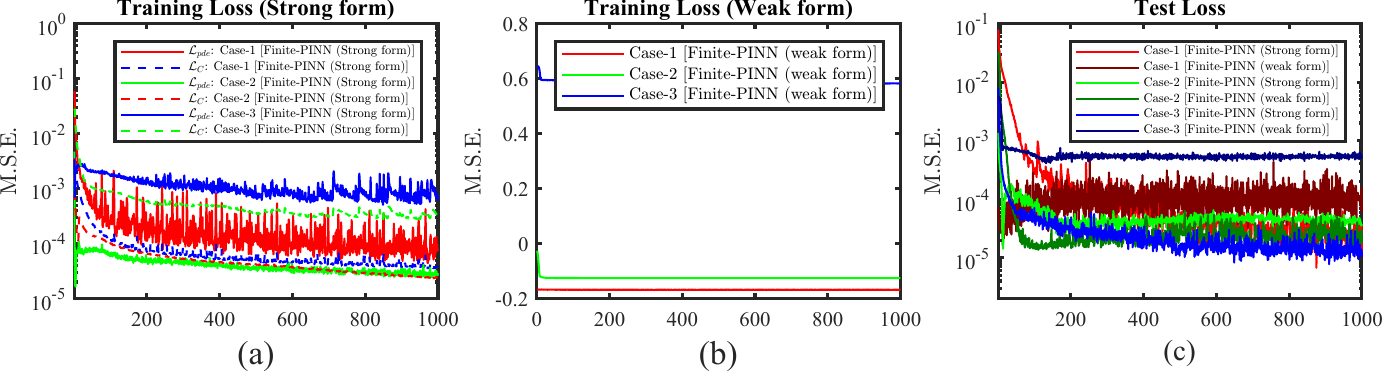}
    \caption{(a) and (b) show the training losses of models using the strong form loss and weak form loss, respectively. (c) shows the evolution of the test loss for all three cases.}
    \label{fig:sqk3}
\end{figure}

The schematic of the boundary conditions and loading for the three cases, along with the calculation results, is illustrated in Fig.\ref{fig:p8}. The evolution of the training and test loss is shown in Fig.\ref{fig:sqk3}. The test loss is calculated by comparing the predictions with reference values at all FEM nodes, as obtained by the FEM, since no supervised data is used in training.

The results and their corresponding errors indicate that the Finite-PINN model achieves results comparable to those of the FEM, effectively handling problems with both Dirichlet or Neumann boundary conditions, using both the weak form loss and the strong form loss.

\subsubsection{Inverse problem}
The inverse problem uses the same benchmarks as those solved in the forward problem, aiming to determine the load applied to the structure in the first two cases. The FEM data concerning the strains at specific points are provided as labelled data to identify the load applied to the structures. The reason for using strain data instead of displacement data is that, in real-world scenarios, it is generally easier to obtain strain information from structures compared to displacement data. In engineering, most load identification problems focus on reconstructing the applied load based on the strains obtained from strain gauges \cite{sekula2013real}. Fig.\ref{fig:p10}(c) shows the locations of the 25 collocation points, which correspond to the exact positions of a $5 \times 5$ mesh and could also be regarded as the positions of strain gauges in practice. The four hybrid eigenfunctions shown in Fig.\ref{fig:sq3}(b) is used in this problem. 

\begin{figure}[ht]
    \centering
    \includegraphics[width=0.95\linewidth]{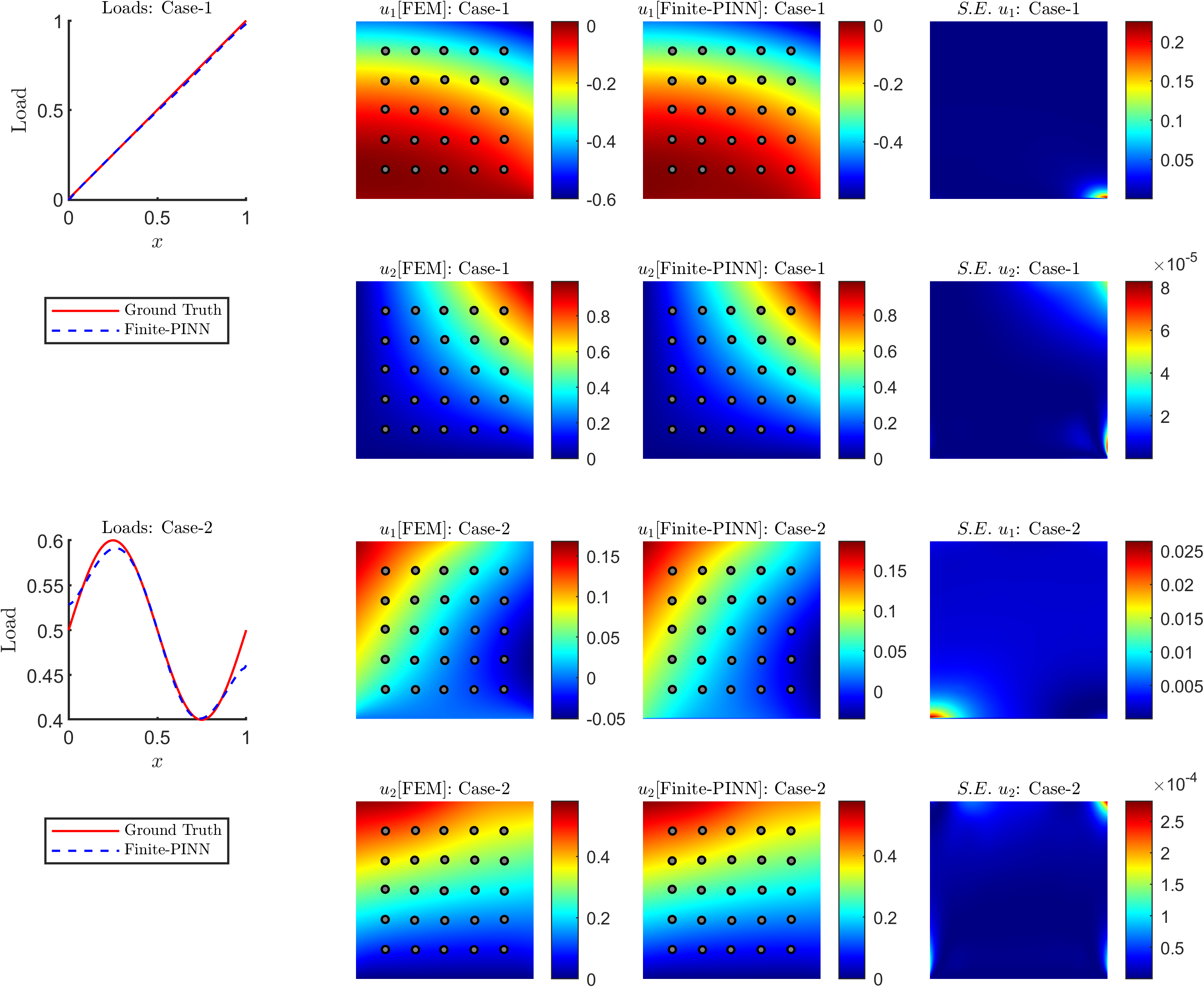}
    \caption{Results of the two inverse problems in Example 2. The left side shows the predictions of the load along with the ground truth. The right side presents the reference field calculated by FEM, the predicted field by the Finite-PINN model, and their corresponding errors.}
    \label{fig:sqi1}
\end{figure}

\begin{figure}[ht]
    \centering
    \includegraphics[width=0.99\linewidth]{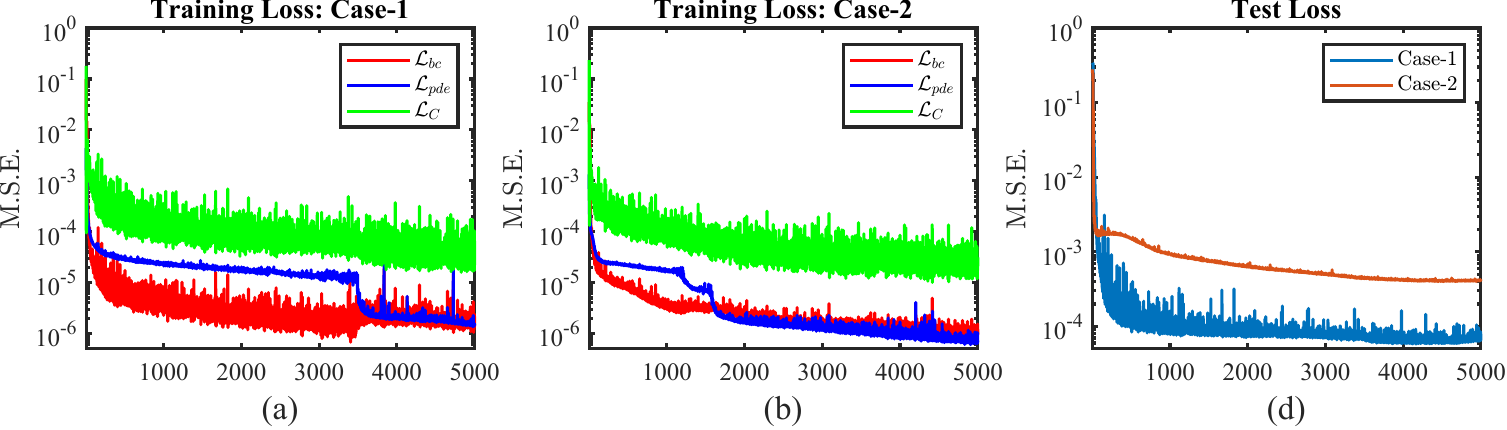}
    \caption{(a)(b) presents the evolutions of the training losses of solving the inverse problems 1 and 2 in Example 2. (c) presents the evolutions of the test loss of the two problems. }
    \label{fig:sqi3}
\end{figure}

The partial differential equations of the inverse problem are the same as the forward problem, and the strong form loss function for training is stated as:
\begin{equation}
\begin{aligned}
\mathcal{L} & = \mathcal{L}_{data} + \mathcal{L}_{pde} + \mathcal{L}_{C} , \\
\mathcal{L}_{data} & = \frac{1}{N_{data}}\sum^{N_{data}}_{i=1}\left(\frac{1}{2}\left(\nabla\boldsymbol{u}\left(\boldsymbol{x}^i\right) + \left(\nabla\boldsymbol{u}\left(\boldsymbol{x}^i\right)\right)^T\right) - \boldsymbol{\varepsilon}^0\right)^2,  & x_1 \in \partial \Omega,\quad x_2 & = L, \\
\mathcal{L}_{pde} & = \frac{1}{N_{pde}}\sum^{N_{pde}}_{i=1}\left( \Delta \boldsymbol{\sigma}\left(\boldsymbol{x}^i\right)\right)^2, & \boldsymbol{x} & \in \boldsymbol{\Omega}, \\
\mathcal{L}_{C} & = \frac{1}{N_{C}}\sum^{N_{C}}_{i=1}\left(\boldsymbol{\sigma}\left(\boldsymbol{x}^i\right) - \mathbb{C}:\frac{1}{2}\left(\nabla\boldsymbol{u}\left(\boldsymbol{x}^i\right) + \left(\nabla\boldsymbol{u}\left(\boldsymbol{x}^i\right)\right)^T\right)\right) ^2, & \boldsymbol{x} & \in \boldsymbol{\Omega},
\end{aligned}
\label{loss2i}
\end{equation}
where $\boldsymbol{\varepsilon}^0$ represents the given strain data. The load identification results and the corresponding reconstructed displacement field are presented in Fig.\ref{fig:sqi1}. As previously stated, $N_{data}=5\times5=25$. The evolution of the training loss and test loss is shown in Fig.\ref{fig:sqi3}. Here, the test loss refers to the difference between the loads predicted by the Finite-PINN model and the ground truth loads applied to the structure. The neural networks and hypreparameters used for the inverse problems are the same as those used in the previous forward problems. 

The results in Fig.\ref{fig:sqi1} show satisfactory performance, validating the capability of the proposed Finite-PINN model in solving inverse problems. {\color{black} Only the results obtained using the strong form loss are presented here, as the training with the weak form loss diverged during the process.}

\subsection{Example 3: a thermal expansion problem}
\begin{figure}[ht]
    \centering
    \includegraphics[width=0.42\linewidth]{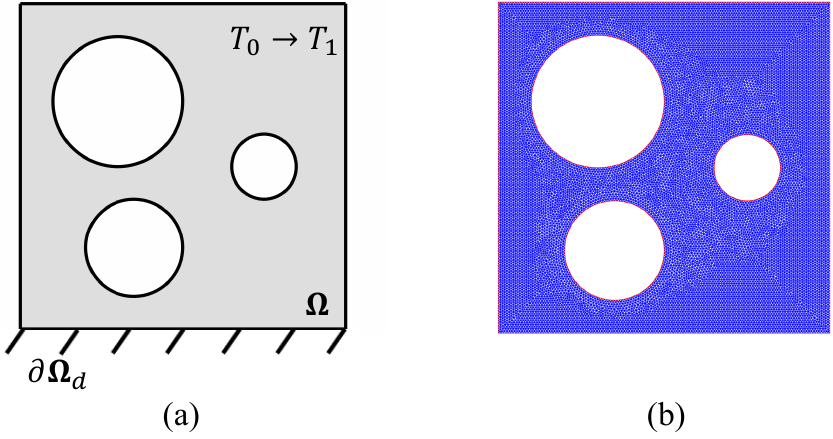}
    \caption{(a) The porous structure used in Example 3. (b) Finite element mesh of the porous structure.}
    \label{fig:p11}
\end{figure}

This example explores a 2D thermal expansion problem on a more complex porous structure. The structure under investigation is a 2D porous square with a side length of 1 and three holes. The bottom side of the structure is fully fixed as a Dirichlet boundary condition. The definition of the problem and the finite element mesh of the structure are illustrated in Fig.\ref{fig:p11}.

The partial differential equation of the thermal expansion problem is defined as:
\begin{equation}
\left\{
\begin{aligned}
& \nabla \cdot \boldsymbol{\sigma}\left(\boldsymbol{x}\right) = 0, & \quad \forall \quad \boldsymbol{x} & \in \mathbf{\Omega}  \\
& \boldsymbol{\sigma}\left(\boldsymbol{x}\right) = \mathbb{C} : \left(\frac{1}{2}\left(\nabla\boldsymbol{u}\left(\boldsymbol{x}\right) + \left(\nabla\boldsymbol{u}\left(\boldsymbol{x}\right)\right)^T\right) - \boldsymbol{\varepsilon}^e \right), & \quad \forall \quad \boldsymbol{x} & \in \mathbf{\Omega}   \\
& \boldsymbol{u}\left(\boldsymbol{x}\right) = \boldsymbol{0}, & \quad \forall \quad \boldsymbol{x} & \in \partial\mathbf{\Omega}_d  
\end{aligned}
\right.
\label{ep1}
\end{equation}
where $\boldsymbol{\varepsilon}^e$ is the thermal strain induced by temperature changes:
\begin{equation}
\varepsilon^e_{kh} = \alpha_T \delta_{k l}\left(T-T_0\right)
\label{ep2}
\end{equation}
where $\alpha_T$ is the coefficient of thermal expansion, $\delta_{kh}$ is the Kronecker delta, and $T$ and $T_0$ are the current temperature and the reference temperature, respectively.

\begin{figure}[ht]
    \centering
    \includegraphics[width=0.69\linewidth]{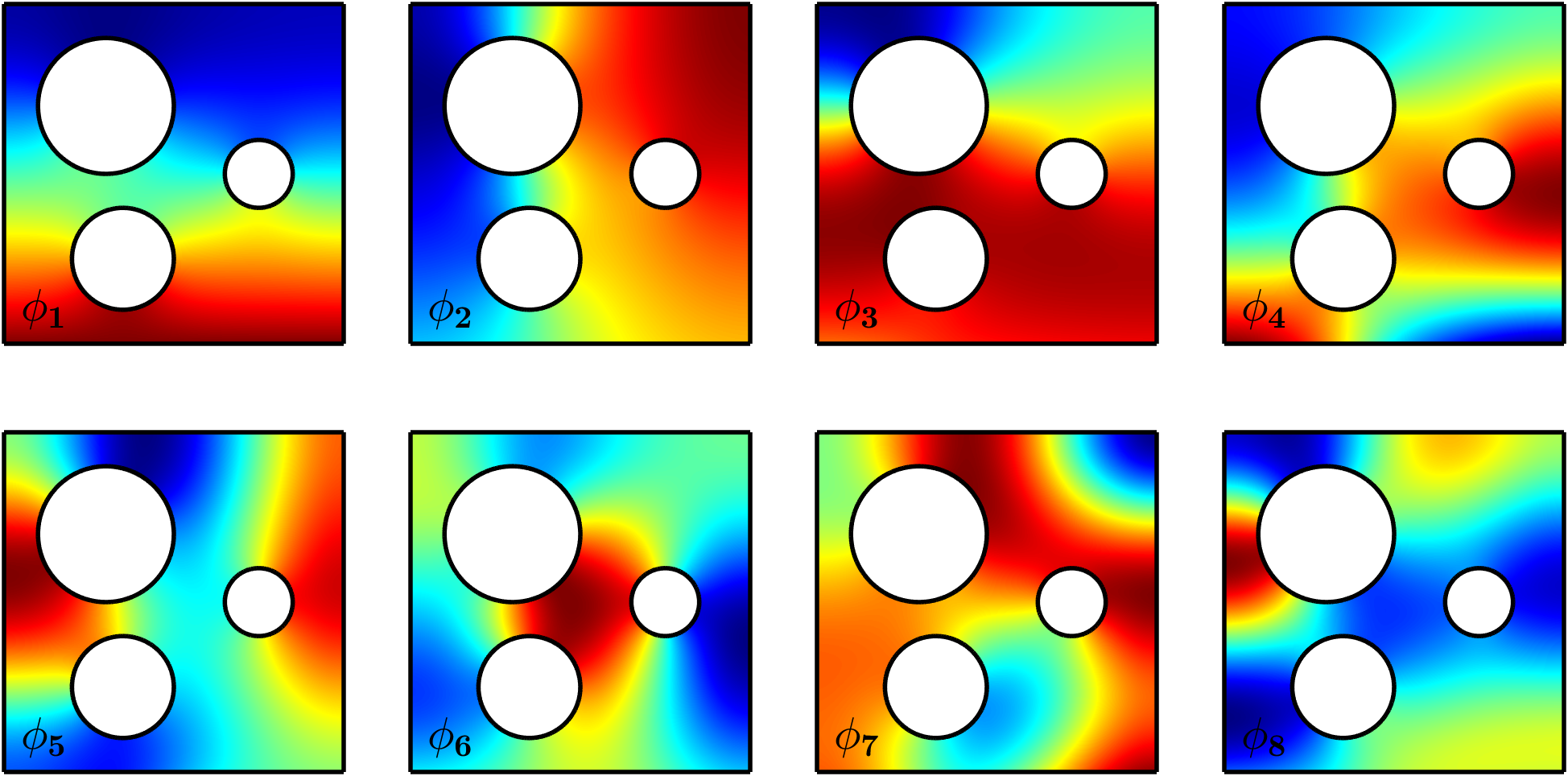}
    \caption{The first eight LBO eigenfunctions of the structure used in Example 2.}
    \label{fig:oh3}
\end{figure}

The parameters in the problem are defined as follows: $\alpha = 1$, $T_0 - T_1 = 1$, $E = 1$, and $\mu = 0.3$. The LBO eigenfunctions of the 2D structure, $\boldsymbol{\phi}$, are calculated using the finite element method, which involves solving Eq.\ref{ef1} with an additional Dirichlet boundary condition as defined in this problem: $u\left(\boldsymbol{x}\right) = \boldsymbol{0}$. Eight LBO eigenfunctions (Fig.\ref{fig:oh3}) are used to approximate both the displacement and stress fields, i.e., $n_u = 8$ and $n_\sigma = 8$.

{\color{black} The Finite-PINN model used to solve this problem is defined in Eqs.~\ref{fpinn} and~\ref{NN}, with both the weak form and strong form loss functions considered in this case.
} Given the number of eigenfunctions, $\mathcal{NN}_\sigma$ is a neural network with an input dimension of 2 and an output dimension of 16, while $\mathcal{NN}_u$ is a neural network with an input dimension of 10 and an output dimension of 2. Both neural networks are specified with 6 hidden layers, each containing 100 nodes. The activation function used is \textit{GELU}. 

\begin{figure}[htbp]
\color{black}
    \centering
    \includegraphics[width=0.99\linewidth]{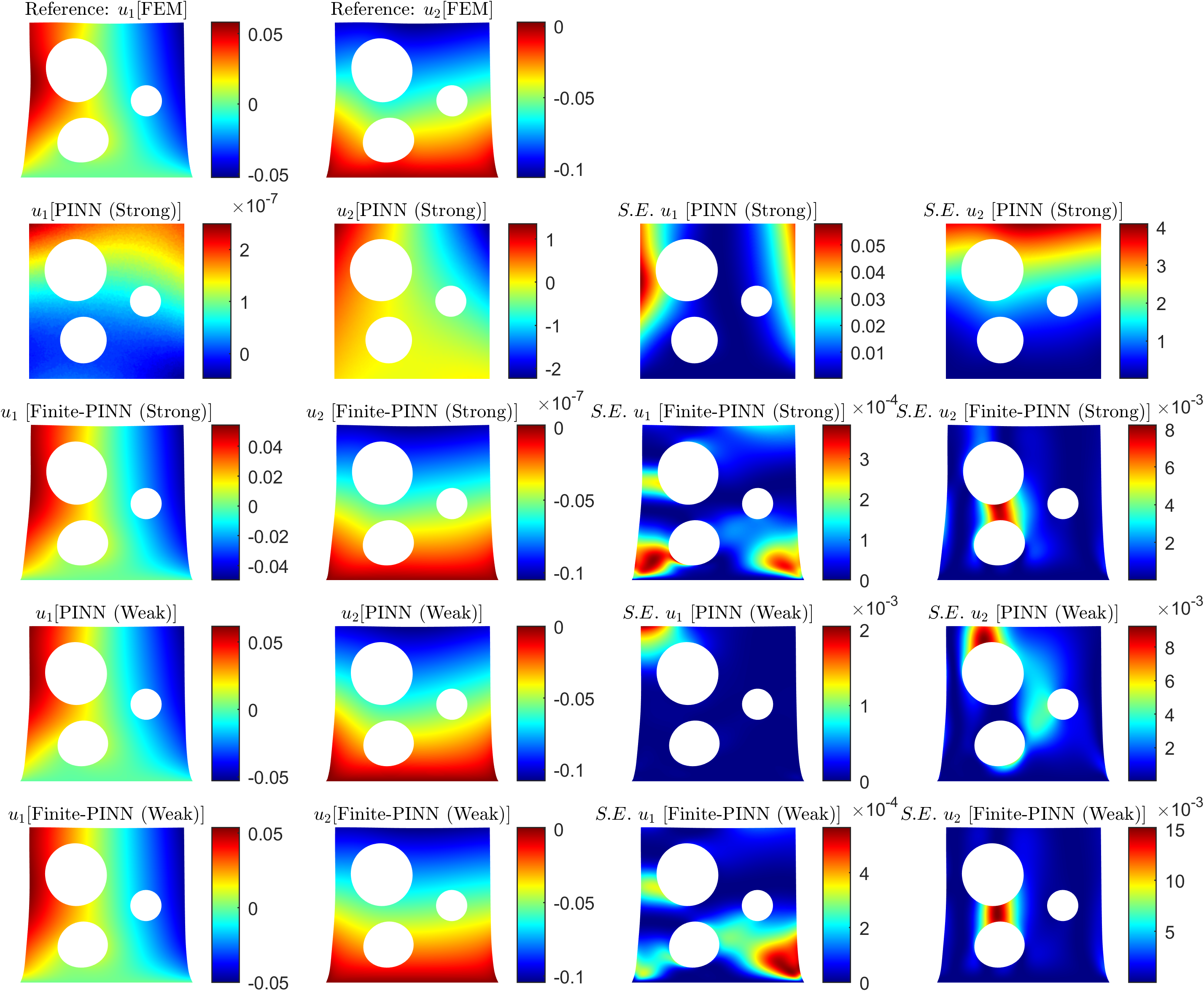}
    \caption{Results of Example 3. The first row shows the reference results calculated by FEM. The other rows present the predictions by the Finite-PINN model and the traditional PINN model, respectively.}
    \label{fig:OH1}
\end{figure}

\color{black}
Similarly, a hard Dirichlet boundary condition is imposed in this problem. The weak form loss function for the training process is defined as follows:
\begin{equation}
\begin{aligned}
\mathcal{L} & = \mathcal{L}_{pde} , \\
\mathcal{L}_{pde} & = \sum^{N_{pde}}_{i=1}v^i \cdot \boldsymbol{\varepsilon}\left(\boldsymbol{x}^i\right) \cdot \left(\mathbb{C} : \boldsymbol{\varepsilon}\left(\boldsymbol{x}^i\right)-\boldsymbol{\varepsilon}^e\right), & \boldsymbol{x}^i & \in \boldsymbol{\Omega}.
\end{aligned}
\label{loss3b}
\end{equation}
\color{black}
and the strong form loss function is:
\begin{equation}
\begin{aligned}
\mathcal{L} & = \mathcal{L}_{pde} + \mathcal{L}_{C} , \\
\mathcal{L}_{pde} & = \frac{1}{N_{pde}}\sum^{N_{pde}}_{i=1}\left( \Delta \boldsymbol{\sigma}\left(\boldsymbol{x}^i\right)\right)^2, & \boldsymbol{x} & \in \boldsymbol{\Omega}, \\
\mathcal{L}_{C} & = \frac{1}{N_{C}}\sum^{N_{C}}_{i=1}\left(\boldsymbol{\sigma}\left(\boldsymbol{x}^i\right) - \mathbb{C}:\left(\frac{1}{2}\left(\nabla\boldsymbol{u}\left(\boldsymbol{x}^i\right) + \left(\nabla\boldsymbol{u}\left(\boldsymbol{x}^i\right)\right)^T\right)-\boldsymbol{\varepsilon}^e \right)\right) ^2, & \boldsymbol{x} & \in \boldsymbol{\Omega}. 
\end{aligned}
\label{loss4}
\end{equation}
The traditional PINN model is also employed to solve the same problem. It uses a neural network similar to the displacement neural network used in the Finite-PINN, with the exception of a different input dimension. In this problem, the numbers of collocation points are $N_{data} = 0$ (no labelled data are used), and $N_{pde} = N_{C} = 14678$. The neural network is trained for 1,000 epochs with a batch size of 200. The learning results are presented in Fig.\ref{fig:OH1}.

\begin{figure}[htbp!]
\color{black}
    \centering
    \includegraphics[width=0.99\linewidth]{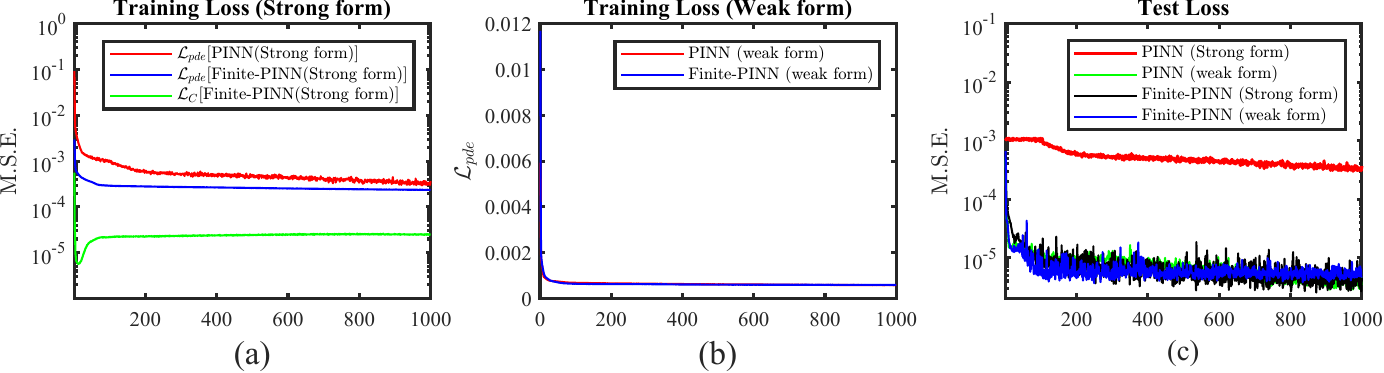}
    \caption{Evolution of training and test losses for solving the problem in Example 3 using the Finite-PINN model. (a) and (b) present the evolution of each loss term in models using the strong form loss and the weak form loss, respectively. (c) depicts the test losses of all models.}
    \label{fig:OH2}
\end{figure}

\color{black}
Fig.~\ref{fig:OH1} shows that, for models using the strong form loss, the Finite-PINN model achieves good agreement with the reference FEM solution. In contrast, the traditional PINN model is unable to accurately learn the field purely from the physics-based loss given the specified neural network architecture and number of collocation points. 
For models using the weak form loss, both the PINN and Finite-PINN models produce satisfactory results compared to the reference, with the Finite-PINN model showing slightly higher accuracy. The evolution of training and testing losses of are shown in Fig.\ref{fig:OH2}. Since the strong form PINN model fails to capture the correct solution so that the loss diverges, Fig.~\ref{fig:OH2}(a) shows only the strong form losses of the Finite-PINN model.
\color{black}

\subsection{Example 4: Approximation problem of a 2D open-notch structure}
\begin{figure}[htbp!]
    \centering
    \includegraphics[width=0.45\linewidth]{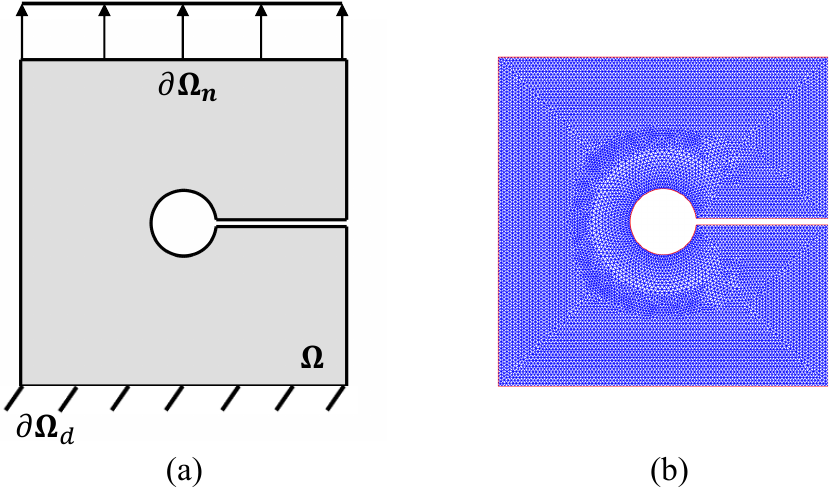}
    \caption{(a) Problem definition for the open-notch structure in Example 4. (b) Finite element mesh of the open-notch structure.}
    \label{fig:p12}
\end{figure}

\begin{figure}[htbp!]
    \centering
    \includegraphics[width=0.69\linewidth]{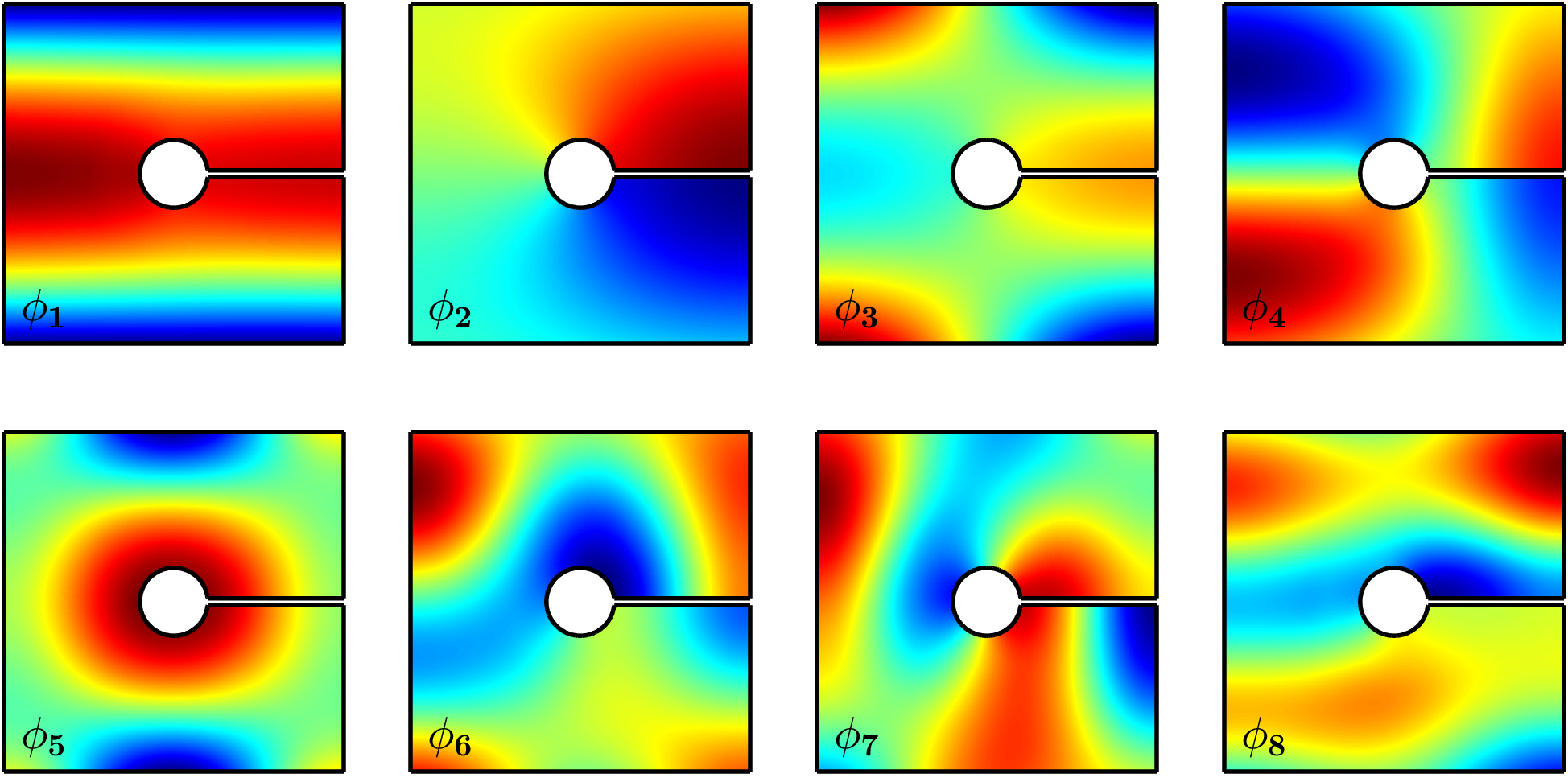}
    \caption{The first 8 hybrid LBO eigenfunctions of the structure used in Example 4.}
    \label{fig:pun}
\end{figure}

This example employs the open-notch structure presented in Fig.\ref{fig:p1}(c) and (d). The geometry consists of a square with a side length of 1, a central hole with a radius of 0.075, and a single side notch on the right side with a width of 0.02. The geometry, along with the assigned boundary conditions and loading for this problem, is illustrated in Fig.\ref{fig:p12}(a). The first 8 hybrid eigenfunctions used here are shown in Fig.\ref{fig:pun}.

\begin{figure}[htbp!]
    \centering
    \includegraphics[width=0.96\linewidth]{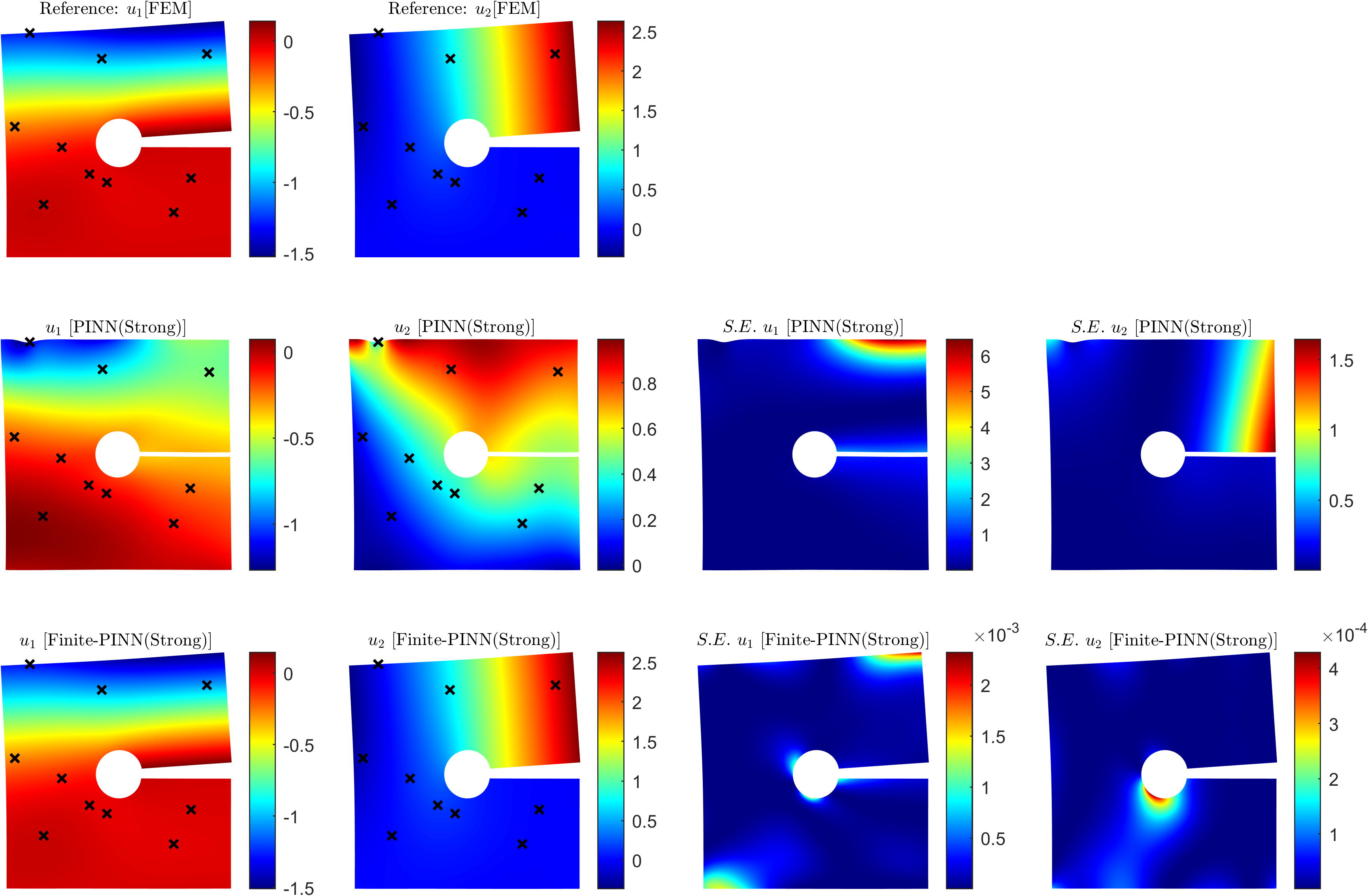}
    \caption{Results of Example 4. The first row shows the reference results calculated by FEM. The second and third rows present the predictions by the traditional PINN model and the Finite-PINN model, respectively. The black 'x' marks the locations of the data used for training.}
    \label{fig:ON1}
\end{figure}

\begin{figure}[htbp!]
    \centering
    \includegraphics[width=0.999\linewidth]{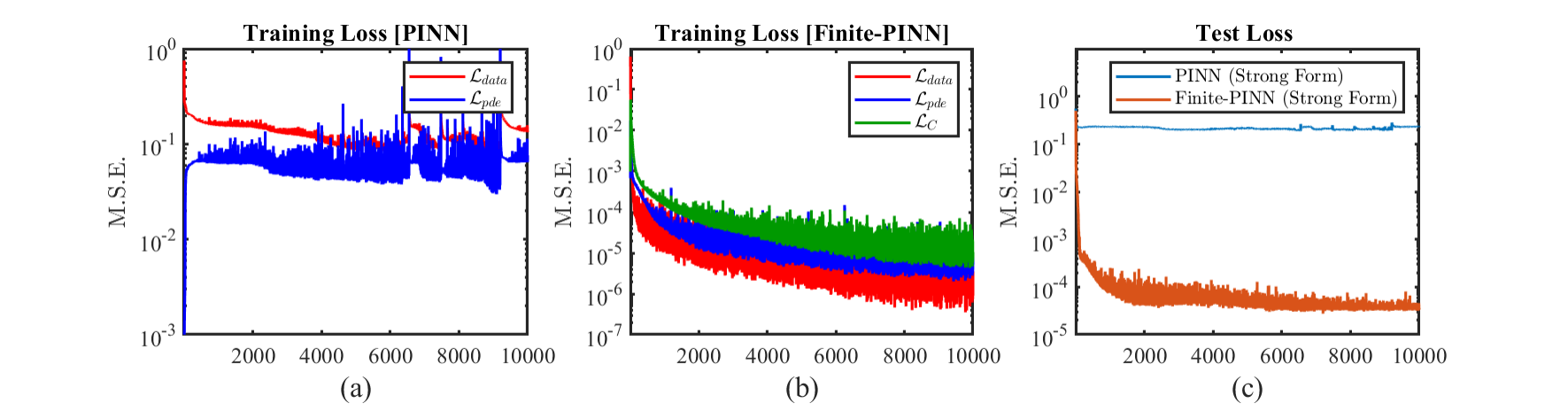}
    \caption{Evolution of training loss for the traditional PINN (a) and Finite-PINN (b) models in Example 4. Evolution of test loss for the traditional PINN (c) and Finite-PINN (d) models in Example 4.}
    \label{fig:ON2}
\end{figure}

The problem is solved using the physics information and a sparse dataset of displacements. {\color{black}Since this is an inverse problem, training is performed using only the strong form loss.} Displacements at only 10 points are randomly selected to be served as the collocation points as for the data loss term, i.e. $N_{data} = 10$. The number of collocation points for the physics loss and the constitutive loss $N_{pde} = N_{C} = 10933$. The Finite-PINN model employs neural networks with 4 hidden layers and 100 nodes per layer for both $\mathcal{NN}_\sigma$ and $\mathcal{NN}_u$. The neural network is trained for 10,000 epochs with a batch size of 1000. The activation function used in these neural networks is \textit{GELU}. The loss function for the Finite-PINN model in this problem is defined as follows:
\begin{equation}
\begin{aligned}
\mathcal{L} & = \mathcal{L}_{data} + \mathcal{L}_{pde} + \mathcal{L}_{C} , \\
\mathcal{L}_{data} & = \frac{1}{N_{data}}\sum^{N_{data}}_{i=1}\left(\boldsymbol{u}\left(\boldsymbol{x}^i\right)-\boldsymbol{u}_0\left(\boldsymbol{x}^i\right)\right)^2, & \boldsymbol{x} & \in \boldsymbol{\Omega}, \\
\mathcal{L}_{pde} & = \frac{1}{N_{pde}}\sum^{N_{pde}}_{i=1}\left(\Delta \boldsymbol{\sigma}\left(\boldsymbol{x}^i\right)-\boldsymbol{0}\right)^2, & \boldsymbol{x} & \in \boldsymbol{\Omega}, \\
\mathcal{L}_{C} & = \frac{1}{N_{C}}\sum^{N_{C}}_{i=1}\left(\boldsymbol{\sigma}\left(\boldsymbol{x}^i\right) - \mathbb{C}:\frac{1}{2}\left(\nabla\boldsymbol{u}\left(\boldsymbol{x}^i\right) + \left(\nabla\boldsymbol{u}\left(\boldsymbol{x}^i\right)\right)^T\right)\right)^2, & \boldsymbol{x} & \in \boldsymbol{\Omega}. 
\end{aligned}
\label{loss4k}
\end{equation}
It is noted that there are no boundary condition loss terms in this loss function, meaning that the displacement values at the 10 collocation points are the only data information provided to the neural network. The results are presented in Fig.\ref{fig:ON1}. The evolution of the training and test loss for both the Finite-PINN model and the traditional PINN model is shown in Fig.\ref{fig:ON2}.

The calculation results in Fig.\ref{fig:ON1} and the evolution of the loss functions in Fig.\ref{fig:ON2} demonstrate that the traditional PINN model fails in this problem because the Euclidean space is significantly less suitable than the topological space for approximating this solution field. On the other hand, using labelled data at 10 points allows the Finite-PINN model to successfully reconstruct the global displacement field.

\color{black}
\subsection{Example 5: Deformation Sensing of a complex geometry}
\begin{figure}[htbp!]
\color{black}
    \centering
    \includegraphics[width=0.35\linewidth]{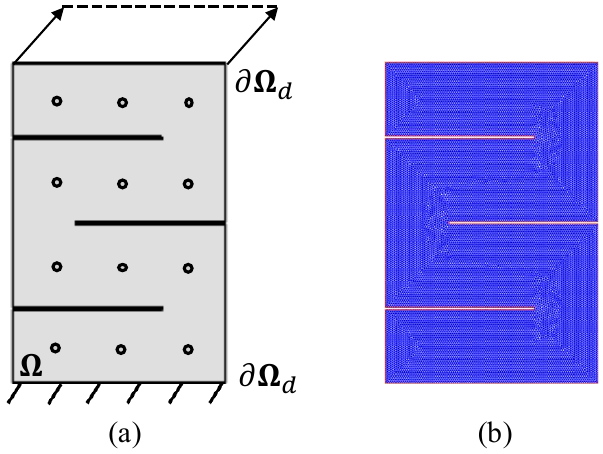}
    \caption{(a) Problem definition of Example 5. (b) Finite element mesh of the structure.}
    \label{fig:GMSd}
\end{figure}

This example aims to solve a geometry sensing problem by using a limited number of strain sensors to reconstruct the deformation, i.e., the displacement field, of a structure. This represents a common type of problem in engineering practice. The geometry of the structure, the locations of the strain sensors, and the applied boundary conditions are illustrated in Fig.\ref{fig:GMSd}. 

The strains at 12 sensor locations, as indicated in Fig.\ref{fig:GMSd}(a), are used as the supervision data. The bottom side of the structure is fully fixed, enforced through a Dirichlet boundary condition loss term. The number of collocation points for both the physics loss and the constitutive loss is set to $N_{pde} = N_{C} = 14,000$. 

The Finite-PINN model employs neural networks with four hidden layers and 128 nodes per layer for both $\mathcal{NN}_\sigma$ and $\mathcal{NN}_u$. The networks are trained for 10,00 epochs with a batch size of $1,000$. The activation function used in these neural networks is \textit{GELU}. 

The strong form loss function for the Finite-PINN model in this problem is defined as follows:
\begin{equation}
\begin{aligned}
\mathcal{L} & = \mathcal{L}_{data} + \mathcal{L}_{pde} + \mathcal{L}_{C} , \\
\mathcal{L}_{data} & = \frac{1}{N_{data}}\sum^{N_{data}}_{i=1}\left(\frac{1}{2}\left(\nabla\boldsymbol{u}\left(\boldsymbol{x}^i\right) + \left(\nabla\boldsymbol{u}\left(\boldsymbol{x}^i\right)\right)^T\right) - \boldsymbol{\varepsilon}^0\right)^2,  & x_1 \in \partial \Omega,\quad x_2 & = L, \\
\mathcal{L}_{pde} & = \frac{1}{N_{pde}}\sum^{N_{pde}}_{i=1}\left(\Delta \boldsymbol{\sigma}\left(\boldsymbol{x}^i\right)-\boldsymbol{0}\right)^2, & \boldsymbol{x} & \in \boldsymbol{\Omega}, \\
\mathcal{L}_{C} & = \frac{1}{N_{C}}\sum^{N_{C}}_{i=1}\left(\boldsymbol{\sigma}\left(\boldsymbol{x}^i\right) - \mathbb{C}:\frac{1}{2}\left(\nabla\boldsymbol{u}\left(\boldsymbol{x}^i\right) + \left(\nabla\boldsymbol{u}\left(\boldsymbol{x}^i\right)\right)^T\right)\right)^2, & \boldsymbol{x} & \in \boldsymbol{\Omega}. 
\end{aligned}
\label{loss5}
\end{equation}

\begin{figure}[htbp!]
\color{black}
    \centering
    \includegraphics[width=0.69\linewidth]{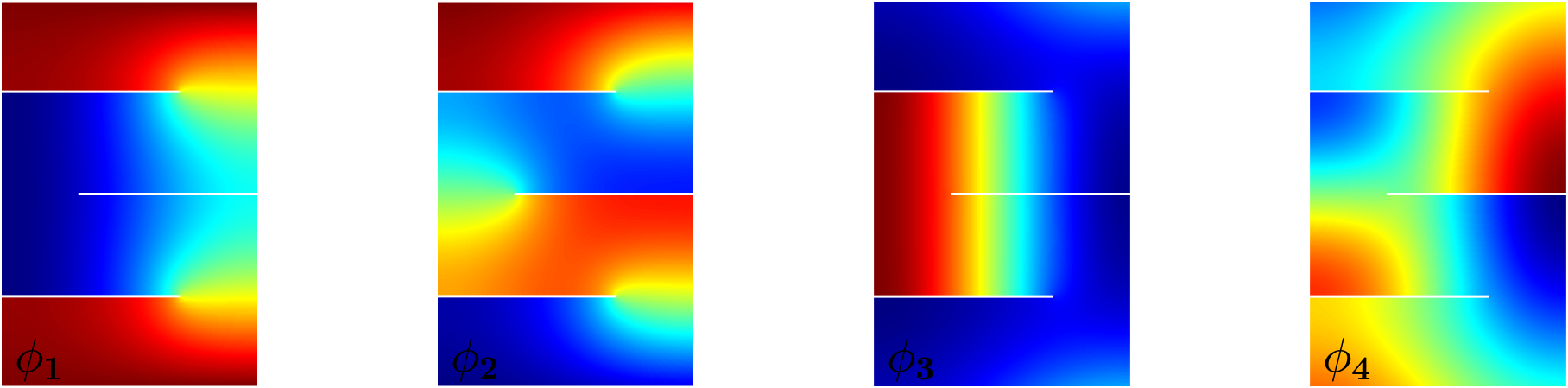}
    \caption{The first 4 hybrid LBO eigenfunctions of the structure used in Example 5.}
    \label{fig:GMSb}
\end{figure}

\begin{figure}[htbp!]
\color{black}
    \centering
    \includegraphics[width=0.99\linewidth]{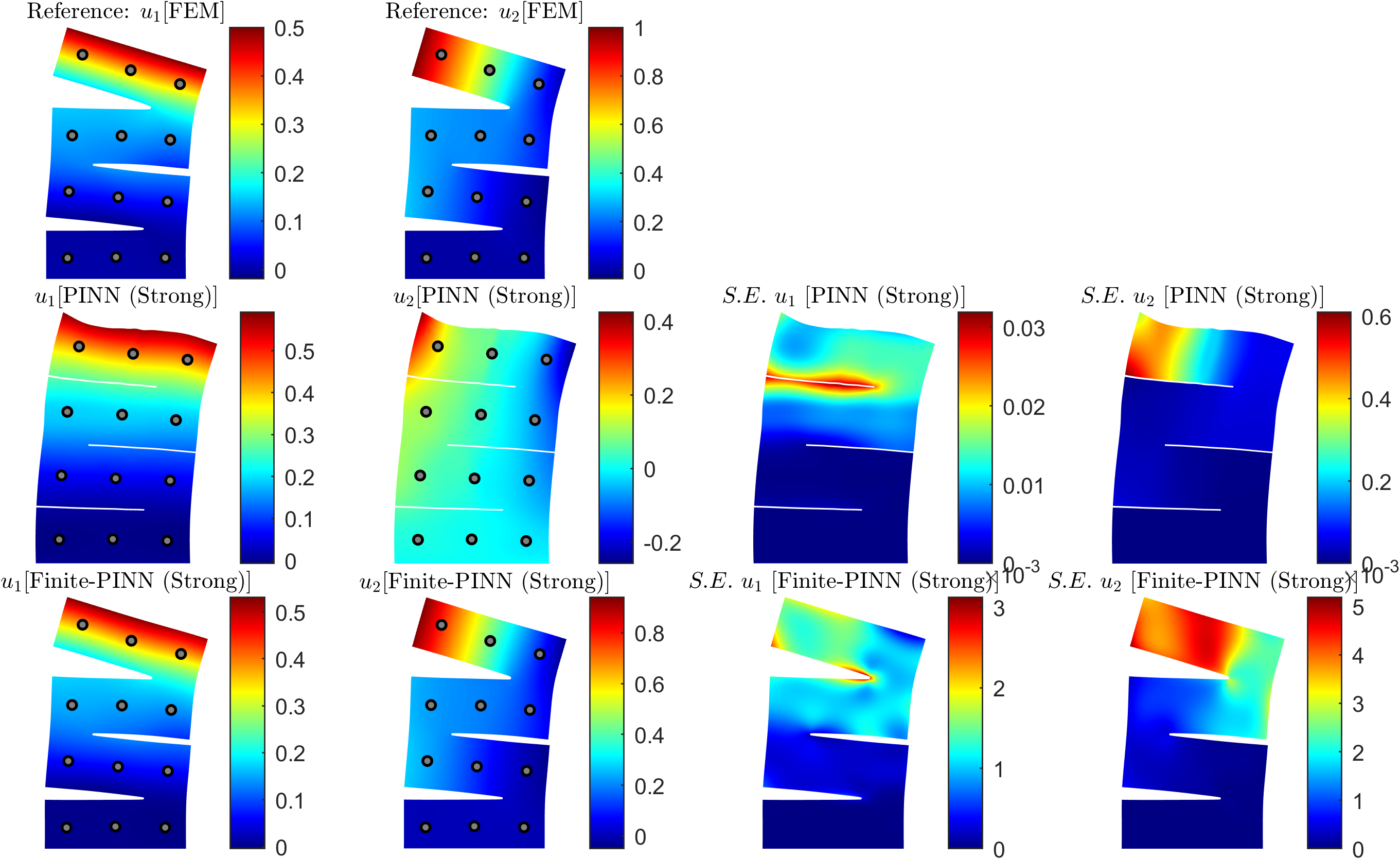}
    \caption{Results of Example 5. The first row shows the reference results calculated by FEM. The second and third rows present the predictions by the traditional PINN model and the Finite-PINN model, respectively. The circles at some figures marks the locations of strain sensors.}
    \label{fig:GS1}
\end{figure}

\begin{figure}[htbp!]
\color{black}
    \centering
    \includegraphics[width=0.85\linewidth]{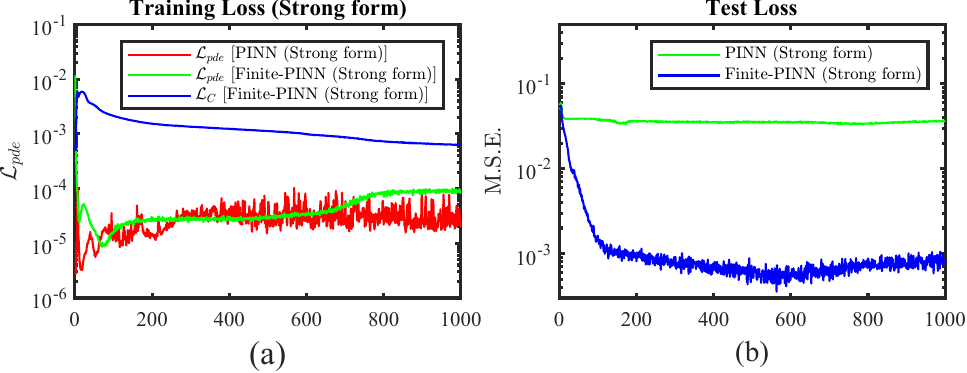}
    \caption{Evolution of training loss (a) and the test loss (b) for the models performed in Example 5.}
    \label{fig:GS2}
\end{figure}

The 4 LBO eigenfunctions used in this examples are shown in Fig.\ref{fig:GMSb}. The results are visualised in Fig.\ref{fig:GS1}, and the evolution of the training and testing losses for both the Finite-PINN model and the traditional PINN model is shown in Fig.\ref{fig:GS2}. The results demonstrate the effectiveness of the Finite-PINN in solving this deformation sensing problem with complex geometries. The finite geometric encoding clearly assists the neural network in addressing such an inverse problem.

\subsection{Example 6: a problem with material nonlinearity on a L-shaped structure}

This example investigates the potential of the Finite-PINN in solving nonlinear solid mechanics problems. An L-shaped two-dimensional structure is used as the geometric domain, and a nonlinear material constitutive model is adopted, where the nonlinear elastic modulus is defined as follows:
\begin{equation}
E = E_0 \cdot \left(1 +  2.0 \cdot \left(\varepsilon_{xx}^2 + \varepsilon_{yy}^2\right)\right)
\label{eqnbv}
\end{equation}
where $E_0$ is the initial elastic modulus, set to $1.0$ in this problem. The Poisson's ratio is fixed at $0.3$ and remains constant throughout the analysis.
\begin{figure}[htbp!]
\color{black}
    \centering
    \includegraphics[width=0.40\linewidth]{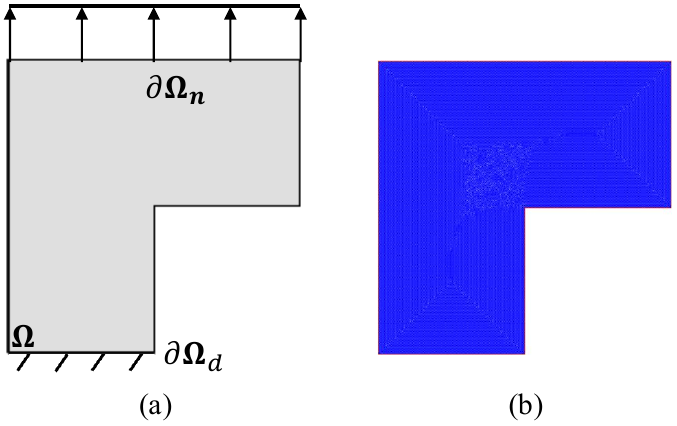}
    \caption{(a) Problem definition of Example 6. (b) Finite element mesh of the structure.}
    \label{fig:LSd}
\end{figure}

\begin{figure}[htbp!]
\color{black}
    \centering
    \includegraphics[width=0.69\linewidth]{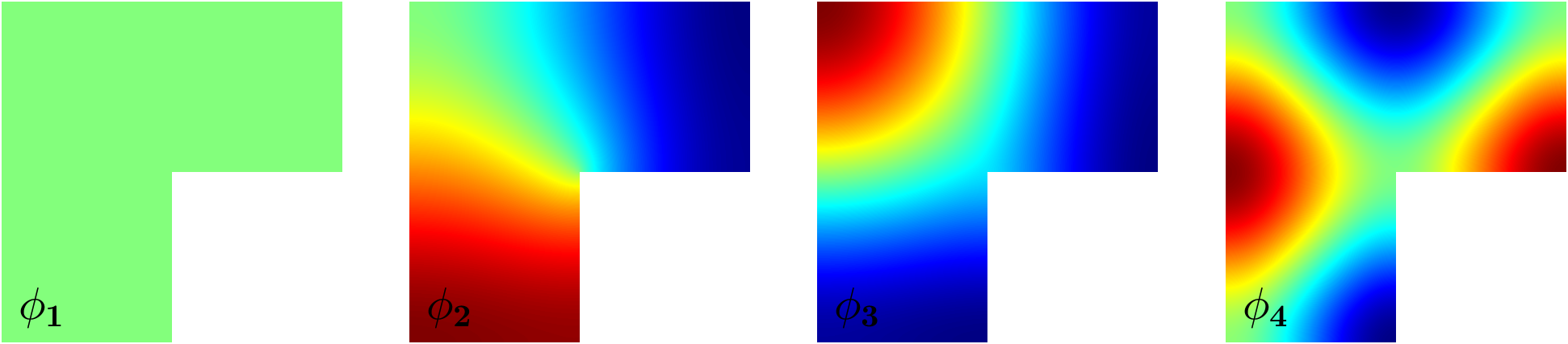}
    \caption{The first 4 LBO eigenfunctions of the structure used in Example 6.}
    \label{fig:LSb}
\end{figure}

\begin{figure}[htbp!]
\color{black}
    \centering
    \includegraphics[width=0.99\linewidth]{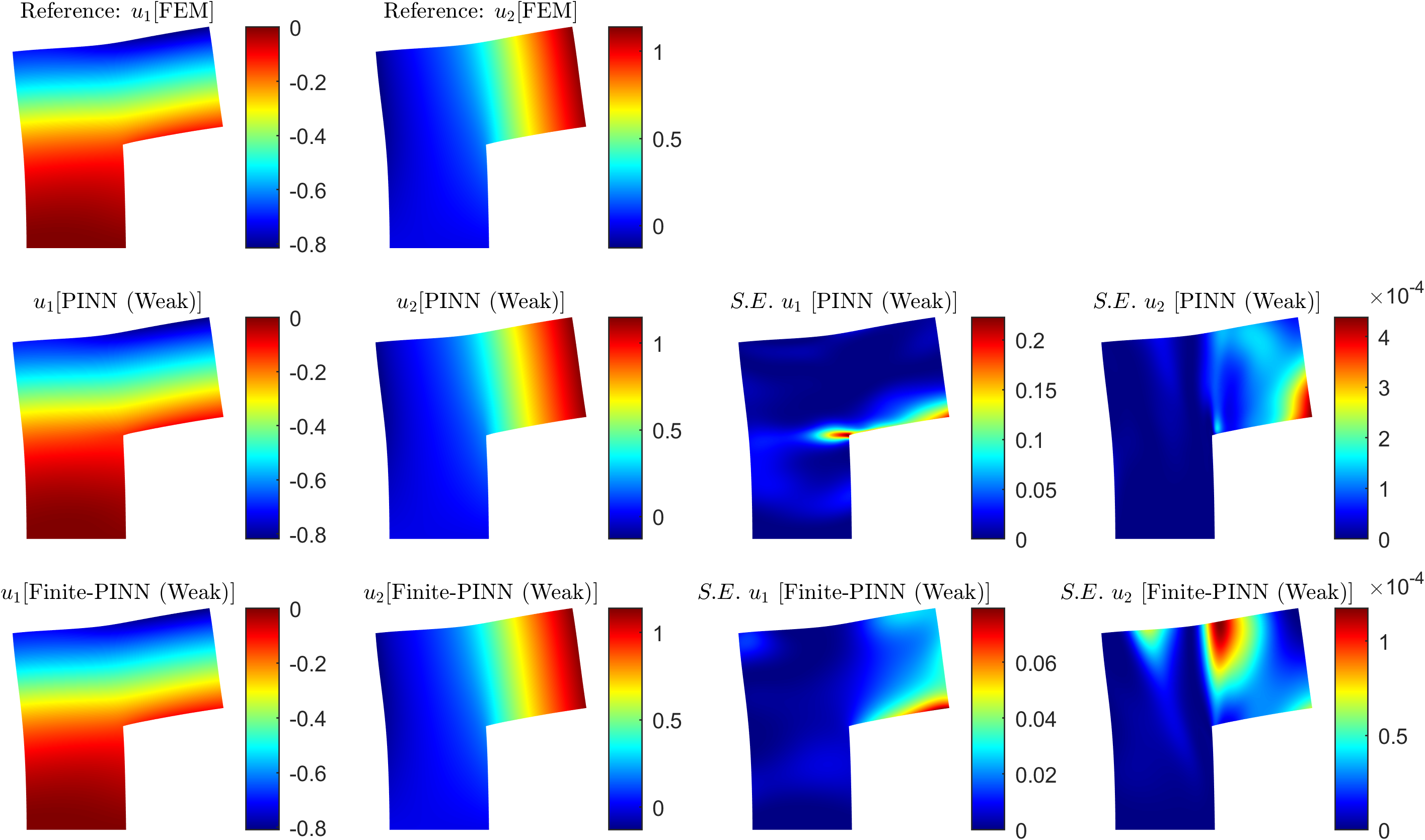}
    \caption{Results of Example 6. The first row shows the reference results calculated by FEM. The second and third rows present the predictions by the traditional PINN model and the Finite-PINN model, respectively.}
    \label{fig:LS1}
\end{figure}

\begin{figure}[htbp!]
\color{black}
    \centering
    \includegraphics[width=0.85\linewidth]{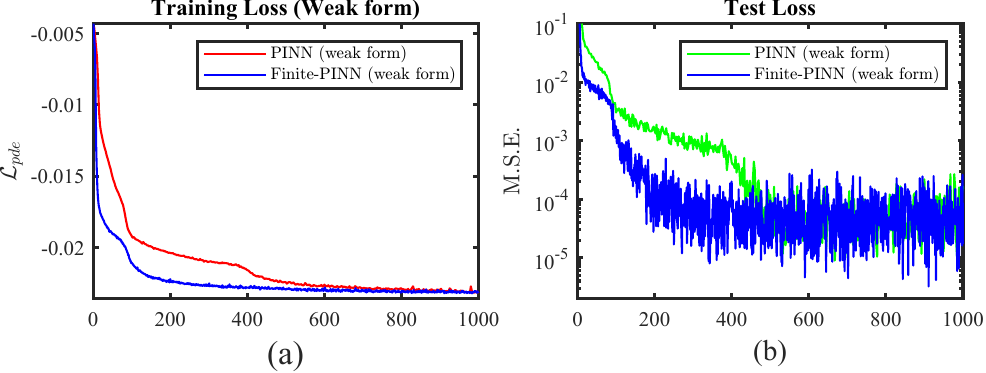}
    \caption{Evolution of training loss (a) and the test loss (b) for the traditional PINN and Finite-PINN models in Example 6.}
    \label{fig:LS2}
\end{figure}

The geometric configuration and applied boundary conditions are illustrated in Fig.\ref{fig:LSd}. The corresponding loss function is defined as follows:
\begin{equation}
\begin{aligned}
\mathcal{L} & = \mathcal{L}_{pde} , \\
\mathcal{L}_{pde} & = \sum^{N_{pde}}_{i=1}v^i \cdot \boldsymbol{\varepsilon}\left(\boldsymbol{x}^i\right) \cdot \left(\mathbb{C} : \boldsymbol{\varepsilon}\left(\boldsymbol{x}^i\right)\right) - \sum^{N_{nbc}}_{i=1} a^i \cdot \overline{\boldsymbol{t}} \cdot \boldsymbol{u}\left(\boldsymbol{x}^i\right), & \boldsymbol{x}^i & \in \boldsymbol{\Omega} .
\end{aligned}
\label{loss3}
\end{equation}
In this case, ${N_{d}} = 0$, ${N_{n}} = 0$, and ${N_{pde}} = 0$. The constitutive tensor $\mathbb{C}$ are calculated based on Eq.\ref{eqnbv}. The used four eigenfunctions are presented in Fig.\ref{fig:LSb}. The results obtained using the Finite-PINN and traditional PINN models are presented in Fig.~\ref{fig:LS1}, while the evolution of the training and testing losses is shown in Fig.~\ref{fig:LS2}. The results demonstrate good agreement between the Finite-PINN predictions and the reference results obtained using the FEM.

\color{black}

\subsection{Example 7: a 3D problem for a elastic spring}
\begin{figure}[htbp]
    \centering
    \includegraphics[width=0.50\linewidth]{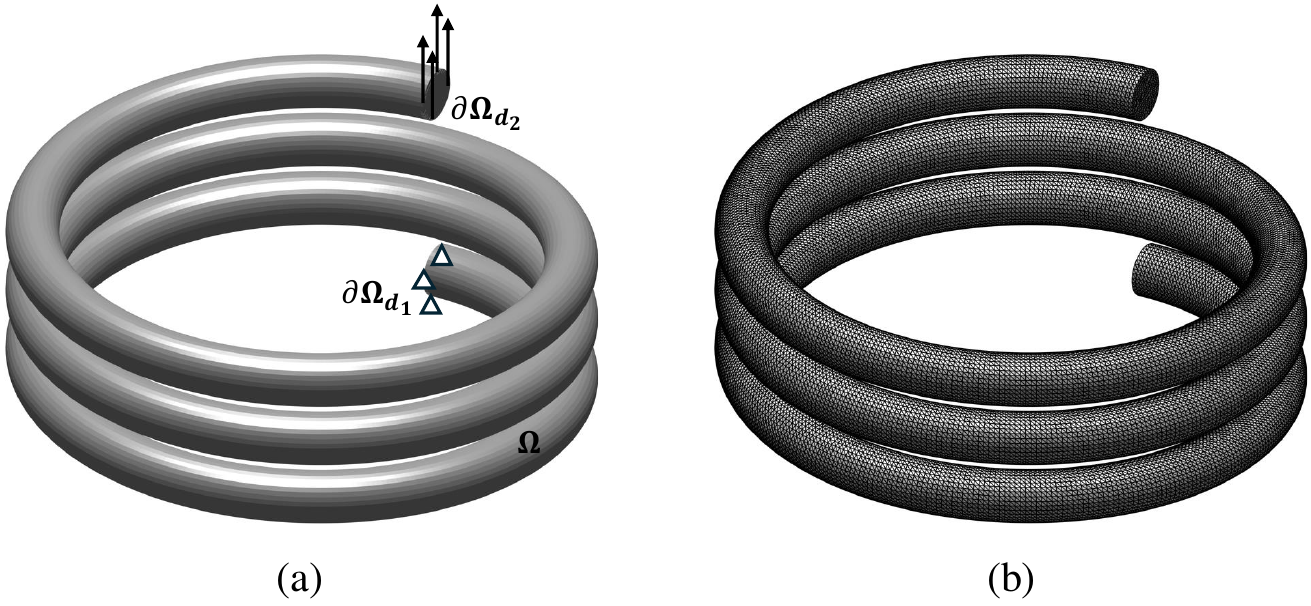}
    \caption{(a) The spring structure and boundary conditions defined in Example 7. (b) Finite element mesh of the spring structure. }
    \label{fig:p13}
\end{figure}

The last example focuses on solving a 3D solid mechanics problem. The structure to be investigated is a 4-loop elastic spring with a spring rod radius of 0.1 and a spiral radius of 0.5. The bottom end of the spring is fully fixed, and an upward displacement is applied to the top end of the spring. The applied boundary conditions and the finite element mesh of this structure are depicted in Fig.\ref{fig:p13}(a).
\begin{figure}[htbp]
    \centering
    \includegraphics[width=0.92\linewidth]{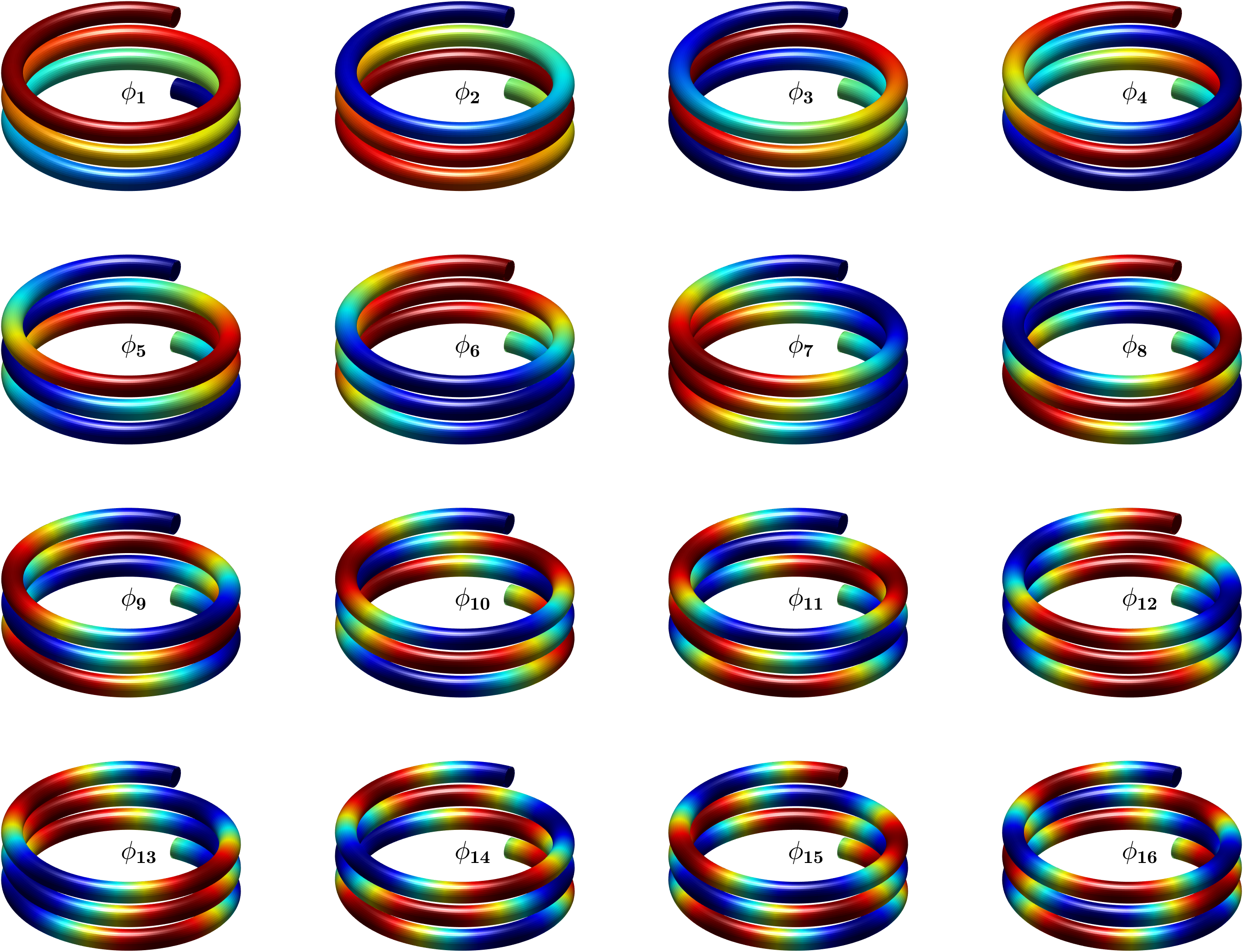}
    \caption{The first 16 LBO eigenfunctions of the spring structure used in Example 7.}
    \label{fig:sp3}
\end{figure}

\begin{figure}[htbp!]
    \centering
    \includegraphics[width=0.99\linewidth]{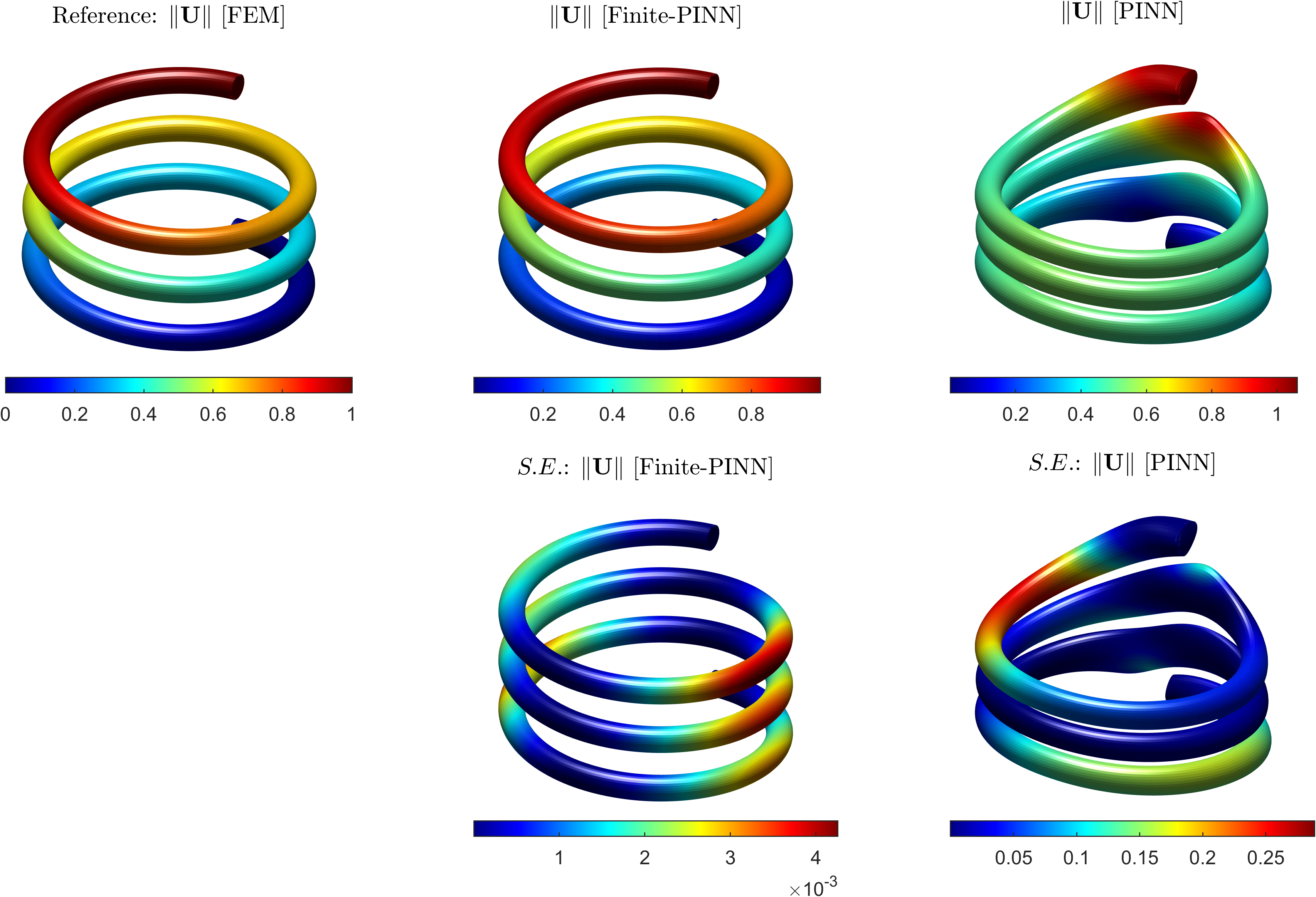}
    \caption{Results of Example 7. The left side shows the reference solution calculated by FEM. The middle and right sides depict the solutions obtained by the Finite-PINN model and the traditional PINN model, respectively.}
    \label{fig:sp1}
\end{figure}

\begin{figure}[htbp!]
    \centering
    \includegraphics[width=0.80\linewidth]{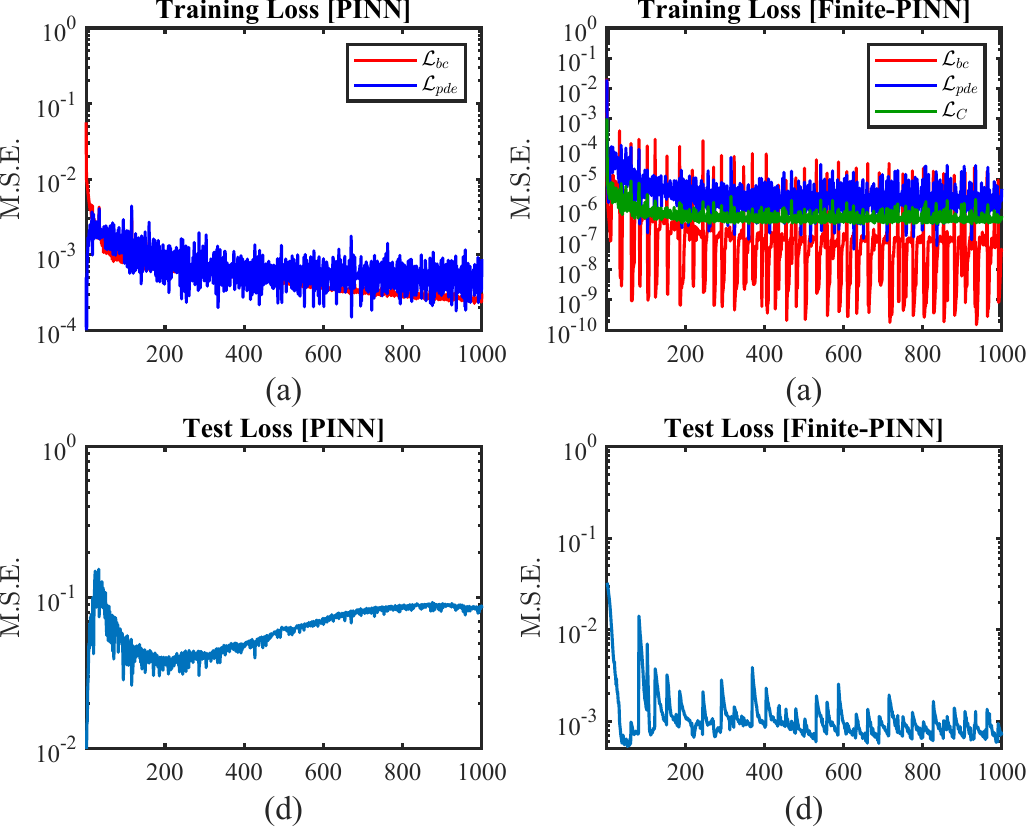}
    \caption{Evolution of training loss for the traditional PINN (a) and Finite-PINN (b) models in Example 7. Evolution of test loss for the traditional PINN (c) and Finite-PINN (d) models in Example 7.}
    \label{fig:sp2}
\end{figure}

This problem aims to solve the displacement field of the spring based on the boundary conditions illustrated in Fig.\ref{fig:p13}(a). One end of the spring is fixed while an upwards load is applied on the other end. The Finite-PINN model employs neural networks with 4 hidden layers and 128 nodes per layer for both $\mathcal{NN}_\sigma$ and $\mathcal{NN}_u$. The activation function used in these neural networks is \textit{GELU}. The first 16 LBO eigenfunctions are used as the input Riemannian manifolds, which are shown in Fig.\ref{fig:sp3}. The weak form loss is used for the problem which is defined as follows:
\begin{equation}
\begin{aligned}
\mathcal{L} & = \mathcal{L}_{bc1} + \mathcal{L}_{bc2} + \mathcal{L}_{pde} + \mathcal{L}_{C} , \\
\mathcal{L}_{bc1} & = \frac{1}{N_{d_1}}\sum^{N_{d_1}}_{i=1}\left(\boldsymbol{u}\left(\boldsymbol{x}^i\right)- \boldsymbol{0}\right)^2 , & \boldsymbol{x} & \in \partial\boldsymbol{\Omega}_{d_1}, \\
\mathcal{L}_{bc2} & = \frac{1}{N_{d_2}}\sum^{N_{d_2}}_{i=1}\left(u_2\left(\boldsymbol{x}^i\right)-1\right)^2, & \boldsymbol{x} & \in \partial\boldsymbol{\Omega}_{d_2}, \\
\mathcal{L}_{pde} & = \sum^{N_{pde}}_{i=1}v^i \cdot \boldsymbol{\varepsilon}\left(\boldsymbol{x}^i\right) \cdot \left(\mathbb{C} : \boldsymbol{\varepsilon}\left(\boldsymbol{x}^i\right)\right), & \boldsymbol{x} & \in \boldsymbol{\Omega}, \\
\mathcal{L}_{C} & = \frac{1}{N_{C}}\sum^{N_{C}}_{i=1}\left(\boldsymbol{\sigma}\left(\boldsymbol{x}^i\right) - \mathbb{C}:\frac{1}{2}\left(\nabla\boldsymbol{u}\left(\boldsymbol{x}^i\right) + \left(\nabla\boldsymbol{u}\left(\boldsymbol{x}^i\right)\right)^T\right)\right)^2, & \boldsymbol{x} & \in \boldsymbol{\Omega}, 
\end{aligned}
\label{loss6}
\end{equation}
where $N_{d_1} = 131$, $N_{d_2} = 129$ and $N_{pde} = N_{C} = 76665$. Still, no labelled data are used to supervise the training in this example. The results obtained by the Finite-PINN and traditional PINN models are depicted in Fig.\ref{fig:sp1}. The evolution of the training and test losses is shown in Fig.\ref{fig:sp2}. The neural network is trained for 1,000 epochs with a batch size of 200. The test loss is calculated by all the node locations from the reference FEM solution.

Similar to the previous example, the traditional PINN model fails to approximate the solution field for the spring structure, while the Finite-PINN model achieves good results. As seen from the LBO basis functions presented in Fig.\ref{fig:sp3}, the constructed Riemannian manifold is actually built upon the topology of the spring structure in this case. It can be observed from the results obtained by the traditional PINN model in Fig.\ref{fig:sp1} that the circles in the deformed spring appear to be "tied" to each other. This is because, in the original Euclidean space used by the traditional PINN model, the distance between any two points is simply the Euclidean distance. For instance, two points on separate circles may still appear close to each other, even if their geodesic distance is quite large.

Additionally, since most of the surfaces of the spring are free surfaces, with boundary conditions only applied to the two ends, applying free-surface Neumann boundary conditions in the traditional PINN model by including additional loss terms becomes difficult and complex. This complexity can also distract the optimizer from focusing on optimizing more important objectives. This corresponds well with the previously introduced theory: since the Finite-PINN model addresses the two major challenges of using PINN for general solid problems, it becomes a suitable model for solving general solid mechanics problems. The Finite-PINN model improves both the implementation and results when solving solid mechanics problems compared to the traditional PINN model.

\section{Discussion}
\label{sec6}
In this section, we investigate the effect of several variables on the approximation results of the Finite-PINN model, using the open-notch structure as the example under investigation.

\subsection{Number of LBO eigenfunctions}
The architecture of the Finite-PINN model differs from that of the original $\Delta$-PINN model \cite{costabal2024delta}, which only uses LBO eigenfunctions as input features. The Finite-PINN model employs both Cartesian coordinates and LBO eigenfunctions to generate a Euclidean-topological-joint space for the approximation. In this way, a much smaller number of LBO eigenfunctions are needed to approximate the required field, as the Cartesian input alone is sufficiently rich to approximate all types of fields based on the universal approximation theorem.

\begin{figure}[htbp!]
    \centering
    \includegraphics[width=0.99\linewidth]{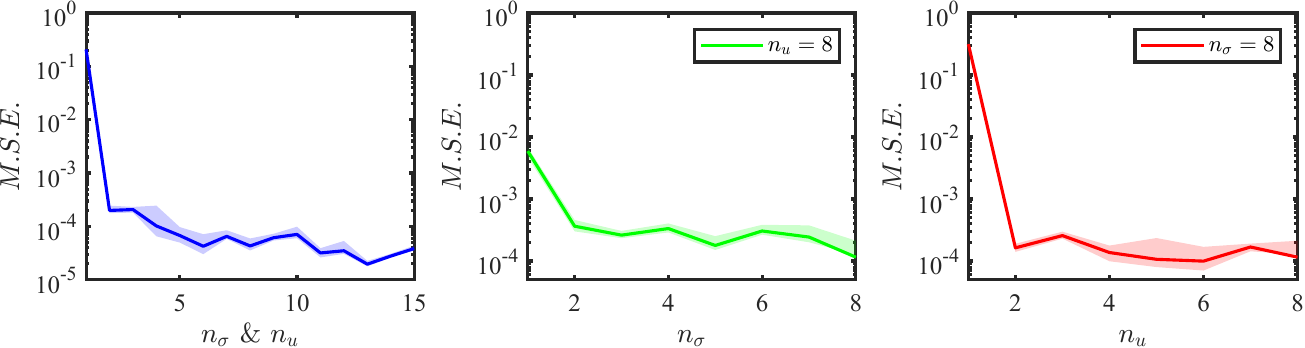}
    \caption{(a) Sensitivity test of the prediction error with respect to the same value of $n_\sigma$ and $n_u$. (b) Sensitivity test of the prediction error with respect to $n_\sigma$ when $n_u = 8$. (c) Sensitivity test of the prediction error with respect to $n_u$ when $n_\sigma = 8$.}
    \label{fig:pzc}
\end{figure}

We investigate the effect of using different numbers of LBO functions on the results of the general Finite-PINN model. Due to the separate architecture of the Finite-PINN model, we need to examine the impact of the number of LBO functions used to approximate the stress field and the displacement field, respectively. Using the open-notch model as the example, we first conduct a coarse investigation into the sensitivity of changing $n_{\boldsymbol{\sigma}}$ and $n_{\boldsymbol{u}}$ by a same value. We run 1000 epochs and 20 times for all models and record the final test loss. The results are plotted in Fig.\ref{fig:pzc} which shows the mean values and variations of the test loss over the 20 times simulations. It is observed that the test loss converges between $10^{-4}$ and $10^{-5}$ when both $n_{\boldsymbol{\sigma}}$ and $n_{\boldsymbol{u}}$ are equal to 5.

In the second step, we set one of the values of $n_\sigma$ or $n_u$ to the previously found converged value, and vary the other to examine the detailed influence of each specific parameter on the results. The test results of changing $n_\sigma$ while keeping $n_u = 8$ are depicted in Fig.\ref{fig:pzc}(b). Similarly, the test results of changing $n_u$ while keeping $n_\sigma = 8$ are also depicted in Fig.\ref{fig:pzc}(c). It is seen that for this open-notch case, the results converge with a low number of $n_\sigma$ and $n_u$; approximately 4 is sufficient.

\subsection{Mesh sensitivity}
\begin{figure}[htbp!]
    \centering
    \includegraphics[width=0.33\linewidth]{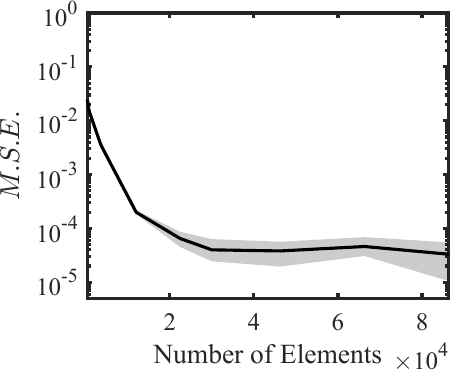}
    \caption{Sensitivity test of the prediction error with respect to the number of elements.}
    \label{fig:meshsen}
\end{figure}

In this work, the finite element method is used to calculate the LBO eigenfunctions numerically. The LBO function thus depends on the mesh quality of the employed finite element model, meaning that the results of the Finite-PINN model are also affected. We investigate this effect by varying the mesh size of the finite element model in the open-notch example. The mesh size is adjusted from 1 to 0.1, resulting in the number of elements changing from 244 to 86440. The test loss is used as the standard to quantify the performance of different models, as demonstrated in Fig.\ref{fig:meshsen}. 

As shown in Fig.\ref{fig:meshsen}, apart from the initial attempts with a very limited number of elements, the results exhibit low sensitivity to the number of elements. This is advantageous in practical applications, as it allows users greater flexibility to select the most suitable mesh size for simulation, balancing accuracy and efficiency according to their specific needs.

\color{black}
\subsection{Uncertainties}
\subsubsection{Noises}
Physics informed neural networks method differs from traditional numerical methods since its capability of data fusion, i.e. the combination of the data-driven approach and physics-driven approach. Such a capability makes itself better using in engineering applications, such as to integrate the observed experimental data into the PDE solving process. In such a case, PINN are required to have the ability to deal with uncertain data input, since noises are absolutely in the experimental observations. 

\begin{figure}[htbp!]
\color{black}
    \centering
    \includegraphics[width=0.95\linewidth]{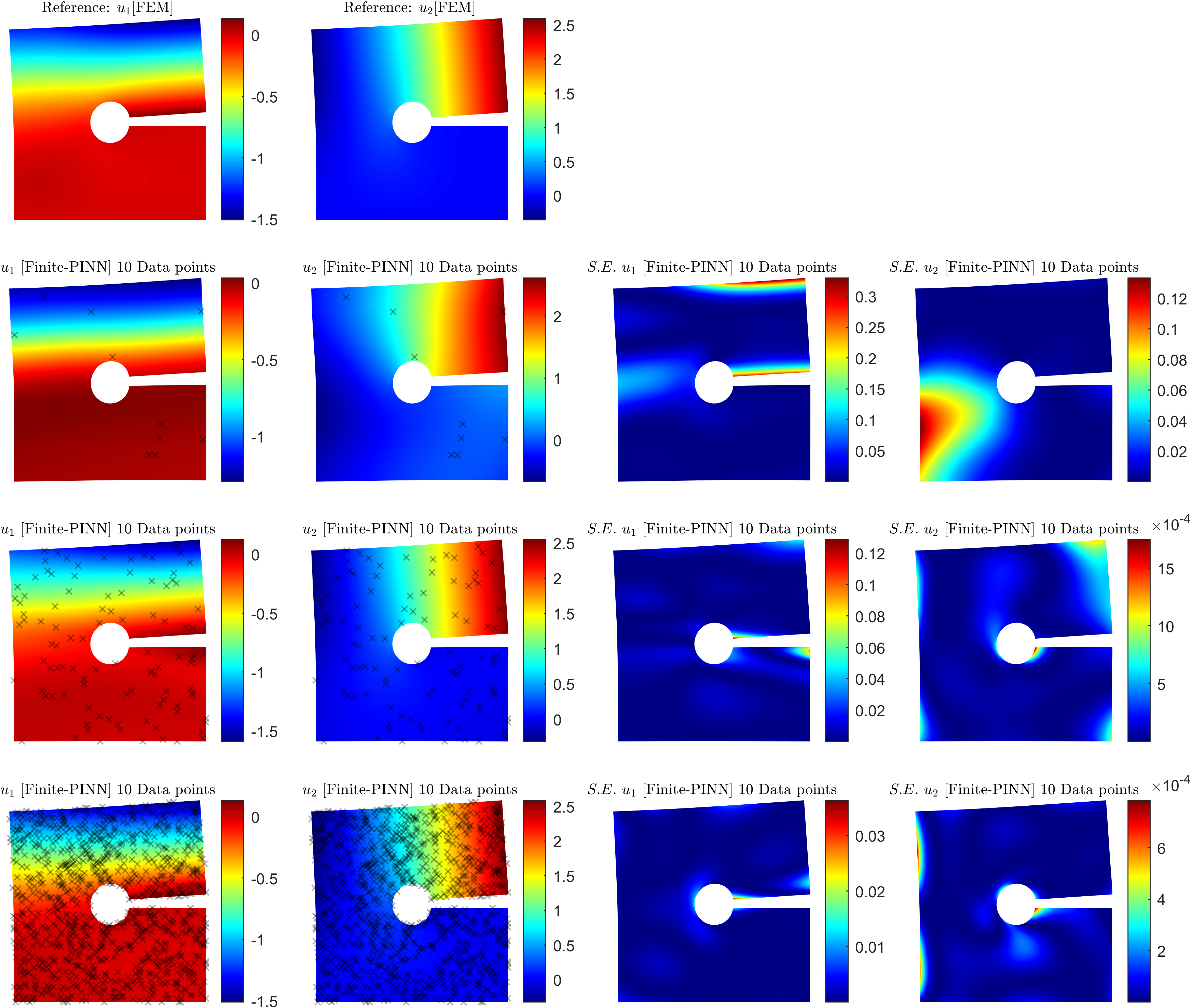}
    \caption{Visualised results of Example 4 using different numbers of sparse data points with 5\% Gaussian noise.}
    \label{fig:ondc}
\end{figure}

We here take the two inverse problems, the approximation problem of the open-notch structure (Example 4) and the shape sensing problem (Example 5), as examples to investigate the ability of Finite-PINN in fighting against noises. In each test, different levels of Gaussian Noises are applied on all the supervising data, i.e. the displacements and the strains data at the collocation points, with different collocation point numbers of the supervision data. The obtained results are given in Table.\ref{tab:ec1} and Table.\ref{tab:ec2}. Results of the open-notch problem with 5\% of data noise added on used different numbers of collocation points are visualised in Fig.\ref{fig:ondc}. The Finite-PINN model demonstrates a relatively good ability to resist noise introduced into the data. When larger amounts of data are used, the model benefits from the physics-based loss function, which acts as a physics-informed regularisation during training, helping to stabilise and improve prediction accuracy. 
\begin{table}[htbp]
\color{black}
\caption{Results of Example 4 performing with data noises.}
\centering
\begin{tabular}{ccccc}
\hline
Gaussian Noises & 0\% & 1\% & 5\% & 10\% \\
\hline
10 data points & 2.4e-4 & 1.5e-3 & 3.0e-3 & 5.2e-3 \\
100 data points & 5.0e-5 & 6.12e-5 & 3.76e-4 & 7.01e-4 \\
1000 data points & 3.7e-5 & 3.9e-5 & 1.26e-4 & 3.00e-4 \\
\hline
\end{tabular}
\label{tab:ec1}
\end{table}
\begin{table}[htbp]
\color{black}
\caption{Results of Example 5 performing with data noises.}
\centering
\begin{tabular}{cccccc}
\hline
Gaussian Noises & 0\% & 1\% & 5\% & 10\% \\
\hline
10 data points & 4.9e-4 & 2.6e-3 & 7.3e-3 & 9.2e-3 \\
100 data points & 9.0e-5 & 6.65e-4 & 3.76e-4 & 7.01e-4 \\
1000 data points & 8.7e-5 & 2.84e-4 & 3.22e-4 & 4.43e-4 \\
\hline
\end{tabular}
\label{tab:ec2}
\end{table}

\subsubsection{Uncertain Boundary conditions}
The inverse problems are usually presented with uncertain boundary conditions, especially uncertain Neumann boundary conditions in engineering. A methods are sometimes required to solve certain inverse problems when the boundary conditions or even the locations of boundary conditions are not known. In such a context, we investigate the potential of the Finite-PINN method under unknown boundary condition locations in those inverse problems.  
We take all the inverse examples utilised above as the examples for the investigation. Instead of using the hybrid LBO eigenfunctions as introduced in Section \ref{sec4} and Appendix \ref{apc}, we only utilise the LBO eigenfunctions obtained with homogeneous Neumann boundary conditions defined on all boundaries. At the same moment, the collocation points at all the boundaries are removed from the loss considerations, i.e. only interior collocation points are remained to be accounted by the physics loss. A quantitive comparison between the results obtained with and without known boundary condition locations are shown in Table.\ref{tab:ect}. 
\begin{table}[htbp]
\color{black}
\caption{Results of inverse problems with unknown boundary conditions and boundary condition positions.}
\centering
\begin{tabular}{p{5.8cm}p{4.2cm}p{5.2cm}}
\hline
Examples & Test loss (M.S.E) & Test loss (M.S.E) \\
& (Hybrid LBO eigenfunctions) & (Homogeneous LBO eigenfunctions) \\
\hline
Load identification (Example 2) & 5.6e-5 & 3.3e-1  \\
Sparse data reconstruction (Example 4) & 8.4e-5 & 1.2e-4 \\
Deformation Sensing (Example 6) & 1.7e-4 & 1.8e-4 \\
\hline
\end{tabular}
\label{tab:ect}
\end{table}

\begin{figure}[htbp!]
\color{black}
    \centering
    \includegraphics[width=0.95\linewidth]{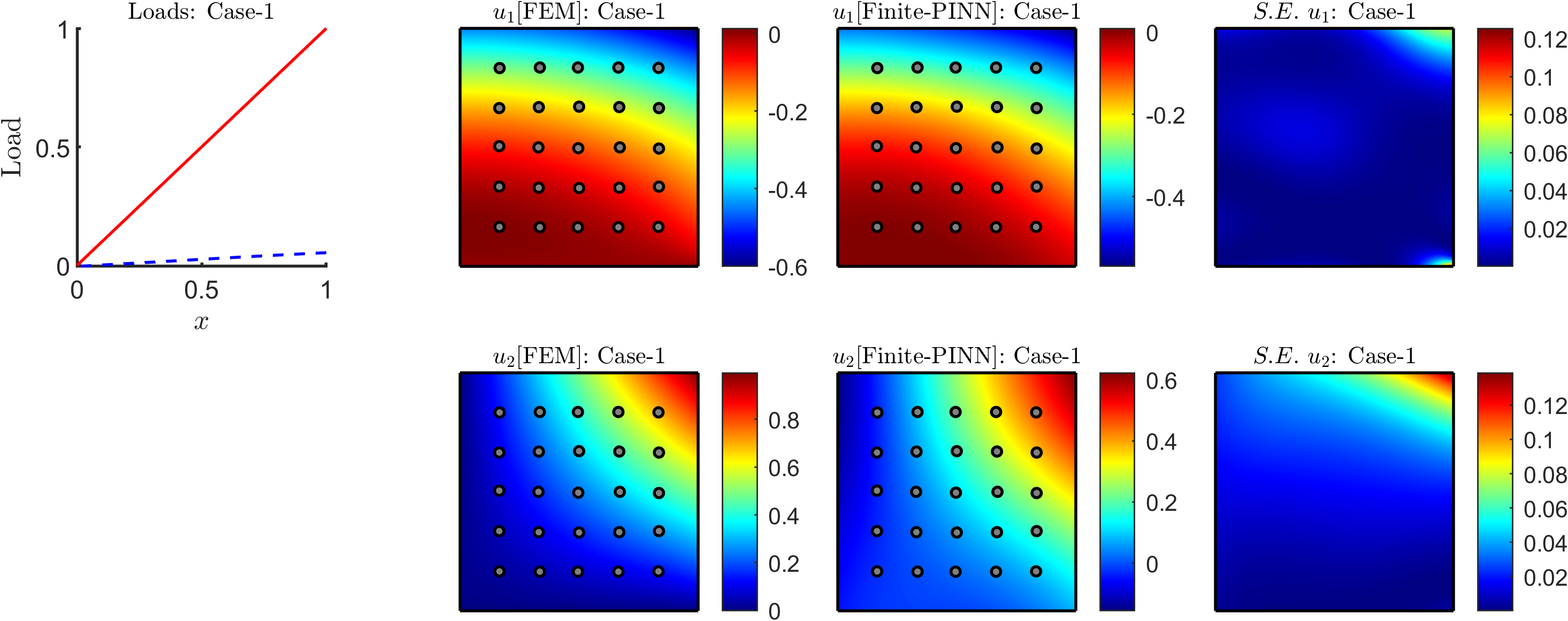}
    \caption{Finite-PINN results for the load identification problem in Example 2 with inadequate boundary definitions. In this case, the Finite-PINN model employs LBO eigenfunctions computed with purely homogeneous Neumann boundary conditions. The left subfigure shows that the Finite-PINN model is unable to identify the correct load.}
    \label{fig:unb}
\end{figure}

From those comparison, it is seen that unknown boundary conditions effect does not affect the final results much in the last two examples. It is shown that for the deformation sensing problem and the sparse field reconstruction problem, the unceratin or unknown boundary conditions didn't bother the solving of the problems much. However, for the load identification problems, the Finite-PINN methods show its limitations in accurately identifying the load applied on the structure (Fig.\ref{fig:unb}).

\subsection{Comparison with other methods} 
The Finite-PINN method is proposed as a new method for solve solid mechanics problems, and the previous studies have indicated its capability of solving both the forward and inverse problems. This subsection aims to evaluate the effectiveness and efficiency of this method compared with other method. We conduct the comparison of Finite-PINN with both traditional numerical methods i.e. the Finite Element Method and deep-learning-based methods i.e. the physical informed neural network and the Deep energy approach that have been introduced in Section \ref{sec3} that served as a good competitor of solving solid mechanics problems. 

To properly evaluate both the efficiency and effectiveness of all methods, we employ both the $L^2$ norm and the $H^1$ semi-norm losses as evaluation metrics. In addition, the energy loss, derived from the energy functional in Eq.~\ref{fnc}, is also used for performance assessment. All the cases studies in Section \ref{sec5} are conducted with this investigations and the obtained results are summarised in Table.\ref{tab:bc}. Note that the PINN model trained with the weak form loss corresponds to the Deep Energy approach in this study. 
\begin{table}[htbp]
\color{black}
\centering
\caption{Comparison of Finite-PINN with other methods}
\begin{tabular}{ccccccccc}
\hline  
\multirow{2}{*}{Cases} & \multirow{2}{*}{Forward}  & \multirow{2}{*}{Methods} 
& \multicolumn{3}{c}{Loss} & & \multicolumn{2}{c}{Computational Time} \\
\cline{4-6} \cline{8-9}
& & & L2 & H1 semi & Energy & & offline & Online \\
\hline
\multirow{3}{*}{\makecell{Example 1: \\ Case-1}} & \multirow{3}{*}{Forward} 
  &   FEM                      & 0.0     & 0.0     & -6.64e-3 & & / & 20.5s \\
  & & PINN(Strong form)        & 3.15e-2 & 9.94e-3 & -1.02e-2 & & / & $\sim$1500s \\
  & & Finite-PINN(Strong form) & 3.09e-2 & 8.79e-3 & -1.10e-2 & & / & $\sim$500s \\
\hline
\multirow{3}{*}{\makecell{Example 2: \\ Case-1}} & \multirow{3}{*}{Forward} 
  &   FEM                      & 0.0     & 0.0     & -1.18e-1 & & /    & 1.4s \\
  & & Finite-PINN(Strong form) & 4.38e-1 & 7.55e-2 & -9.57e-2 & & 1.7s & 27.3s \\
  & & Finite-PINN(weak form)   & 5.12e-1 & 8.26e-2 & -1.17e-1 & & 1.7s & 3.2s \\
\hline
\multirow{3}{*}{\makecell{Example 2: \\ Case-2}} & \multirow{3}{*}{Forward} 
  &   FEM                      & 0.0     & 0.0     & -1.25e-1 & & /    & 1.4s \\
  & & Finite-PINN(Strong form) & 3.99e-1 & 6.63e-2 & -8.44e-2 & & 1.7s & 34.9s \\
  & & Finite-PINN(weak form)   & 3.78e-1 & 5.16e-2 & -1.25e-1 & & 1.7s & 3.9s \\
\hline
\multirow{3}{*}{\makecell{Example 2: \\ Case-3}} & \multirow{4}{*}{Forward} 
  &   FEM                      & 0.0     & 0.0     & 6.0e-1 & & /    & 1.3s \\
  & & Finite-PINN(Strong form) & 3.32e-1 & 1.12e-2 & 6.2e-1 & & 1.7s & 29.9s \\
  & & Finite-PINN(weak form)   & 9.93e-1 & 8.14e-2 & 5.9e-1 & & 1.7s & 1.5s \\
\hline
\multirow{3}{*}{\makecell{Example 2: \\ Case-1}} & \multirow{3}{*}{Inverse} 
  &   FEM                      & \multicolumn{6}{l}{failed} \\
  & & Finite-PINN(Strong form) & 1.75e0 & 1.17e-2 & 1.12e-3  & & 1.7s & 24.3s \\
  & & Finite-PINN(weak form)   & \multicolumn{6}{l}{diverged}  \\
\hline
\multirow{3}{*}{\makecell{Example 2: \\ Case-2}} & \multirow{3}{*}{Inverse} 
  &   FEM                      & \multicolumn{6}{l}{failed} \\
  & & Finite-PINN(Strong form) & 2.23e0 & 1.12e-1 & -8.6e-2 & & 1.7s & 25.6s \\
  & & Finite-PINN(weak form)   & \multicolumn{6}{l}{diverged}  \\
\hline
\multirow{5}{*}{Example 3} & \multirow{5}{*}{Forward} 
  &   FEM                      & 0.0 & 0.0 & 1.12e-3 & & / & 1.4s \\
  & & PINN(Strong form)        & 2.91e-1 & 1.33e-1 & 7.12e-3 & & / & diverged \\
  & & PINN(weak form)          & 7.74e-2 & 2.32e-4 & 7.65e-4 & & / & 2.9s \\
  & & Finite-PINN(Strong form) & 7.29e-2 & 2.71e-4 & 1.78e-3 & & 1.6s & 30.5s \\
  & & Finite-PINN(weak form)   & 7.84e-2 & 2.96e-4 & 7.64e-4 & & 1.6s & 3.6s \\
\hline
\multirow{5}{*}{Example 4} & \multirow{5}{*}{Inverse} 
  &   FEM                      & 0.0 & 0.0 & -1.80e-3 & & /  & 1.4s \\
  & & PINN(Strong form)        & 1.44e0  & 2.95e0  & 2.73e-2  & & / & diverged \\
  & & PINN(weak form)          & \multicolumn{6}{l}{diverged} \\
  & & Finite-PINN(Strong form) & 3.78e-1 & 2.71e-4 & -1.72e-3 & & 1.6s & 16.7s \\
  & & Finite-PINN(weak form)   & \multicolumn{6}{l}{diverged} \\
\hline
\multirow{5}{*}{Example 5} & \multirow{5}{*}{Inverse} 
  &   FEM                      & 0.0 & 0.0 & -1.55e-2 & & /  & 1.9s \\
  & & PINN(Strong form)        & 1.12e1 & 2.95e0  & 8.89e0  & & / & diverged \\
  & & PINN(weak form)          & \multicolumn{6}{l}{diverged} \\
  & & Finite-PINN(Strong form) & 1.01e0 & 8.42e-3 & -1.56e-2 & & 1.6s & 16.7s \\
  & & Finite-PINN(weak form)   & \multicolumn{6}{l}{diverged} \\
\hline
\multirow{4}{*}{Example 6} & \multirow{4}{*}{Forward} 
  &   FEM                      & 0.0 & 0.0 & -2.17e-02 & & 0.0 & 16.6s \\
  & & PINN(Strong form)        & \multicolumn{6}{l}{diverged} \\
  & & PINN(weak form)          & 5.79e-2 & 1.71e-3 & -2.34e-02 & & / & 26.3s \\
  & & Finite-PINN(Strong form) & \multicolumn{6}{l}{diverged} \\
  & & Finite-PINN(weak form)   & 5.86e-2 & 1.83e-3 & -2.34e-02 & & 2.8s & 17.3s \\
\hline
\multirow{4}{*}{Example 7} & \multirow{4}{*}{Forward} 
  &   FEM                      & 0.0 & 0.0 & -2.17e-02 & & / & 16.6s \\
  & & PINN(Strong form)        & 1.50e2  & 1.75e1  & -2.17e-02 & & /  & diverged \\
  & & PINN(weak form)          & 8.70e1  & 6.60e2  & -2.34e-02 & & / & 39.6s \\
  & & Finite-PINN(Strong form) & 3.22e-1 & 8.83e-4 & -2.17e-02 & & 2.8s  & 98.0s \\
  & & Finite-PINN(weak form)   & 2.21e-1 & 1.27e-3 & -2.34e-02 & & 2.8s & 19.4s \\
\hline
\end{tabular}
\label{tab:bc}
\end{table}

Several observations can be drawn from the table. Comparing the Finite-PINN and traditional PINN models, it is evident that the Finite-PINN generally achieves better performance in both accuracy and efficiency. Specifically, for problems defined on complex geometries, such as Examples 4, 5, and 6, the traditional PINN models fail to capture the correct solution direction, resulting in divergence. In contrast, the Finite-PINN consistently produces accurate results in these cases.

In terms of computational time, the Finite-PINN model typically exhibits faster convergence than the traditional PINN model for both forward and inverse problems. However, the Finite-PINN requires an additional offline process to compute the LBO eigenfunctions and their derivatives, which increases the total time required to solve a PDE. On the other hand, the LBO eigenfunctions are specific to a given geometry, meaning that once calculated for a particular geometry or structure, they can be reused for any problem defined on that geometry when using the Finite-PINN model.

Comparing the use of strong form and weak form losses, it is observed that the weak form loss performs better in solving forward problems. Training with the weak form loss achieves faster convergence in these cases compared to the strong form loss. However, for inverse problems, attempts to use the weak form loss consistently fail to produce satisfactory results. This is primarily because inverse problems often involve unknown or uncertain boundary conditions and PDE parameters, rendering the energy-based variational loss ineffective in accurately representing the PDE in functional space. This highlights the necessity of using the strong form loss in PINNs for inverse problems, where the results show good agreement with the reference solutions.

Additionally, when comparing the PINN models with conventional numerical methods such as FEM, it is found that the PINN and Finite-PINN models using the weak form loss achieve comparable solution times to FEM in forward problems. However, for inverse problems, FEM also struggles to provide accurate solutions. This is due to the fact that FEM, like the weak form PINNs, is based on the weak form of the governing PDE, which limits its effectiveness in solving inverse problems.

\color{black}

\section{Conclusion}
\label{sec7}
This work proposes a novel neural network architecture aimed at solving solid mechanics problems. An investigation into using PINN for solving solid mechanics problems is conducted through a comparison with the finite element method (FEM). Two main challenges that limit the application of PINN in general solid mechanics problems are identified: a) PINN generates solutions defined over an infinite domain, which conflicts with the finite domain of most solid structures; and b) The Euclidean space is not a suitable space for PINN to effectively learn or solve the solution field of a solid mechanics problem. The proposed physics-informed neural network adopts a novel two-part architecture that separates the learning of the stress field and the displacement field, demonstrating excellent performance in addressing different solid mechanics problems. Both forward and inverse solid mechanics problems are tested using the Finite-PINN method, and it is found that the Finite-PINN model is either superior in efficiency for simpler problems, provides higher approximation accuracy, or is capable of solving problems with complex structures where the traditional PINN model fails.

The proposed Finite-PINN model retains the original representation of the initial PINN model, where the displacement field to be solved is still represented by a function, but now exists within a Euclidean-Topological space rather than the original Euclidean space. This allows the Finite-PINN model to maintain the key benefits of the traditional PINN model, such as easier implementation, reduced memory requirements, data fusion capabilities. Trainings with Both the strong form and weak form losses are included in this work. The examples and case studies presented in the paper demonstrate the feasibility of the Finite-PINN model. Training with both the strong form and weak form losses is included in this work. The examples and case studies presented demonstrate the feasibility of the Finite-PINN model. The weak form Finite-PINN model performs well in solving forward problems, while the strong form Finite-PINN model is capable of solving inverse problems where traditional numerical models and even conventional PINNs fall short.

Despite these advantages, the Finite-PINN model has some drawbacks. It requires an additional offline stage to prepare the LBO eigenfunctions and their derivatives for use in the online stage. This prepared dataset is specific to a particular structure, limiting the generalisation ability of the Finite-PINN model to problems defined on the same domain. Additionally, for the strong form model, certain inverse problems with unknown boundary condition locations—such as load identification problems—expose weaknesses in the Finite-PINN’s ability to solve them effectively.

In addition, this paper utilised the simplest finite element method to calculate the required LBO eigenfunctions and their derivatives using linear triangular or tetrahedral elements as an initial attempt. However, this approach is not ideal, as the elements are only first-order, which can sometimes lead to unsatisfactory second-order derivatives of the LBO eigenfunctions. In this context, employing higher-order finite element types could potentially improve the performance of the Finite-PINN model. Furthermore, to overcome the mesh sensitivity issue inherent in the finite element method, researchers could consider using other numerical approaches, such as mesh-free methods, to calculate the required values \cite{li2002meshfree,chen2006meshless} as introduced in Appendix \ref{apb}. The primary objective of the numerical calculation is to obtain the LBO eigenfunction values and their derivatives at the collocation points in the offline stage. Regardless of the numerical methods used, the online stage is solely dedicated to training the neural network, i.e. to solve PDE problems.

Returning to the main point, similar to the initial PINN model, the proposed Finite-PINN model presents not only a neural network architecture but also an innovative approach focusing on using deep learning in solid mechanics problems. The idea behind the Finite-PINN model incorporates not only the PDE information but also the topological information of the studied structure. In this context, the finite-element-inspired architecture may also be applicable to other problems involving deep learning models for solving solid mechanics problems. For example, the Finite-PINN model could potentially be extended to operator learning models to learn solutions for a class of solid mechanics problems, focusing on a specific solid structure and fixed loading positions, with various inhomogeneous boundary conditions. These developments are beyond the scope of this paper but may be explored in future work.

\section*{Acknowledgments}
The project has been funded by the European Union Program Horizon Europe under grant agreement no. 101079091.

\appendix
\section{Finite element Formulation of the LBO eigenproblem}
\label{apa}
The weak formulation of Eq.\ref{LBO} is similar to that of the Poisson equation. By rearranging the terms in Eq.\ref{LBO} and integrating after multiplying by a trial function from the functional space over the entire domain, we obtain:
\begin{equation}
-\int_{\Omega} \boldsymbol{v}\left(\boldsymbol{x}\right)\left(\Delta u\left(\boldsymbol{x}\right)+\lambda u\left(\boldsymbol{x}\right)\right) \mathrm{d} \boldsymbol{x}=0,
\label{ef3}
\end{equation}
where $\boldsymbol{v}\left(\boldsymbol{x}\right)$ is a trial function, and since Gauss's Theorem, 
\begin{equation}
\int_{\partial \Omega} \boldsymbol{n} \nabla \boldsymbol{u} \boldsymbol{v} d \boldsymbol{x} = \int_{\Omega} \operatorname{div}\left(\nabla \boldsymbol{u} \boldsymbol{v} \right) d \boldsymbol{x}=\int_{\Omega} \Delta \boldsymbol{u} \boldsymbol{v} + \nabla \boldsymbol{u} \nabla \boldsymbol{v} d \boldsymbol{x},
\label{ef4}
\end{equation}
the weak form of Eq.\ref{LBO} can be obtained as
\begin{equation}
\int_{\Omega}\left(\nabla \boldsymbol{v} \cdot \nabla \boldsymbol{u}-\lambda \boldsymbol{v} \boldsymbol{u}\right) \mathrm{d} \boldsymbol{x} - \int_{\partial \Omega} \boldsymbol{v} \nabla \boldsymbol{u} \cdot \boldsymbol{n} \mathrm{d} \boldsymbol{x}=0,
\label{ef5}
\end{equation} 
which can be finally simplified to:
\begin{equation}
\int_{\Omega}\left(\nabla \boldsymbol{v} \cdot \nabla \boldsymbol{u}-\lambda \boldsymbol{v} \boldsymbol{u}\right) \mathrm{d} \boldsymbol{x} =0,
\label{ef6}
\end{equation} 
since the last term vanishes due to the applied free Neumann boundary condition on all boundaries. The solution of Eq.\ref{ef6} is defined within the Sobolev space. The variational problem can then be expressed using bilinear forms as follows:
\begin{equation}
\text { find } u \in H_{0}^{1}: \quad
a\left(\boldsymbol{v}, \boldsymbol{u}\right) - \lambda b\left(\boldsymbol{v}, \boldsymbol{u}\right) = 0 \quad \forall \boldsymbol{v} \in H_{0}^{1},
\label{ef7}
\end{equation} 
where $a\left(\cdot,\cdot\right)$ and $b\left(\cdot,\cdot\right)$ are two utilised bilinear forms where $a\left(\boldsymbol{v}, \boldsymbol{u}\right) = \int_{\Omega} \nabla \boldsymbol{v} \cdot \nabla \boldsymbol{u} \, \mathrm{d} \boldsymbol{x} $ and $ b\left(\boldsymbol{v}, \boldsymbol{u}\right) = \int_{\Omega} \boldsymbol{v} \boldsymbol{u} \, \mathrm{d} \boldsymbol{x} $, and the solution is in the Hilbert space $H_{0}^{1}$. Eq.\ref{ef7} can be solved using the Galerkin method, which transforms the problem into a finite-dimensional subspace $V_{h}$, where $V_{h} \subset H_{0}^{1}$:
\begin{equation}
\text { find } \boldsymbol{u}_{h} \in V_{h}: \quad a \left(\boldsymbol{v}_{h}, \boldsymbol{u}_{h}\right) = \lambda b \left(\boldsymbol{v}, \boldsymbol{u} \right) \quad \forall \boldsymbol{v}_{h} \in V_{h}
\label{ef8}
\end{equation} 

In this work, the finite element method is employed to solve Eq.\ref{ef8} by approximating the solution by selected bases in $V_{h}$: 
\begin{equation}
\boldsymbol{N} = N_1, N_2, N_3, ... N_i, ... N_h.
\label{ef9}
\end{equation} 
The number of basis functions $h$ is equal to the dimension of the functional space $V_{h}$, which also corresponds to the number of nodes in a finite element model. By expressing both $\boldsymbol{u}$ and $\boldsymbol{v}$ as linear combinations of these basis functions, the finite element problem can be formulated as:
\begin{equation}
\text { find } \boldsymbol{u} \in \mathbb{R}^{h}: \quad a\left(\sum_{i} N_{i} u_{i} , \sum_{j} N_{j} v_{i} \right) = \lambda b\left(\sum_{i} N_{i} u_{i} , \sum_{j} N_{j} v_{i} \right) \quad \forall v_{i} \in \mathbb{R}^{h}
\label{ef10}
\end{equation} 
The Galerkin method takes the basis functions themselves as trial functions:
\begin{equation}
\text { find } \boldsymbol{u} \in \mathbb{R}^{h}: \quad a\left(\sum_{i} N_{i} u_{i} , N_{j} \right)=  \lambda b\left(\sum_{i} N_{i} u_{i} , N_{j} \right) \quad \forall j \in 1, \ldots, h
\label{ef11}
\end{equation} 
And since $a\left(\cdot,\cdot\right)$ and $b\left(\cdot,\cdot\right)$ are bilinear functions, the problem can be rewritten as
\begin{equation}
\text { find } \boldsymbol{u} \in \mathbb{R}^{h}: \quad \sum_{i=1}^{N} a\left(N_{i}, N_{j}\right) \boldsymbol{u}_{i} = \lambda \sum_{i=1}^{N} b\left(N_{i}, N_{j}\right) \boldsymbol{u}_{i}
\label{ef12}
\end{equation} 
To rewrite the equation in vector form, we obtain the final eigenvalue problem to solve:
\begin{equation}
\quad \mathbf{K} \boldsymbol{u}= \lambda \mathbf{M} \boldsymbol{u}
\label{ef13}
\end{equation} 
where $\mathbf{K} = \sum_{i=1}^{N} a\left(N_{i}, N_{j}\right)$ represents the stiffness matrix, and $\mathbf{M} = \sum_{i=1}^{N} b\left(N_{i}, N_{j}\right)$ represents the mass matrix. The mass matrix can be set as identical matrix in this problem. In the classical finite element method, $\boldsymbol{N}$ denotes the shape functions of the elements.

The LBO eigenfunctions $\boldsymbol{\phi}$ for a general geometry or structure are obtained by solving the eigenvalue problem given in Eq.\ref{ef13}. The resultant LBO eigenfunctions are represented in a discrete form by their values at the nodes in a finite element model, denoted as $\phi_{ki}$, where $i = 1, \ldots, h$ and $k = 1, \ldots, n$, with $h$ being the number of nodes and $n$ being the number of basis functions. The required first-order and second-order derivatives of $\boldsymbol{\phi}$ can then also be obtained:
\begin{equation}
\begin{aligned}
\phi_{k,j} = & N_{i,j} \phi_{ki}, \quad k=1,\ldots,n, \quad k=1,\ldots,h \\
\phi_{k,jj} = & N_{i,j} \phi_{ki,j}, \quad k=1,\ldots,n, \quad i=1,\ldots,h  \\
\end{aligned}
\label{ef14}
\end{equation} 
$N_{i,j}$ represents the gradient tensor of the shape function, which is typically denoted by $\boldsymbol{B}$:
\begin{equation}
\boldsymbol{B} = \nabla \boldsymbol{N}
\label{ef15}
\end{equation} 
The definitions of shape functions vary for different types of elements. In this work, the linear triangular element is used for the 2D formulation, and the tetrahedral element is used for the 3D formulation.

\color{black}
\section{RBF-based Meshfree method Formulation of the LBO eigenproblem}
\label{apb}
Similar to the Finite element formulation, the meshfree method could also be employed to solve Eq.\ref{ef8} by approximating the solution using radial basis functions selected in the approximation space \( V_{h} \):
\begin{equation}
\boldsymbol{R} = R_1, R_2, R_3, \ldots, R_i, \ldots, R_h,
\label{mf1}
\end{equation}
where each \( R_i \) is an RBF centred at node \( i \), such as Gaussian, multiquadric, or inverse multiquadric functions, centred at nodal points, replacing the classical finite element shape functions.

The multiquadric radial basis function \( R_i(\mathbf{x}) \) centred at node \( \mathbf{x}_i \) is defined as:
\begin{equation}
R_i(\mathbf{x}) = \sqrt{ \left(\mathbf{x} - \mathbf{x}_i \right)^2 + c^2 },
\label{mq_rbf}
\end{equation}
where \( \mathbf{x} \) is the evaluation point, \( \mathbf{x}_i \) is the centre of the basis function, and \( c > 0 \) is a shape parameter controlling the flatness of the function.

The number of basis functions \( h \) corresponds to the number of nodes (or centres) in the meshfree model. By expressing both \( \boldsymbol{u} \) and \( \boldsymbol{v} \) as linear combinations of these RBFs, the meshfree formulation of the problem can be written as:
\begin{equation}
\text{find } \boldsymbol{u} \in \mathbb{R}^h: \quad a\left(\sum_i R_i u_i, \sum_j R_j v_j \right) = \lambda b\left(\sum_i R_i u_i, \sum_j R_j v_j \right) \quad \forall v_j \in \mathbb{R}^h
\label{mf2}
\end{equation}
Using the Galerkin method, the basis functions themselves are chosen as test functions:
\begin{equation}
\text{find } \boldsymbol{u} \in \mathbb{R}^h: \quad a\left(\sum_i R_i u_i, R_j \right) = \lambda b\left(\sum_i R_i u_i, R_j \right) \quad \forall j = 1, \ldots, h
\label{mf3}
\end{equation}
Due to the bilinearity of \( a(\cdot,\cdot) \) and \( b(\cdot,\cdot) \), the above can be expressed as a system of algebraic equations:
\begin{equation}
\sum_{i=1}^h a(R_i, R_j) u_i = \lambda \sum_{i=1}^h b(R_i, R_j) u_i
\label{mf4}
\end{equation}
Similar to the FEM, the eigenvalue problem becomes the follows in matrix form:
\begin{equation}
\mathbf{K} \boldsymbol{u} = \lambda \mathbf{M} \boldsymbol{u}
\label{mf5}
\end{equation}
where the stiffness matrix \( \mathbf{K} = \left[a(R_i, R_j)\right] \) assembled using the meshfree RBF basis functions, and the mass matrix $\mathbf{M}$ can be set as identical matrix in this problem.

The first-order derivative of the radial basis function Eq.\ref{mf2} is:
\begin{equation}
R_{i,j}(\mathbf{x}) = \frac{x_j - x_{ij}}{R_i(\mathbf{x})}
\label{mq_rbf2}
\end{equation}

\section{Hybrid LBO eigenfunctions}
\label{apc}
The entire domain boundary of a given problem, $\partial \boldsymbol{\Omega}$, is composed of three parts as defined in Fig.\ref{fig:p3}(b):
\begin{equation}
\partial \boldsymbol{\Omega} = \partial \boldsymbol{\Omega}_d \cup \partial \boldsymbol{\Omega}_n \cup \partial \boldsymbol{\Omega}_f,
\end{equation}
and
\begin{equation}
\partial \boldsymbol{\Omega}_d \cap \partial \boldsymbol{\Omega}_n \cap \partial \boldsymbol{\Omega}_f = \varnothing.
\end{equation}
where $\partial \boldsymbol{\Omega}_d$ represents the boundary segment assigned with Dirichlet boundary conditions, $\partial \boldsymbol{\Omega}_n$ represents the boundary segment applied with Neumann boundary conditions, and $\partial \boldsymbol{\Omega}_f$ denotes the free boundaries. The hybrid LBO eigenfunctions are computed by combining the solutions of two LBO eigenvalue problems: a general problem and a specific problem. The general problem is as defined by Eq.\ref{LBO}, and the specific problem is defined as:
\begin{equation}
\left\{
\begin{aligned}
-\Delta u\left(\boldsymbol{x}\right)=\lambda u\left(\boldsymbol{x}\right) & \quad \forall \quad \boldsymbol{x} \in \mathbf{\Omega}, \\ 
-\nabla u \cdot \boldsymbol{n}=\boldsymbol{0}  & \quad \forall \quad  \boldsymbol{x} \in \partial \mathbf{\Omega}_f, \\
u = \boldsymbol{0}  & \quad \forall \quad  \boldsymbol{x} \in \partial \mathbf{\Omega}_d \cup \partial \mathbf{\Omega}_n.
\end{aligned}
\right.
\label{LBO2}
\end{equation}
with respect to a specific structure and specific locations of applying boundary conditions. If the solution of Eq.\ref{LBO} is denoted by $\leftindex^{g}{\boldsymbol{\phi}}$ and the solution of Eq.\ref{LBO2} is denoted by $\leftindex^{s}{\boldsymbol{\phi}}$, the hybrid LBO eigenfunctions $\boldsymbol{\phi}$ are then obtained as:
\begin{equation}
\boldsymbol{\phi} = \leftindex^{g}{\boldsymbol{\phi}} + \leftindex^{s}{\boldsymbol{\phi}}
\label{phi}.
\end{equation}
It can be observed that the specific problem Eq.\ref{LBO2} is formulated by assigning zero Dirichlet boundary conditions on the non-free boundaries. The hybrid LBO eigenfunctions can handle problems where arbitrary boundary conditions are applied at fixed boundary locations. The proof is as follows:
\begin{proof}
The solution of Eq.\ref{LBO}, $\leftindex^{g}{\boldsymbol{\phi}}$ maintains all its boundaries as free boundaries, i.e.:
\begin{equation}
-\nabla \leftindex^{g}{\boldsymbol{\phi}} \cdot \boldsymbol{n}=0  \quad \forall \quad  \boldsymbol{x} \in \partial \mathbf{\Omega}.
\label{cpn1}
\end{equation}
The solution of Eq.\ref{LBO2}, $\leftindex^{s}{\boldsymbol{\phi}}$, satisfies the following boundary constraints due to the assigned Neumann and Dirichlet boundary conditions:
\begin{equation}
\begin{aligned}
  & -\nabla \leftindex^{s}{\boldsymbol{\phi}} \cdot \boldsymbol{n}=\boldsymbol{0}  & \quad \forall & \quad  \boldsymbol{x} \in \partial \mathbf{\Omega}_f, \\
-\nabla \leftindex^{s}{\boldsymbol{\phi}} \cdot \boldsymbol{n}=\boldsymbol{0} \ \ \quad \& & \quad -\nabla u \cdot \boldsymbol{\tau}=\boldsymbol{0}   & \quad \forall & \quad \boldsymbol{x} \in \partial \mathbf{\Omega}_d \cup \partial \mathbf{\Omega}_n,  \\
\end{aligned}
\label{cpn2}
\end{equation}
where $\boldsymbol{\tau}$ is the tangent vector on the boundary, orthogonal to the normal vector $\boldsymbol{n}$. The second constraint in Eq.\ref{cpn2} holds because the boundaries $\mathbf{\Omega}_d$ are assigned fixed Dirichlet boundary conditions. By taking the intersection of the two constraints, we obtain the constraints that are always satisfied by the hybrid solution $\boldsymbol{\phi} = \leftindex^{g}{\boldsymbol{\phi}} + \leftindex^{s}{\boldsymbol{\phi}}$:
\begin{equation}
\begin{aligned}
 -\nabla \left(\leftindex^{g}{\boldsymbol{\phi}} + \leftindex^{s}{\boldsymbol{\phi}} \right) \cdot \boldsymbol{n}=\boldsymbol{0} & \quad \forall & \quad  \boldsymbol{x} & \in \partial \mathbf{\Omega}_f \cap \partial \mathbf{\Omega} \\
 -\nabla \left(\leftindex^{g}{\boldsymbol{\phi}} + \leftindex^{s}{\boldsymbol{\phi}} \right) \cdot \boldsymbol{\tau}=\boldsymbol{0} & \quad \forall & \quad  \boldsymbol{x} & \in \varnothing \cap \left(\partial \mathbf{\Omega}_f \cap \partial \mathbf{\Omega}\right)\\
\end{aligned}
\label{cpn3}
\end{equation}
which can be obtained as:
\begin{equation}
\begin{aligned}
 -\nabla \boldsymbol{\phi}  \cdot \boldsymbol{n}=\boldsymbol{0} & \quad \forall & \quad  \boldsymbol{x} & \in \mathbf{\Omega}_f, \\
 -\nabla \boldsymbol{\phi}  \cdot \boldsymbol{\tau}=\boldsymbol{0} & \quad \forall & \quad  \boldsymbol{x} & \in \varnothing. \\
\end{aligned}
\label{cpn4}
\end{equation}
This implies that the hybrid LBO eigenfunctions exclusively satisfy the condition $-\nabla \boldsymbol{\phi} \cdot \boldsymbol{n} = 0$ on the free boundaries, without affecting other derivative properties.
\end{proof}
\color{black}

\bibliographystyle{unsrt}  
\bibliography{templateArxiv}

\end{document}